\documentclass[11pt]{article}

\usepackage{amsmath, amsthm, amssymb, mathtools}
 \usepackage{graphicx}

\usepackage{setspace}

\usepackage[colorlinks, citecolor=blue, linkcolor=blue]{hyperref}
\usepackage{natbib}
\usepackage{fullpage}
\usepackage{dcolumn}
\usepackage{booktabs}

\usepackage{colortbl}

\usepackage{xcolor}

\usepackage{color}

\usepackage{fvextra}

\usepackage{csquotes}

\usepackage{relsize}

\newcommand{\blue}{\textcolor{blue}}
\newcommand{\red}{\textcolor{red}}
\newcommand{\orange}{\textcolor{orange}}
\newcommand{\green}{\textcolor{green}}

\newcommand{\is}{$i$'s }

\newcommand{\sq}{\ensuremath{\subseteq}}

\newtheorem{Result}{Result}
\newtheorem{Definition}{Definition}

\newtheorem{Theorem}{Theorem}
\newtheorem{Proposition}{Proposition}

\newtheorem{Hypothesis}{Hypothesis}

\usepackage{eurosym}

\usepackage{subcaption}

\usepackage{setspace}

\usepackage{makecell}
\usepackage{booktabs}
\usepackage{array}
\usepackage[utf8]{inputenc} 
\usepackage[T1]{fontenc}
\usepackage{ltxtable}

\usepackage{fancyvrb}
\usepackage{pdfpages}

\begin{document}

\title{Information Aggregation with AI Agents
\thanks{I would like to thank Andis Sofianos, Diego Marino-Fages, Michael Vlassopoulos, Alessandro Mennuni, Christos Ioannou and participants at Durham, Southampton, the Durham Edinburgh York workshop, DIRDI Impact and Invention Conference, and the AI in Finance and Accounting workshop for useful comments. This research is based on work funded under ESRC grant ES/V004425/1.}\\ 
\vspace{0.9cm}
\author{Spyros Galanis\thanks{Department of Economics, University of Durham, spyros.galanis@durham.ac.uk.}}
}

\maketitle

\begin{abstract}

Can Large Language Models (AI agents) aggregate dispersed private information through trading and reason about the knowledge of others by observing price movements? We conduct a controlled experiment where AI agents trade in a prediction market after receiving private signals, across four information structures of increasing complexity. We find that although the median market is effective at aggregating information in the easy information structures, performance deteriorates in the harder structures, suggesting that AI agents struggle in environments where more than two levels of interactive reasoning are required, a ceiling close to the one documented in human subjects. Consistent with our theoretical predictions, market accuracy does not improve from allowing cheap talk communication, changing the duration of the market, or strategic prompting; initial price has little average effect but matters in the very hard structure. We also find that ``smarter'' AI agents perform better at aggregation and are more profitable. Surprisingly, giving them feedback about past performance does not improve aggregation. A further wave of markets, run three months later with capability-frontier models, aggregates information more often in the three easier structures but not in the hardest one, where higher capability replaces markets that are confidently wrong with markets that hedge near 0.5. 



 \vspace{0.2cm}


\end{abstract}

\section{Introduction}
Recent advancements in Large Language Models (LLMs) have enabled the development of AI agents capable of autonomously executing complex tasks. They can gather private information and reason about necessary steps based on their prompts, execute actions by invoking external tools and collaborating with other agents, and evaluate the success of their actions. Soon, they will be tasked to trade securities in financial markets, leveraging their private information.%
\footnote{Robinhood, one of the biggest retail trading platforms in the U.S., announced in May 2026 that customers can connect their own AI agents to their account to automate trading and credit card purchases.}
\cite{shengEtAl2026} show that hedge funds that adopt generative AI earn 2-4\% higher annualized abnormal returns than non-adopters.
Firms are already delegating substantive work to them, and the evidence suggests that the gains are uneven across tasks. In a field experiment inside a large consulting firm, \cite{dellacquaEtAl2026} find sizeable improvements in quality and speed on tasks that lie inside the capability frontier of the model, but a deterioration in performance on tasks that lie just outside it. 

However, a critical open question remains: to what extent can they reason about the private information of others (humans or AI agents) when observing their actions (buy or sell orders)? This is fundamental for deploying AI agents in financial markets, otherwise information may not get aggregated and prices may diverge from the real value of assets. As \cite{hayek45} argues, a well-functioning pricing system that aggregates information is necessary to achieve efficient outcomes. 
Moreover, prices in secondary markets affect real outcomes through their informational role: managers, capital providers, and regulators learn from prices and use them to guide their decisions, hence less informative prices feed back into worse outcomes \citep{bondEtAl2012}.

We study information aggregation by running a controlled experiment with AI agents who receive private information and trade a security in a prediction market, which pays 0 or 1 based on the outcome of a Yes/No question. The conjunction of everyone's information reveals the true value of the security. Is trading activity enough to drive the price of the security close to its true value, so that information aggregates? 

The intuition behind information aggregation is simple. If the price of a security is high and a trader thinks its value is low, he will sell, otherwise he will buy. The other traders observe the price movements and try to infer the private information that led to the buy or sell orders. After incorporating this (now public) information into their own private information, traders will buy or sell and the process continues, until all private information is aggregated. However, this aggregation requires that traders are sophisticated enough to form higher order beliefs about what others know, what they know about what others know, and so on, in order to interpret their actions and update their knowledge and beliefs.

Reasoning about the private information of others by observing their actions is a fundamental human ability, but it is not obvious that AI agents are trained to emulate it. 
On the one hand, they already demonstrate impressive capabilities on several domains and Large Language Models (LLMs) can be prompted to optimise \citep{yangEtAl2023}. On the other hand, partly because they are trained using Reinforcement Learning from Human Feedback (RLHF), they can emulate human behaviour in games \citep{parkEtAl2023} and experiments \citep{horton2023}, which is not always sophisticated,  and suffer from the same behavioral biases as humans \citep{biniEtAl2025}.

We vary several conditions in the experiment and measure their impact on information aggregation, or market accuracy. The first is the complexity of the information and payoff structure. There are three traders and three signals, and we consider four levels of difficulty.  In the easiest, each trader needs to reason about the signal of only one other trader to determine whether the answer to the question is Yes or No. The hardest is a version of the well-known ``dirty faces'' puzzle, where each trader gets two signals so interactive reasoning is more complex \citep{weber2001}. The other conditions are allowing AI agents to post public comments, to measure the effect of cheap talk communication; prompting them to be strategic (forward-looking) or myopic to measure whether AI agents can implement a strategy over many periods; changing the initial price (0.3,0.5,0.7) to examine whether price manipulation has an adverse effect; and altering the duration of the market (3, 6, and 9 rounds) to examine if more trading helps or confuses AI agents.%
%
These treatments generate a total of 144 different market configurations, and we run each configuration at least 12 times, with a diverse set of AI agents, generating 1772 prediction markets in total. We then run an information provision treatment, where we inform AI agents, before they trade, about the qualitative results from the first wave, on which factors were effective on information aggregation and profits, generating another 1728 prediction markets.

For all four information structures, we construct securities which are `separable'.%
\footnote{Informally, a security is separable if there is no prior belief and candidate price at which every trader's conditional expectation of the security equals that price while its value still varies across the states the prior considers possible; at any uninformative price, some trader has information that moves his expectation away from it. Appendix \ref{separability} gives the formal definition.}
Using a result from \cite{ostrovsky12}, this allows us to have a sharp theoretical prediction: if AI agents are rational and sophisticated then information will get aggregated in all Nash equilibria, for any initial price and irrespective of whether the traders are myopic or strategic, or whether they communicate.

We employ eight LLMs for our main analysis: Claude Haiku 3.5 and 4.5, Gemini 2.5 and 3 Flash, GPT-4o and GPT-5 mini,  gemma3:4b, and qwen3:8b. As a measure of each model's capabilities, we adopt the Artificial Analysis Intelligence Index \citep{ArtificialAnalysisTeam2025}, one of the most comprehensive publicly available syntheses of model capabilities. The intelligence index integrates ten evaluation suites, combining performance across reasoning, mathematics, coding, and agentic workflow tasks, among others. We form twelve teams of three traders. Eight teams are homogeneous, comprising traders using the same model.  The remaining four teams feature a variety of models, enabling us to test whether ``diversity'' of intelligence has an impact on information aggregation or profits. 

Our first main finding (Result \ref{res:info deteriorates}) is that, although the median market  is effective in the two easy structures, the more complex environments we study are associated with worse information aggregation.%
\footnote{The average market fares very poorly across all structures, but this is driven mostly by the performance of the upper 20\% tail of the log-error distribution.}
In particular, the median market across all structures assigns probability 0.91 to the correct outcome, that is, it prices the winning security (the one that eventually pays 1, whether Yes or No) at 0.91. In the Easy and Medium structures this probability is almost 1, suggesting that AI agents understand them completely. In the Hard structure, it drops to 0.73, hence noticeably worse but still better than random guessing. In the Very Hard structure, however, it drops to 0.5, which is completely uninformative.  Given that the securities are separable in all structures, our hypothesis that complexity does not influence information aggregation is rejected.  These findings align with related human experiments \citep{nagel1995, weber2001, kneeland2015} demonstrating that most participants perform up to two steps of iterated reasoning.

%
%


We conducted our baseline experiment in January 2026. However, given the rapid pace of innovation in artificial intelligence, model architectures are updated almost quarterly. To assess external validity, we conducted an out-of-sample extension of 576 markets in April 2026 using three newly released capability-frontier models: GPT-5.4, Claude Opus 4.6, and Gemini 3.1 Pro (Section \ref{sec:robustness frontier}). Although we find that frontier models are better at aggregating information in the three easier structures, in the Very Hard structure they perform similarly to the first-wave models, where capability changes the composition of the failures rather than their frequency: markets that were confidently wrong are replaced by markets that hedge near 0.5 (Result \ref{res:frontier shape}). This suggests that the deterioration we document in Result \ref{res:info deteriorates} is not an artifact of the models available when the first wave was run. 

Our second main finding is to confirm our hypothesis that the following factors do not have a statistically significant impact on information aggregation: prompting AI agents to be strategic versus myopic, cheap talk communication, and changing the duration of the market (Results \ref{result: strategic}, \ref{result: comments}, \ref{result: duration}). The initial price has no detectable average effect, although it matters in the Very Hard structure (Result \ref{result:initial price}).

Our third finding concerns the role of agent intelligence in information aggregation and profits. Given that interactive reasoning requires sophisticated agents, our hypothesis was that intelligence would positively impact information aggregation. Indeed, we find that the average intelligence of the group is associated with a lower average log error (Result \ref{result: average intelligence}). To quantify this result, we ran two quantile regressions, using the median and the upper 20\% tail of the log-error distribution. Although the tail log error is reduced, the impact on the median is statistically insignificant. This result suggests that the role of intelligence is not to improve the performance of the typical market, but to avoid the few markets that misprice the security completely,  trading at 1 (0) when the true value is 0 (1). Moreover, we find that the individual profits of an AI agent increase with his intelligence  but decrease as the average intelligence of the group increases. These results are consistent with previous experiments with humans \citep{corgnetEtAl2018}, yet the Intelligence Index does not specifically test for interactive reasoning tasks, which is surprising.

An important question is whether AI agents can leverage feedback from past play to improve future performance. To examine this, we run a later-wave information provision treatment in which a new wave of AI agents receives qualitative summaries based on the empirical outcomes of the previous 1,772 markets, using the same model versions as in the first wave. Contrary to recent literature suggesting that LLMs successfully improve when given performance histories \citep{yangEtAl2023}, we do not find evidence that this information provision is associated with improved information aggregation in our setting (Result \ref{result: information provision}). 

Given our result that strategic prompting does not influence information aggregation, we further examine whether this is because AI agents act strategically but there is no impact on information aggregation, as predicted by economic theory,  or they are just unable to act strategically. We do this by examining the public messages they post and the private messages that are only visible by their future selves. AI agents could act strategically if the public and private messages differ. For example, they could post public messages in order to lie about their signal or withhold information, when prompted to be strategic. If they understand the dynamic nature of the market,  they could use their private messages to instruct their future selves on how to implement a multi-round strategy, or be more inclined to publicly reveal their signal in the last round they are trading. 

We introduce three measures of communication strategy: semantic alignment between private and public messages (cosine similarity), information hoarding (word gap between public and private), and direct deception, measured by the information revealed about the trader's signal from his public message, as judged by an AI agent ex post. We find that while the strategic treatment has no effect, the agents' behavior shows that they do have some understanding of the strategic nature of the environment. In 99\% of markets, agents on average hoard information, generating public announcements that are significantly shorter and semantically detached from their internal private reasoning. Furthermore, making agents aware of prior experimental outcomes widens the word gap and increases direct deception.

Our text analysis also reveals what initially appears to be a sophisticated inter-temporal communication strategy. Rather than maintaining a constant rate of deception, AI agents display a distinct ``sawtooth'' pattern of revelation: they reveal little in the opening rounds, but their probability of truthful revelation increases at specific intervals (Rounds 3, 6, and 9). At first glance, this mimics a sophisticated agent who selectively reveals more only when the financial penalty for honesty drops to zero. However, closer analysis shows that this sawtooth pattern is not consistent with dynamic planning, as it persists even when we restrict our sample to 9-round markets. Agents prematurely drop their deception in Rounds 3 and 6, acting as if the game is ending, before reverting to high-deception baseline behaviors in the subsequent rounds. Moreover, because the trading order is fixed, each trader has his own final trading opportunity (for example, Rounds 4 and 5 for the first two traders in six-round markets), at which revealing his signal is costless; yet no trader increases revelation there. This suggests that AI agents do not fully grasp  the dynamic nature of the market and they cannot plan ahead.  Rather than executing a cohesive inter-temporal plan via backward induction, the models appear constrained by pattern matching and limitations in multi-step reasoning tasks \citep{dziriEtAl2023}. 

The inability to plan ahead aligns with recent findings in the computer science literature demonstrating that LLMs lack the capacity for autonomous multi-step planning and dynamic state tracking \citep{valmeekamEtAl2023, kambhampatiEtAl2024}. It can also explain why the strategic prompting does not produce any change in behavior. We instruct AI agents to be strategic by telling them to maximise their payoff in the current and all future rounds where they trade. If they cannot plan ahead, then they cannot devise a multi-period strategy.

We conclude by motivating our choice of the prediction market as the most suitable experimental microstructure to study information aggregation with AI agents.
First, these mechanisms are widely used in the real world for forecasting and asset pricing. Historically employed to predict political events (Iowa Electronic Markets) and used by major corporations \citep{CowgillZitzewitz15} and governments \citep{economist21}, prediction markets have recently experienced rapid growth in adoption and volume. Platforms such as Polymarket and Kalshi now facilitate substantial trading volumes, serving as real-time aggregators of public sentiment. Notably, \cite{diercksEtAl2026} demonstrate that Kalshi markets achieve accuracy comparable to, or exceeding, professional forecasts for CPI and GDP releases. Furthermore, the pricing mechanism powering these markets, the Market Scoring Rule (MSR), is mathematically equivalent to the Constant Function Market Maker, a special case of the Automatic Market Maker (AMM) used extensively in decentralized finance (DeFi) and blockchain trading \citep{frongilloEtAl24}. \cite{leharParlour2025} show theoretically and empirically that AMMs have some advantages over limit order markets, such as providing more stable liquidity during extreme market events, and exhibiting smaller price impact with lower volatility.

Second, the MSR provides a structured environment that eliminates standard market microstructure frictions. In traditional continuous double auctions, an agent must submit a limit order and endure execution risk while waiting for a counterparty. The MSR, conversely, acts as an automated, inventory-based market maker, deterministically adjusting prices and guaranteeing instant liquidity for any buy or sell order.%
\footnote{See \cite{ostrovsky12} and \cite{galanisEtAl24} for examples.}
This design is well suited to evaluating AI agents. Because LLMs lack persistent endogenous memory and cannot asynchronously ``wait'' for an order to fill, sequential trading against a deterministic AMM matches their autoregressive processing. By guaranteeing liquidity and removing execution latency, our design isolates the price discovery process from standard market-microstructure frictions, so that failures to aggregate information cannot be attributed to execution risk, counterparty search, or liquidity timing. Residual non-reasoning channels, such as prompt comprehension or the mapping of beliefs into trades, are mitigated by providing each trader with explicit price-impact calculations for candidate trades (Section \ref{preliminaries}).




Finally, the MSR framework provides a benchmark that does not rely on the presence of exogenous noise traders or strategic market makers. In traditional sequential trade models or strategic auction frameworks \citep{kyle85, glostenMilgrom85}, information revelation is inherently confounded by the necessity of noise or liquidity trader volume to prevent market failure and provide strategic camouflage for informed agents. In contrast, our model provides tractable predictions regarding price paths, trading volume, and agent profits under myopic behavior, while theoretically predicting information aggregation across all Nash equilibria.

\subsection{Literature}

Our paper contributes to several strands of the literature. The first studies under which conditions information gets aggregated. \cite{ostrovsky12} and \cite{chenEtAl12} show that in a market with either myopic or strategic traders, separable securities are both necessary and sufficient for information aggregation, using both the model of \cite{kyle85}  and the Market Scoring Rule (\cite{mckelveyPage90}, \cite{hanson03, hanson07}), which is directly applicable to prediction markets. Information aggregation is based on the ``we cannot agree to disagree'' and ``no trade'' theorems of \cite{aum76},  \cite{milSto82}, and \cite{geaPol82}. 

\cite{plottSunder88} provide one of the first laboratory studies of information aggregation, showing that markets with human traders aggregate diverse private information in some environments, such as with a complete set of contingent claims, but not in others. 
\cite{pageSiemroth2021} find that prices incorporate almost all public information but less than 50\% of private information held by traders, using 664 experimental markets drawn from five studies. 
\cite{rasoolyRozzi25} conduct a field experiment with humans, where prices are randomly shocked in 817 prediction markets, finding that the effect was persistent even after 60 days. 
\cite{bossaertsEtAl2024} study field prediction markets that use the same logarithmic market scoring rule as the current paper, and identify informed traders from their trading behaviour alone. They find that informed traders are a small minority of participants, that only these traders earn the profits an informed trader should earn in this mechanism, and that markets containing them forecast better than markets that do not. 
\cite{galanisEtAl24} show theoretically and experimentally that ambiguity aversion can lead to no information aggregation with separable securities, but a new class, strongly separable securities, overcomes this limitation. We contribute to this literature by conducting the first, to our knowledge, experiment that studies information aggregation with AI agents that trade in a prediction market.

A related experimental literature measures the depth of interactive reasoning that human subjects display, using guessing games \citep{nagel1995, costaCrawford2006}, the cognitive hierarchy model \citep{camererEtAl2004}, the dirty faces game \citep{weber2001}, and ring games \citep{kneeland2015}, and consistently finds that most subjects perform no more than two steps of interactive reasoning. We similarly find that in the Very Hard structure, which emulates the dirty faces game and requires more than two steps of interactive reasoning, even the most capable models struggle to aggregate information. 

%

The second strand studies the impact of algorithmic and AI powered trading on market efficiency. An established empirical literature finds that algorithmic and high-frequency trading improve liquidity and price discovery \citep{hendershottEtAl2011, brogaardEtAl2014}, although \cite{weller2018} shows that algorithmic trading can reduce information acquisition, so that prices reflect less fundamental information. Closer to our setting, \cite{colliardEtAl2025} show using experiments that algorithmic market makers struggle to learn competitive pricing strategies, resulting in less competitive market outcomes. \cite{douEtAl2025} show with simulation experiments that AI speculators, the integration of algorithmic trading with reinforcement-learning algorithms, collude with each other, undermining competition and market efficiency. \cite{dugastEtAl2025} show that data abundance increases price informativeness. The current paper shows that AI agents struggle to reason interactively as the environment becomes more complex, hence market efficiency is reduced.


The third strand applies machine learning and LLMs to financial markets. \cite{guEtAl2020} demonstrate significant economic gains for investors who use machine learning forecasts, whereas \cite{lopezliraTang2025} show that ChatGPT 4 can predict stock market reactions from news headlines, and that performance improves with model size. 
Closest to our paper are experiments that place LLM agents in laboratory asset markets designed to elicit bubbles with human participants. \cite{henningEtAl2025} find that, in a classic experimental finance paradigm, LLM traders are ``textbook-rational'': unlike human subjects, who consistently generate bubbles, they price the asset near its fundamental value. \cite{saxenaEtAl2026} document a wider spectrum of behaviors, ranging from coordination on the fundamental value to speculative bubbles, with advanced models adapting to, and exploiting, less sophisticated counterparts. 
Our results also qualify the textbook-rationality finding. In our setting, AI agents fail to behave according to the textbook prediction, even though the Very Hard structure is a version of the dirty faces puzzle, whose solution is almost certainly part of their training data. 
One explanation is that if the game is not presented in a format similar to a text they have read, they cannot make the connection.

%

The fourth strand is at the intersection of economics and computer science. One branch studies how AI agents behave in economically interesting problems (\cite{chenEtAl2023}, \cite{biniEtAl2025}), whereas another studies how experimentation with LLMs can provide insights about human behavior (\cite{charnessEtAl2023}, \cite{korinek2023}, \cite{bail2024}, \cite{manningEtAl2024}). We contribute to this strand by conducting one of the first experiments with prediction markets and LLMs. 

%

The paper is organised as follows. Section \ref{preliminaries} describes the main elements of the prediction market and the teams of AI agents that are used in the experiment. Section \ref{experimental design} discusses the experimental design, including the four information structures, and Section \ref{sec:general hypotheses} lists our general hypotheses. Section \ref{sec:data} describes the data and we present our results in Section \ref{sec:results}. 
Section \ref{conclusion} concludes. In the Supplementary Appendix, we provide the full prompts and run robustness checks.

\section{Preliminaries}
\label{preliminaries}

This section introduces the information and payoff structure, the prompt design, the AI agents and the notion of information aggregation. The next section details our experimental design.

\subsection{Information and Payoff Structure}
In each prediction market there are three agents, who trade sequentially for 3,6, or 9 rounds. Agent 1 trades in rounds 1,4, and 7,  agent 2 trades in  rounds 2,5, and 8, and agent 3 trades in rounds 3,6, and 9. The information structure is determined by 3 signals, $d_a, d_b, d_c$, each with two possible realisations, 0 (`No') and 1 (`Yes'). A state of nature $\omega$ is determined by the realisation of these three signals. Hence,  the state space $\Omega = \{a,b,c,d,e,f,g,h\}$ consists of 8 states, shown in Table \ref{tab: all info structures}. For example, state $a = (1,1,1)$ realises when all three signals resolve to Yes. In all treatments, traders have a common uniform prior on $\Omega$ and each signal resolves to 1 with probability $0.5$. The draws of each signal are independent.  There are two tradable assets, which are complementary. The first asset, $X: \Omega \rightarrow {\mathbb R}$ (`betting on Yes', or Yes shares), pays 1 if the answer to the prediction market question is Yes, and $0$ otherwise. The second asset, $X'$ (`betting on No', or No shares), pays 1 if the answer to the question is No, and 0 otherwise. Therefore, the price of $X$ is always 1 minus the price of $X'$. An information and payoff structure (structure in short) determines which signals are received by which trader, and on which states the answer to the question is Yes, so that $X$ pays 1. Section \ref{sec:information and payoffs} describes the four structures we use in the experiment.
%
%

\subsection{Logarithmic Market Scoring Rule}

At the beginning of the market an initial price between 0 and 1 is set for $X$ (the Yes shares) by the market maker. The price of the No shares is set accordingly. Each trader is endowed with £1000 and they trade sequentially in each round. When it is his turn, a trader can buy or sell Yes and No shares, as well as do nothing (Hold).  The pricing mechanism uses the Logarithmic Market Scoring Rule (LMSR). The price of Yes is $p_Y = \frac{e^{\eta q_Y}}{e^{\eta q_Y} + e^{\eta q_N}}$, where $\eta$ is a liquidity parameter and $q_Y, q_N$ denote the number of outstanding Yes and No shares.%
\footnote{The liquidity parameter (we use $\eta = 0.01$) determines how quickly the price of $X$ changes when a trader buys or sells.  See \cite{cultivatelabs21} for an explanation of how the logarithmic MSR is implemented in practice and \cite{schlegelEtAl2026} for axiomatic foundations.}
The LMSR is a special case of a Market Scoring Rule (MSR) (\cite{mckelveyPage90}, \cite{hanson03, hanson07}).   Because the LMSR uses the Logarithmic Scoring Rule, which is proper, it has the following property. If the trader is risk neutral and myopic, so that he only maximises the expected value of his payoff for the current round, then he will trade up to the point where the price of  the Yes shares is equal to his posterior belief that the answer to the prediction market question is Yes. We call this the myopically optimal price.


\subsection{Information Aggregation}

Traders receive their private information before trading, by learning the realisation of their signals. The four information and payoff structures we employ are described in detail in Section \ref{sec:information and payoffs}. In all structures, the conjunction of the knowledge of the three traders reveals the true value of $X$.%
\footnote{This assumption is made by several papers in the literature, such as \cite{plottSunder88} and \cite{ostrovsky12}.}
This means that if traders communicated truthfully, the true value of Yes (0 or 1) would be revealed.
If traders are myopic and this is common knowledge, then the true value of $X$ is revealed in 3 rounds, after everyone has traded once, as we show in Section \ref{sec:myopic prices}. However, agents can be strategic and although in some treatments they are allowed to exchange information publicly, this is cheap talk. We say that {\bf information gets aggregated} at state $\omega$ if the price of $X$ in the final round is equal (or very close) to the true value of Yes, which is $X(\omega)$. 

The main purpose of this paper is to test our theoretical predictions about the market characteristics that improve or hinder information aggregation. Our measure of success for a market is therefore the logarithmic error, $-[y \ln (p)+(1-y) \ln(1-p)]$,  between the true value of $X$ at state $\omega$, denoted $y = X(\omega)$, and the final price of $X$ in the market, denoted $p$.%
\footnote{When calculating the logarithmic error in the data, $p$ is restricted to be in $[\epsilon, 1-\epsilon]$, $\epsilon = 10^{-15}$, so that we avoid having an infinite error. The maximum error with this restriction is around $34.5$.}
We also examine the trade volume and profitability of traders in markets.

\subsection{AI Agents}

We conducted our experiment with a variety of Large Language Models (LLMs). For the first two waves, we employed eight distinct models. The first six were accessed through an Application Programming Interface (API) and are closed-weight models: Claude Haiku 3.5, Claude Haiku 4.5, Gemini 2.5 Flash, Gemini 3 Flash, GPT-4o, and GPT-5 mini. The last two were open-weight models and they were run locally:  gemma3:4b, and qwen3:8b.%
\footnote{In open-weight models the parameters are available for modification and they can be run locally.  However, they are not open-source, because the code or the training data are not necessarily released.}
We formed twelve teams of three traders each, which participated in every treatment at least once. Eight teams were homogeneous, each comprising traders using the same model.  The remaining four teams featured a variety of models; in these heterogeneous teams, the models in Table \ref{tab:all ai models} are listed in trading order (the first listed model is trader 1, and so on), and the assignment of models to positions is fixed across all markets.  As a measure of each model's capabilities, we adopted the Artificial Analysis Intelligence Index \citep{ArtificialAnalysisTeam2025}, accessed on January 6, 2026.%
\footnote{The intelligence scores were accessed at \url{https://artificialanalysis.ai/evaluations/artificial-analysis-intelligence-index}. Note that the scores change over time as new models and new evaluations are added. The intelligence index encompasses many studies which evaluate the performance of LLMs across reasoning, knowledge, mathematics, coding, instruction following, long-context reasoning and agentic workflow tasks. Some of these are: MMLU-Pro \citep{wangEtAl2024} , GPQA Diamond \citep{reinEtAl2024}, HLE \cite{phanEtAl2025}, AIME 2025, SciCode \citep{tianEtAl2024}, LiveCodeBench \citep{jainEtAl2024}, IFBench \citep{pyatkinEtAl2025}, Terminal-Bench Hard, and $\tau^2$-Bench Telecom \citep{barresEtAl2025}.}
This is one of the most comprehensive and publicly available syntheses of model capabilities. It combines performance across reasoning, knowledge, mathematics, coding, instruction following, long-context reasoning and agentic workflow tasks.%
\footnote{See Appendix A in \cite{kimEtAl2025} for details.}
See Table \ref{tab:all ai models} for details, where we also report the average intelligence and standard deviation for all heterogeneous teams. Finally, for all models we set temperature at 1. This is the default, neutral setting in many APIs, so that text generation is neither too rigid (deterministic) nor chaotic. In the Supplementary Appendix, Figures \ref{fig:graph_mse_by_model} and \ref{fig:graph_profits_by_model} show the performance of each model in terms of log error and profits.   

In Section \ref{sec:robustness frontier}, we report the results from a third wave of 576 markets that we ran in April 2026, with three frontier models: GPT-5.4, Claude Opus 4.6, and Gemini 3.1 Pro, to check for robustness of our main results and in particular Result \ref{res:info deteriorates}. The Artificial Intelligence Indices, accessed in April 2026,  are not comparable with those accessed in January 2026 for the previous LLMs, as different tests and evaluations of models are added over time. Therefore, we keep the analysis separate. 
\begin{table}[!htbp] \centering
  \caption{AI Models, Team Compositions and Intelligence} 

\setlength{\tabcolsep}{4pt}
\begin{tabular}{lc}
\hline \hline
\textbf{Team Composition} & \textbf{Average Intelligence (Standard Deviation)} \\
\hline
\multicolumn{2}{l}{\textit{Homogeneous Teams (3 Identical Agents)}} \\
3x Gemini 3 Flash & 46 (0) \\
3x GPT-5 mini & 41 (0) \\
3x Claude Haiku 4.5 & 30 (0) \\
3x Gemini 2.5 Flash & 21 (0) \\
3x GPT-4o & 19 (0) \\
3x qwen3:8b & 15 (0) \\
3x Claude Haiku 3.5 & 12 (0) \\
3x gemma3:4b & 7 (0) \\
\\
\multicolumn{2}{l}{\textit{Heterogeneous Teams (Mixed Agents)}} \\
Claude Haiku 4.5, Gemini 3 Flash, GPT-5 mini & 39 (8.1) \\
Gemini 3 Flash, GPT-4o, qwen3:8b & 26.6 (16.8) \\
Gemini 2.5 Flash, GPT-4o, Claude Haiku 3.5 & 17.3 (4.7) \\
gemma3:4b, qwen3:8b, Claude Haiku 3.5 & 11.3 (4) \\
\\
\multicolumn{2}{l}{\textit{Frontier Teams, April 2026 (Section \ref{sec:robustness frontier})}} \\
3x GPT-5.4 & 57 (0) \\
3x Gemini 3.1 Pro & 57 (0) \\
3x Claude Opus 4.6 & 53 (0) \\
GPT-5.4, Claude Opus 4.6, Gemini 3.1 Pro & 55.7 (2.3) \\
\hline
\hline \\[-1.8ex]
\multicolumn{2}{p{\linewidth}}{\textit{Notes:} GPT-4o corresponds to the ``gpt-4o-2024-08-06'' version, GPT-5 mini to the ``gpt-5-mini-2025-08-07'' version, Claude Haiku 4.5 to the ``claude-haiku-4-5-20251001'' version,  and Claude Haiku 3.5 to the ``claude-3-5-haiku-20241022'' version, and gemini 2.5 to the ``gemini-2.5-flash-preview-09-2025''. Gemini 3 Flash corresponds to the ``gemini-3-flash-preview" version that was released on December 17, 2025.  Models gemma3:4b and qwen3:8b are open-weight models that were downloaded and run locally, whereas the other models were accessed through an API. All eight models have the same version number throughout the experiments. The frontier teams correspond to the ``gpt-5.4-2026-03-05'', ``claude-opus-4-6'', and ``gemini-3.1-pro-preview'' versions; their intelligence scores were accessed in April 2026 and are not comparable to the January 2026 scores of the other teams, as evaluations are added to the index over time.} \\
\end{tabular} 
\label{tab:all ai models}
\end{table}

\subsection{Prompt Design}

When chatting back and forth with an LLM, it appears as if it can recall the conversation and answer accordingly. However, inherently an LLM has no memory, it just `reads' the whole conversation every time it is called to answer. This means that  whenever we call an AI agent to trade in a round, we need to dynamically generate a prompt that describes the prediction market, lists the trading history and previous private or public comments, calculates the trader's portfolio and price impact of various trades, specify the trader's goal and ask for a reply. 

The prompt we construct in each round contains the following parts. Part one provides the details of the market, such as the question, whether comments are allowed, who participates, how many rounds there are in the market and what is the current round. Part two describes the public information, which is shared with all traders, and the private information that is shared with the current trader. The third part provides an explanation of prediction markets, including what are the Yes and the No shares. Part four provides an objective for the trader, to be either myopic or strategic  (see Section  \ref{sec: myopic vs strategic horizons}).  


Part five provides the history of trades up to now, the public comments that have been posted, and the current portfolio of the trader. As we do not rely on LLMs doing their own mathematical calculations, the prompt  informs the current trader about the maximum number of Yes and No shares he can buy and sell, as well as the price impact from various trades, for example buying 25\% of the maximum shares he can buy. This provides a comprehensive description of how the price will move after the AI agent trades. Part six enumerates the qualitative results from the first wave of the experiment, and it only appears in the Experiment Disclosure treatment.  The last part asks the trader for his trading decision (buy, sell, or do nothing), a private justification, and a public comment (if it is allowed by the market).  See Supplementary Appendix \ref{sa:sec: prompts} for an example of a prompt.


\subsection{Market implementation}

We ran our experiments using a prediction market platform, which was originally developed for running private prediction markets with humans. 
For this paper, we implemented in Python an Application Programming Interface (API) which allowed the programmatic execution of trades, as well as providing access to past trades and calculating the price impact for hypothetical trades to inform AI agents. We then created a program which creates a new market for each combination of the given parameters: teams of AI agents, number of rounds, initial price, structure, strategic, comments allowed. It then orchestrates the trading in rounds, invoking the LLMs using the relevant APIs, retrieving past trades and comments, calculating price impact for various trades, generating prompts dynamically for the current trader and executing trades through the API. The final output for each market is a text file containing all prompts and decisions of the traders, and a CSV file containing all trades which we use for our quantitative analysis.  In the Supplementary Appendix, Figure \ref{fig:flow_chart} provides a graphical representation.

\section{Experimental Design}
\label{experimental design}


Our experimental design focused on six dimensions. The first was the information and payoff structure that was presented to the traders. We considered four structures (t3s111y2, t3s110, t3s111, t3s111o2ye2) that are successively more complex,  as we explain in Section \ref{sec:information and payoffs}. The second dimension specified the number of trading rounds: three, six and nine. Given that there are three traders in all markets,  each AI agent trades once if there are three rounds, twice with six rounds and three with nine rounds. The order of trading is always fixed, so trader 1 trades first, then trader 2 and trader 3. The third dimension related to whether we prompted the AI agents to be myopic, so that they are instructed to maximise the current round's payoff, or strategic, so that they maximise the sum of payoffs from all rounds where they trade. In principle, they are able to carry out a strategy across rounds as they can post a private message to their future self.

The fourth dimension was whether public comments were allowed in the market, so that traders can communicate their private information. The fifth dimension related to the initial price of Yes, set by the uninformed market maker: we implemented three initial prices, 0.3, 0.5, and 0.7. Recall that the true value of Yes is either 0 or 1. Running this first wave, 4x3x2x2x3 experimental design, generated results that we included in the last, ``Experiment Disclosure'' treatment. In this treatment, AI agents were informed about the qualitative results of the first wave, before trading. Note that the disclosure wave was necessarily run after the first wave had concluded, so it is a sequential, later-wave treatment rather than a contemporaneously randomized factor; we discuss the implications for interpretation in Section \ref{sec:information provision}.  In summary, we applied a 4x3x2x2x3x2 experimental design to examine the impact on information aggregation of the difficulty in reasoning about the private information of others, the initial price, the length of trading, the communication, the information provision, and explicitly prompting traders to be strategic or myopic. 

\subsection{Information and payoffs}
\label{sec:information and payoffs}

In this section, we describe the four information and payoff structures that we presented to the traders. See the Supplementary Appendix for the exact wording of the generated prompts. There are three signals, $d_a, d_b, d_c$, that take two values $\{0,1\}$, and they are drawn independently, each with probability 0.5. Each signal is framed as a Yes/No question. For example, $d_a$ is the question ``Will sales in country A exceed 1 million?''. The prediction market question is ``Will Company X post next quarter profits that exceed 1 million'', and the answer depends on the realisations of the three signals. There are therefore 8 possible states, denoted $\{a,b,c,d,e,f,g,h\}$, uniquely determined by the realisations of these three signals. For example, state $a = (1,1,1)$ specifies that all signals resolve to Yes. The first part of Table \ref{tab: all info structures} depicts the realisations of the three signals and the 8 states.  We say that security $X$ pays 1 if the answer to the prediction market question is Yes, and 0 otherwise.

A structure specifies the information partition of each trader, the value of $X$ at each state, and the realisations of the three signals that determine the true state. The four structures we consider have the following common characteristics. First, in all states, if the three traders could talk truthfully and combine their private information, the true value of $X$ would become common knowledge.  Second, if traders are myopic, so that the price of $X$ is always the expected value of the last person who traded, and everyone being myopic is common knowledge, then the price of $X$ becomes equal to the true value of $X$ at the end of the third round, when everyone has traded once. Third, the information and payoff structure (but not the true state) are publicly announced. Finally, the structures become successively harder as traders need to reason about what others know and how public information evolves as they trade.

The four structures are t3s111y2, t3s110, t3s111, and t3s111o2ye2.%
\footnote{The naming of the structures follows this logic: ``t3'' stands for 3 traders, whereas ``s111'' and ``s110'' specify the realisations of the three signals and therefore the true state. In the first three structures an agent is informed about his own signal (e.g. trader 1 is informed about $d_a$, trader 2 about $d_b$), except in the last structure where ``o2'' specifies that an agent is informed about the two other signals (e.g. trader 1 is informed about $d_b$ and $d_c$). The ``y2'' specifies that at least 2 signals must resolve to Yes so that $X$ pays 1, whereas ``ye2'' specifies that exactly 2 signals must resolve to Yes. In structures t3s110 and t3s111 all three signals must resolve to Yes for $X$ to pay 1. Note that these names are never revealed to the AI agents.}
The first three specify the same information structure for the three traders, depicted in Table \ref{tab: all info structures}. In particular,  trader 1 is privately informed about $d_a$, trader 2 is privately informed about $d_b$, and trader 3 is privately informed about $d_c$. The partition cells for each trader are depicted in red and black. Trader 1's information partition is $\{\{a,b,c,d\}, \{e,f,g,h\}\}$, the two cells denoted in red and black, and similarly for traders 2 and 3.  

The information structure of t3s111o2ye2 specifies that trader 1 is privately informed about $d_b$ and $d_c$, trader 2 is privately informed about $d_a$ and $d_c$, and trader 3 is privately informed about $d_a$ and $d_b$. Note that each trader has four partition cells, depicted in the last part of Table \ref{tab: all info structures}. For example, the information partition of Trader 2 is $\{\{a,c\}, \{b, d\}, \{e, g\}, \{f, h\}\}$. This feature makes interactive reasoning much harder than in the other three structures.

The true value of $X$ in each state is given in the last part of Table \ref{tab: all info structures}, for each structure. In structure t3s111y2, $X$ pays 1 if at least two signals resolve to Yes, whereas in t3s110 and t3s111 all three signals need to resolve to Yes. In the last structure, t3s111o2ye2, $X$ pays 1 only when exactly two signals resolve to Yes. We denote with blue the true state in each structure.

\begin{table}
\caption{Information Structures}
\label{tab: all info structures}
\medskip

\centering

\begin{tabular}{lcccccccc}

\toprule
 States & a & b & c & d & e & f & g & h \\
\midrule

Signals &
\multicolumn{8}{c}{Realisations} \\

$d_a$ & 1 & 1 & 1 & 1 & 0 & 0 & 0 & 0\\
$d_b$   & 1 & 1 & 0 & 0 & 1 & 1 & 0 & 0\\
$d_c$  & 1 & 0 & 1 & 0 & 1 & 0 & 1 & 0\\

\underline{t3s111y2, t3s110, t3s111} &
\multicolumn{8}{c}{Information Structure} \\

Trader 1 & \red{a} & \red{b} & \red{c} & \red{d} & e & f & g & h \\
Trader 2 & \red{a} & \red{b} & c & d & \red{e} & \red{f} & g & h \\
Trader 3 & \red{a} & b & \red{c} & d & \red{e} & f & \red{g} & h \\

\addlinespace

\underline{t3s111o2ye2} &
\multicolumn{8}{c}{Information Structure} \\
Trader 1  & \red{a} & \orange{b} & \green{c} & d & \red{e} & \orange{f} & \green{g} & h\\
Trader 2  & \red{a} & \orange{b} & \red{c} & \orange{d} & \green{e} & f & \green{g} & h\\
Trader 3 & \red{a} & \red{b} & \orange{c} & \orange{d}& \green{e} & \green{f} & g & h\\

\bottomrule
\addlinespace

 & \multicolumn{8}{c}{Payoff of Yes ($ X = $)} \\
t3s111y2  & \blue{1} & 1& 1 & 0 & 1 & 0&0 & 0\\
t3s110  & 1 & \blue{0} & 0 & 0 & 0 & 0&0 & 0\\
t3s111 & \blue{1} & 0 & 0 & 0 & 0 & 0&0 & 0\\
t3s111o2ye2 & \blue{0} & 1& 1 & 0 & 1 & 0&0 & 0\\

\end{tabular}

\end{table}

\subsection{Myopic Prices}
\label{sec:myopic prices}

To provide a benchmark and understand how the structures compare in terms of complexity, we derive the equilibrium prices in the first three rounds for each structure, in the case where all traders are myopic and this is common knowledge. Because the logarithmic scoring rule that we employ in the prediction market is a proper scoring rule, it has the following property. If a trader is myopic, so that he only cares about the payoff of the current round, the optimal trading is to move the price so that it is equal to his expected value of $X$, which in this case is equal to his posterior belief that the answer to the question is Yes.  Moreover, this optimal behavior does not depend on what the previous price is.

Table \ref{tab:myopic prices} denotes the true state in blue and the payoff function of $X$, for each structure. For example, t3s111y2 describes that the answer to the prediction market question is Yes if at least two signals resolve to Yes, hence the payoff function is $X = (\blue{1},1,1,0,1,0,0,0)$. We denote with blue the true state, which is $a = (1,1,1)$ in this structure. Trader 1 is informed that his signal $d_a$ is Yes, hence he considers states $\{a,b,c,d\}$ to be possible. His posterior belief about Yes is 0.75, hence he trades Yes/No shares up until the price of Yes is 0.75.%
\footnote{Whether he buys or sells Yes or No shares depends on the previous price. For example, if the initial price of the market was 0.9 for Yes, he would buy No shares as he has no Yes shares to sell. If the initial price was 0.4, he would buy Yes shares. In both cases, his trades would drive the price of Yes to 0.75.}
A price of 0.75 reveals to everyone that $\{e,f,g,h\}$ are impossible, because in that case the price would be 0.25. In other words, the public information that is revealed in Round 1 is $\{a,b,c,d\}$. Trader 2 combines this public information with his private information, $\{a,b,e,f\}$, to deduce that the true state is in $\{a,b\}$. As both pay 1, trader 2 trades so that the price of Yes is 1. Trader 3 deduces that the true value of Yes is 1 and his trades do not move the price. In all subsequent rounds (if there are any), the price does not change.

The myopic prices for the other structures are computed similarly. We say that t3s111y2 is the easiest structure because each trader needs to only reason about the signal of one other trader, in order to determine the price of $X$. In structures t3s110 and t3s111, the traders need to reason about the signals of everyone else. The only exception is trader 3 in structure t3s110, who receives a 0 signal and therefore immediately deduces that the answer is No, independently of what others have traded. For that reason, we consider t3s110 to be easier than t3s111. 

Structure t3s111o2ye2 is the most difficult in terms of reasoning about others. First, the partition cells for each trader  are four, instead of two. Second, reasoning about which state is no longer possible requires complex counterfactuals. Finally, the myopic price stays the same from round 1 to round 2; however, the public information set shrinks. This may be difficult for the LLMs to process when reasoning about the private information of others. The true state is $a$ and it is mutual knowledge (all three traders know it) that states $d$ and $h$ are impossible. In the first round, the myopic price of 0.5 makes it common knowledge that $d$ and $h$ are impossible, otherwise the price would be 0. In round 2, the myopic price stays the same at 0.5, which makes it common knowledge that $b$ is impossible, otherwise (and given that $d$ is impossible), the price would be 1. State $f$ is also excluded because the price would be 0 otherwise.

Although the structure is complex, it is derived from the following well-known ``dirty faces'' (or girls with red hats) puzzle \citep{gea92, weber2001}. Its popularity suggests it is likely part of the training sets of LLMs used in the experiment.%
\footnote{When we asked each LLM, on a separate session, to describe the puzzle, they were fairly accurate.}
However, by looking at the private and public justifications of the AI agents, it seems that no one made the connection.

It is described as follows. There are three girls who wear either a red or a white hat. They can only see the hat of the other two girls but not their own. The true state is that all hats are red and it is therefore mutual knowledge that there is at least one red hat. When asked, each girl announces she does not know the colour of her hat. The teacher then announces that there is at least one red hat, a fact that is mutual (but not common) knowledge. Then, the first girl announces that she does not know the colour of her hat, and so does the second girl. Although these announcements do not change after the teacher's announcement, the public knowledge shrinks and the third girl (if she is very sophisticated) deduces that her hat is red. The teacher's announcement makes it common knowledge that the state `all hats are white' is impossible. This prompts the gradual reduction in the public knowledge, so that in the third round the action changes as well.

\begin{table}[htbp]
\caption{Myopic Price of Yes and Public Information }

\medskip
\centering

\label{tab:myopic prices}

\begin{tabular}{@{}llcccc@{}}
\toprule
 &  & Myopic price & Public information & Volume & Profits \\
\midrule

\multicolumn{4}{l}{\textbf{t3s111y2 - Easy}} \\
\multicolumn{4}{l}{Payoff: $X = (\blue{1},1,1,0,1,0,0,0)$} \\
\addlinespace[0.3em]

Trader 1 &  & 0.75 & $(a,b,c,d)$ & 110 & £46 \\
Trader 2 &  & 1    & $(a,b)$ & 880 & £29\\
Trader 3 &  & 1    & $(a,b)$ & 0 & £0\\
\addlinespace[0.8em]

\multicolumn{4}{l}{\textbf{t3s110 - Medium}} \\
\multicolumn{4}{l}{Payoff: $X = (1,\blue{0},0,0,0,0,0,0)$} \\
\addlinespace[0.3em]

Trader 1 &  & 0.25 & $(a,b,c,d)$ & 110 & £46\\
Trader 2 &  & 0.5  & $(a,b)$ & 110 & -£41\\
Trader 3 &  & 0    & $b$ & 990 & £69\\
\addlinespace[0.8em]

\multicolumn{4}{l}{\textbf{t3s111 - Hard}} \\
\multicolumn{4}{l}{Payoff: $X = (\blue{1},0,0,0,0,0,0,0)$} \\
\addlinespace[0.3em]

Trader 1 &  & 0.25 & $(a,b,c,d)$ & 110 & -£64\\
Trader 2 &  & 0.5  & $(a,b)$ & 110 & £69\\
Trader 3 &  & 1    & $a$ & 990 & £69\\
\addlinespace[0.8em]

\multicolumn{4}{l}{\textbf{t3s111o2ye2 - Very Hard}} \\
\multicolumn{4}{l}{Payoff: $X = (\blue{0},1,1,0,1,0,0,0)$} \\
\addlinespace[0.3em]

Trader 1 &  & 0.5 & $(a,e),(b,f),(c,g)$ & 56 & £6\\
Trader 2 &  & 0.5 & $(a,c),(e,g)$ & 0 & £0\\
Trader 3 &  & 0   & $a$ & 990 & £69\\
\hline \\[-1.8ex] 

\end{tabular}

\vspace{1em} 
\begin{minipage}{0.9\textwidth} 
\footnotesize
\textit{Notes}: The table depicts the myopically optimal price per round, the public information that is revealed, the volume (number of shares) and the profit of the AI agent who trades at that round. We denote with blue the true state. The myopic price of $X$ is the expected value of $X$ given the private and public information for each trader.  It reveals information to all other traders, who use it to form their updated beliefs and trade, and it does not depend on the previous price, however the profits and volume do. For trader 1, the volume and profits are averaged over the three possible initial prices, 0.3, 0.5, and 0.7. For the other two traders, volume and profits are the same because trader 1 always moves the price to the same level, irrespective of the initial price. Where the myopic price is 0 or 1, benchmark volumes and profits are computed by truncating the price at 0.00005 and 0.99995, since reaching the boundary exactly requires unbounded share purchases under LMSR.
\end{minipage}

\end{table}

\subsection{Myopic vs. Strategic Horizons}
\label{sec: myopic vs strategic horizons}

The number of periods can influence whether an AI agent acts strategically but the prompt itself may also have that power. We altered the prompt to test whether the AI agents can act strategically if they are instructed to do so, and whether this has an impact on information aggregation. We provided the following prompts.

\begin{itemize}
    \item \textbf{Myopic treatment:} Use reasoning to determine your belief q, then choose your action (Buy, Sell, or Hold: Yes and No shares). Maximize your expected payoff in this round only, based on your belief q and the current price p, ignoring future rounds.
    \item \textbf{Strategic treatment:} Use reasoning to determine your belief q, then choose your action (Buy, Sell, or Hold: Yes and No shares). Maximize the sum of your expected payoffs over all trading rounds, based on your belief q and the current price p. Consider how your current trade affects the price and the beliefs of others in future
rounds.
\end{itemize}

\section{General hypotheses}
\label{sec:general hypotheses}

In order to formulate our hypotheses, we use the following result by \cite{ostrovsky12}, adapted to our setting.

\begin{Theorem} \label{thm: ostrovsky}
If security $X$ is separable under information structure $\Pi$, then for any prior distribution $\mu$, for any strictly proper scoring rule $s$, initial price $y_0$, and
discount factor $\gamma \in (0,1]$, in any Nash  equilibrium, information gets aggregated.
\end{Theorem}

In Appendix \ref{separability}, we formally define the notion of separability and show that the securities in all four structures are separable. Given that the logarithmic is a proper scoring rule, this theorem says that information should get aggregated in all four structures, irrespective of their complexity.%
\footnote{The implicit assumption of full rationality is indispensable. For example, \cite{galanisKotronis21} show that information aggregation may fail if traders are boundedly rational and unaware of some dimensions of their uncertainty.}

\begin{Hypothesis}
Information aggregation is unaffected by the complexity of the structure.
\end{Hypothesis}

The  theorem specifies that whether traders are myopic or strategic does not matter. In the strategic treatment we instruct traders to maximise their payoffs over all future rounds. This is consistent with a discount factor of $\gamma = 1$, although we do not mention this term in the prompt. The myopic treatment is consistent with $\gamma = 0$. Although the theoretical model excludes a zero discount factor, it is straightforward to show that as $\gamma$ converges to 0, traders will behave almost myopically and reveal their expectations almost truthfully. 

\begin{Hypothesis}
Information aggregation is unaffected by prompting AI agents to be strategic or myopic.
\end{Hypothesis}

The theoretical model does not allow traders to post comments. However, it is straightforward to show that this should not hinder information aggregation. Roughly, the reason is that if there is no information aggregation in the long term, separability implies that at least one trader is able to achieve strictly positive profits, irrespective of what other traders do or say.

\begin{Hypothesis}
Information aggregation is unaffected by allowing AI agents to post public comments.
\end{Hypothesis}

A direct application of Theorem \ref{thm: ostrovsky} ensures that the initial price should not impact information aggregation.

\begin{Hypothesis}
Information aggregation is unaffected by manipulating the initial price of the market.  
\end{Hypothesis}

An important difference from the theoretical model is that it specifies infinitely many rounds of trading, whereas in the current experiment we have up to 9. Although it is impossible to have infinitely many rounds in an experiment, we could simulate an infinite horizon by specifying that, in every round, the game ends with some probability $p$.%
\footnote{See \cite{galanisEtAl24} for such an experiment with humans in a prediction market.}
We chose not to include this extra treatment for the following two reasons. First, the prompt is already long and complicated, so it is not clear that AI agents would be able to understand and act differently if we added this detail. Second, we considered that it would be more interesting and practical to understand the effect of lengthening the duration of the market on information aggregation.

\begin{Hypothesis}
Information aggregation is unaffected by the duration of the market.
\end{Hypothesis}

We now turn to the role of intelligence on market outcomes. 
We conjecture that information aggregation improves as AI agents become smarter, as they should be able to reason better about the information of others. Furthermore, since greater accuracy leads to higher profits, individual profits should rise with intelligence. \cite{corgnetEtAl2018} show in a market experiment with human participants that `smarter' traders are also more profitable. They measure intelligence with three metrics. Fluid intelligence measures their ability to compute correctly and draw statistical inferences. Cognitive reflection measures their ability to avoid behavioral biases and update their beliefs when observing market orders. Theory of the Mind measures whether they can correctly assess the informational content of trades.

\begin{Hypothesis}
Information aggregation improves as the average intelligence of the group increases. 
\end{Hypothesis}

\begin{Hypothesis} \label{hypothesis: profits correlated with intelligence}
Individual profits are positively correlated with an agent's own intelligence, but negatively correlated with the average intelligence of the market.
\end{Hypothesis}

The following hypotheses are derived by assuming that traders play the myopic equilibrium. Table \ref{tab:myopic prices} describes the myopically optimal prices, trading volume, and profits, for each of the four structures and for each trader, using the formulas derived in Supplementary Appendix \ref{sec:profit_calculations}.
Recall that the initial price is 0.3, 0.5, or 0.7, which influences both the trading volume and the profits of the first trader. The other two traders are unaffected because the first trader always moves the price to the same level, irrespective of the initial price.  By computing the average trading volume for each structure and the average profits for each trader, we formulate the next two hypotheses. Table \ref{tab:myopic_volume_profits_average} summarises all calculations. Note that the following hypotheses implicitly assume that all traders are myopic, hence all markets resolve in round 3 and there is no trading in subsequent rounds. 

The average trading volume across all initial prices is as follows: Easy (990), Medium (1210), Hard (1210), Very Hard (1046). We therefore have the following hypothesis.

\begin{Hypothesis} \label{hypothesis: volume}
Trading volume is ordered from lowest to highest as follows: Easy (t3s111y2) < Very Hard (t3s111o2ye2) < Medium (t3s110) = Hard (t3s111).
\end{Hypothesis}

We also compute the average profits for each trader, across all structures and initial prices: Trader 1 (9), Trader 2 (14), and Trader 3 (52).

\begin{Hypothesis} \label{hypothesis: positive profits}
Profits are positive for all traders and ordered, from lowest to highest, as follows: Trader 1 < Trader 2 < Trader 3.
\end{Hypothesis}

A distinctive feature of the prediction market is that it is `constant-sum' in profits: the sum of the profits of all traders depends only on the initial price $p_0$, the final price $p_1$, and the realized outcome, irrespective of how prices fluctuate in between. Under the myopic benchmark the final price equals the true value of the security in every structure, hence total profits are identical across structures and average profits are always 25, as shown in Table \ref{tab:myopic_volume_profits_average}. If aggregation fails, the final price diverges from the true value and total profits fall, so the following hypothesis is a joint test of the myopic full-aggregation benchmark.

\begin{Hypothesis} \label{hypothesis:profits correlated}
Average profits are not correlated with the difficulty of the structure. 
\end{Hypothesis}

Can AI agents improve their market performance by receiving feedback from historical play? \cite{yangEtAl2023} propose an information provision technique that leverages LLMs as iterative optimizers: by updating an LLM's prompt with a history of past actions and their relative success, the model self-corrects, outperforming human-designed prompts by up to 50\%. \cite{galanisMikhalishchev26} show that market accuracy improves as the cost of acquiring more information decreases.

Building on these mechanisms, we postulate that providing AI agents with aggregated historical data about market dynamics will enable them to form more accurate predictions, thereby strictly improving information aggregation. However, the anticipated effect on individual profitability is ambiguous. Because this informational shock is provided uniformly across all agents in the market, no single trader gains an asymmetric informational advantage.

\begin{Hypothesis} \label{hypothesis: info on accuracy}
The effect of information provision on information aggregation is positive. 
\end{Hypothesis}

\begin{Hypothesis} \label{hypothesis: info on profits}
The effect of information provision on individual profits is zero.
\end{Hypothesis}

%

\section{Data}
\label{sec:data}


The dataset consists of observations from 4,076 prediction markets, comprising 12,228 trader-market observations (three traders per market). Our main sample pools the first wave and the disclosure wave (3,500 markets) from January 2026; the third wave of 576 markets with the April 2026 frontier models is kept separate and analyzed in Section \ref{sec:robustness frontier}. Table \ref{tab:descriptive_stats_both_waves} summarises the key market-level variables, where the upper panel describes the first wave (1772 markets), the middle panel describes both waves, and the bottom panel describes the third wave.%
\footnote{With 12 teams of AI agents, the first wave should have consisted of 4x3x2x2x3x12 = 1728 markets. The extra 44 markets were due to a coding error. We decided to keep these markets for completeness.}

We focus on the first wave because the statistics do not change when the second wave is added. A defining characteristic of the data is the extreme skewness in market performance metrics. While the median logarithmic score is low (0.089), indicating that the typical market converges to a high-probability estimate of the correct outcome, the mean score is substantially higher (5.18), driven by a small subset of markets with extreme errors, where the logarithmic penalty creates scores as high as 34.54. A similar pattern is observed in the Squared Error, where the mean (0.266) is significantly larger than the median (0.007).

Trading activity varied substantially across sessions. The average market generated a volume of approximately 1,859 shares traded, though this ranged from inactive markets (0 volume) to very active markets reaching over 14,290 shares. The teams of AI agents had a mean intelligence score of 23.56 (SD = 12.2), spanning a range from 7 to 46. 

Supplementary Appendix Table \ref{tab:treatment_outcomes_both_waves} breaks down market outcomes by information structure and duration, for the first wave, for both waves, and for the third wave of 576 markets with the April 2026 frontier models. The data for the first wave (and similarly for both waves) reveal a strict hierarchy of difficulty, consistent with our experimental design. In the Easy structure (t3s111y2), markets performed reliably well, with a Mean Squared Error (MSE) as low as 0.07 (for 6 rounds) and a `Crash Rate' (markets with a log error above 20) of only 1.4\%. As complexity increases, performance degrades sharply and monotonically. In the Medium structure (t3s110), the average error nearly triples (MSE $\approx 0.2$), and by the Very Hard structure (t3s111o2ye2), the market fails to aggregate information: the MSE reaches 0.43, suggesting a final price of 0.65 when the true value of the security is 0, whereas the Crash Rate rises to over 19\%. The table also serves as a randomization check. The Average Intelligence (Avg IQ) column shows that agent capabilities are evenly distributed across all treatment cells in the first two waves (varying tightly between 22.9 and 23.8), indicating that the observed differences in performance are driven by the structural difficulty of the task and not by chance imbalances in agent composition.%

\begin{table}[!htbp] \centering
  \caption{Descriptive Statistics}
  \label{tab:descriptive_stats_both_waves}
\makeatletter
\begin{tabular}{@{\extracolsep{5pt}}lcccccc}
\\[-1.8ex]\hline
\hline \\[-1.8ex]
Statistic (First Wave) & \multicolumn{1}{c}{N} & \multicolumn{1}{c}{Mean} & \multicolumn{1}{c}{St. Dev.} & \multicolumn{1}{c}{Min} & \multicolumn{1}{c}{Median} & \multicolumn{1}{c}{Max} \\
\hline \\[-1.8ex]
Squared Error & 1,772 & 0.266 & 0.382 & 0 & 0.007 & 1 \\ 
Logarithmic Score & 1,772 & 5.181 & 11.770 & 0 & 0.089 & 34.540 \\ 
Trading Volume & 1,772 & 1,859 & 1,824 & 0 & 1,298 & 14,290 \\ 
Avg. Intelligence (Group) & 1,772 & 23.562 & 12.205 & 7 & 19 & 46 \\ 
Intelligence SD & 1,772 & 2.875 & 5.030 & 0 & 0 & 16.862 \\ 
Initial Squared Error & 1,772 & 0.274 & 0.164 & 0.090 & 0.250 & 0.490 \\ 
Initial Logarithmic Score & 1,772 & 0.745 & 0.347 & 0.357 & 0.693 & 1.204 \\ 
\hline \\[-1.8ex]
\\[-1.8ex]\hline
\hline \\[-1.8ex]
Statistic (Both Waves) & \multicolumn{1}{c}{N} & \multicolumn{1}{c}{Mean} & \multicolumn{1}{c}{St. Dev.} & \multicolumn{1}{c}{Min} & \multicolumn{1}{c}{Median} & \multicolumn{1}{c}{Max} \\
\hline \\[-1.8ex]
Squared Error & 3,500 & 0.277 & 0.390 & 0 & 0.010 & 1 \\ 
Logarithmic Score & 3,500 & 5.571 & 12.152 & 0 & 0.105 & 34.540 \\ 
Trading Volume & 3,500 & 1,898 & 1,828 & 0 & 1,353 & 14,290 \\ 
Avg. Intelligence (Group) & 3,500 & 23.668 & 12.216 & 7 & 19 & 46 \\ 
Intelligence SD & 3,500 & 2.847 & 4.993 & 0 & 0 & 16.862 \\ 
Initial Squared Error & 3,500 & 0.275 & 0.164 & 0.090 & 0.250 & 0.490 \\ 
Initial Logarithmic Score & 3,500 & 0.748 & 0.348 & 0.357 & 0.693 & 1.204 \\ 
\hline \\[-1.8ex]
\\[-1.8ex]\hline
\hline \\[-1.8ex]
Statistic (Third Wave, Frontier) & \multicolumn{1}{c}{N} & \multicolumn{1}{c}{Mean} & \multicolumn{1}{c}{St. Dev.} & \multicolumn{1}{c}{Min} & \multicolumn{1}{c}{Median} & \multicolumn{1}{c}{Max} \\
\hline \\[-1.8ex]
Squared Error & 576 & 0.149 & 0.275 & 0 & 0 & 1 \\ 
Logarithmic Score & 576 & 2.303 & 8.099 & 0 & 0 & 34.540 \\ 
Trading Volume & 576 & 1,105 & 935.245 & 0 & 1,164 & 6,271 \\ 
Avg. Intelligence (Group) & 576 & 55.667 & 1.634 & 53 & 56.333 & 57 \\ 
Intelligence SD & 576 & 0.577 & 1.001 & 0 & 0 & 2.309 \\ 
Initial Squared Error & 576 & 0.277 & 0.165 & 0.090 & 0.250 & 0.490 \\ 
Initial Logarithmic Score & 576 & 0.751 & 0.349 & 0.357 & 0.693 & 1.204 \\ 
\hline \\[-1.8ex]
\end{tabular}
\makeatother
\parbox{0.95\linewidth}{\textit{Notes:} The third-wave panel reports the 576 markets run in April 2026 with the frontier models. Its intelligence scores use the April 2026 vintage of the Artificial Analysis Intelligence Index, which is not comparable to the January 2026 scores in the first two panels.}
\end{table}

\subsection{Methodological Considerations}
\label{sec:methodological}

For our primary OLS specifications, we use CR2 cluster-robust standard errors clustered at the AI team composition level, featuring small-sample degrees of freedom adjustments \citep{pustejovskyTipton2018}. This approach conservatively estimates treatment effects by accounting for intra-cluster correlation within specific AI teams. 
For our auxiliary quantile regressions in the main text, which assess typical market behavior (Q50) and worst-case tail risk (Q80), we report bootstrapped standard errors (B=200 resamples) rather than analytical cluster-robust standard errors. Estimating analytical clustered variance matrices for quantile regressions requires local error density estimation that is usually unstable and fragile in finite samples. Consequently, we employ standard XY-pair bootstrapping for these specific models. We emphasize that the primary purpose of these quantile specifications is to demonstrate the behavioral stability of the point estimates across the error distribution, rather than to serve as exact clustered inference.

Furthermore, because several of our core theoretical propositions predict a null result, relying solely on conservative CR2 adjustments could theoretically favour our hypotheses by making it artificially easy to fail to reject the null. We address this concern in two complementary ways. First, Section \ref{sec:power} reports an ex-ante power analysis that computes the minimum detectable effect (MDE) for each treatment null by Monte Carlo simulation through the exact CR2/Satterthwaite estimator used in the paper, so that each failure to reject can be read against the smallest effect the design could have detected. 

Second, we stress-test the findings under strictly less conservative variance assumptions: for market-level outcomes we re-estimate the OLS models with unclustered HC1 standard errors, and for individual-level outcomes with CR1 errors clustered at the unique market level, which removes the panel-level team penalty while retaining intra-market dependence. The corresponding tables are reported in the Supplementary Appendix. The primary null results for communication, strategic prompting, the average initial-price contrasts, and extended market duration (9 rounds) hold even under these less restrictive assumptions; the Very Hard $\times$ low-initial-error interaction of Result \ref{result:initial price} likewise remains significant under HC1 ($\beta = 2.861$, $p<0.01$). Conversely, when the team-level clustering penalty is removed, the 6-round duration, experiment disclosure, and the Medium information structure cross the threshold into statistical significance in the information aggregation OLS, and the order of trading becomes statistically significant for individual profits. Because these alternative specifications assume independence across separate market sessions, ignoring the intra-team correlation inherent in our multi-round, agent-based panel design, they naturally produce smaller standard errors. We therefore interpret these differences as evidence that the affected estimates are sensitive to the assumed correlation structure, rather than as a basis for stronger causal claims.

An alternative method for addressing the issue of a small number of clusters would be to significantly expand the array of LLMs employed, from 12 to more than 40. However, this approach introduces two limitations. First, we intentionally selected models exclusively from the leading frontier laboratories (Anthropic, Google, and OpenAI) alongside two capable open-weight models. Expanding the baseline to a much larger quantity of models would have necessitated including peripheral, lower-tier agents. Had we done so, any observed breakdown in information aggregation could be easily dismissed as a failure to exclude less competent LLMs. By restricting our sample strictly to established models, we ensure that failures in information aggregation represent genuine limitations of frontier models. Second, expanding the sample to over 40 teams would cause the number of simulated markets to jump from 3,500 to more than 12,000. This scaling would significantly increase computational overhead and compromise our ability to perform rigorous manual auditing.


Finally, we deliberately refrain from estimating pooled regressions that combine the January and April cohorts. While pooling would ostensibly increase statistical power, doing so requires controlling for baseline model capabilities across generations. However, traditional standardized benchmark indices that comprise the Artificial Intelligence Index (e.g., MMLU, HumanEval) fail to maintain reliable cross-generational consistency between early-2025 and mid-2026 models. Because these static benchmarks are non-stationary, using them to construct a unified intelligence metric across differing vintages would introduce substantial measurement error. By restricting our analysis to cohort-specific distributions and matched relative performance tests, we ensure our findings remain robust to the rapid deprecation of static AI benchmarks.



\section{Results}
\label{sec:results}

We first report our results from the first wave (1772 markets), for information aggregation (Section \ref{sec:information aggregation}), and trading volume and profits (Section \ref{sec:trading volume profits}), ignoring the information provision treatment. In Section \ref{sec:information provision}, we describe the information provision treatment, using the full sample of 3500 markets.  It turns out that most of our results are robust when running the same regressions in the full sample. The graphs that we show use the full sample. In Section \ref{sec:public private comments}, we examine how AI agents use the public and private comments, whereas in Section \ref{sec:robustness frontier} we conduct a robustness check with the three frontier models from April 2026.

\subsection{Information Aggregation}
\label{sec:information aggregation}

We present our main results for information aggregation in Table \ref{tab:quantile_regressions_robust}. 
Column (1) presents an OLS regression where the dependent variable is log error (the logarithmic scoring rule error). Because the log error has a wide range (from 0 to 34.5), the mean estimator is highly sensitive to outliers, when markets crash by pricing the security $X$ at 0 when its true value is 1.%
\footnote{Recall that the log error between actual (y) and predicted (p) is $-(y  \ln(p)+(1-y) \ln(1-p))$, where the predicted probability is bounded, $p \in [\epsilon, 1-\epsilon]$, $\epsilon = 10^{-15}$, to avoid having an infinite log error.}
Columns (2) and (3) present Quantile Regressions that allow us to inspect specific parts of the error distribution. Column (2) estimates the model at the median ($\tau = 0.5$), representing the typical market outcome. This estimator is robust to outliers, filtering out the effect of crashes to reveal how the typical agent behaves under varying conditions. Column (3) estimates the model at the 80th percentile ($\tau=0.8$), representing the tail risk. 

We employ a mixed-contrast coding specification across all three models. For the primary design variables, Structure and Market Duration, we use treatment contrasts, setting the easiest structure (t3s111y2) and the shortest duration (3 Rounds) as the reference baseline. Consequently, coefficients for these variables represent the marginal penalty of increasing complexity or duration relative to this baseline. For the other conditions, Initial Log Error, Strategic, and Comments, we employ sum contrasts (deviation from the mean). In this specification, coefficients represent the deviation of a specific condition (e.g., Medium Initial Error) from the global average of that variable, rather than from an arbitrary reference group. Thus, the model intercept captures the predicted log-error of the baseline market design (Easy Structure, Short Duration) under average starting conditions for the sum-contrast variables, evaluated at zero group intelligence; predictions for the average baseline market are obtained by adding the intelligence terms at their sample means.%
\footnote{An alternative specification would be to have the Squared Error as the dependent variable, instead of the Log Error. However, because our prediction market uses the logarithmic scoring rule, AI agents try to minimise Log Error to maximise their profits, hence the Log Error seems more appropriate. The disadvantage of this approach is the sensitivity of the mean estimator to outliers, hence we also employ the two quantile regressions.}

\begin{table}[!htbp] \centering 
 \caption{Information Aggregation Without Disclosure} 
 \label{tab:quantile_regressions_robust} 
\small 
\resizebox{\linewidth}{!}{%
\begin{tabular}{@{\extracolsep{5pt}}lccc} 
\\[-1.8ex]\hline 
\hline \\[-1.8ex] 
& \multicolumn{3}{c}{\textit{Dependent variable:}} \\ 
\cline{2-4} 
\\[-1.8ex] & \multicolumn{3}{c}{Log Error} \\ 
& Mean (OLS) & Median (Q50) & Tail Risk (Q80) \\ 
\\[-1.8ex] & (1) & (2) & (3)\\ 
\hline \\[-1.8ex]
 Constant & 7.548$^{***}$ (2.185) & 0.018 (0.019) & 3.170$^{**}$ (1.138) \\ 
  Comments Allowed & $-$0.191 (0.319) & $-$0.001 (0.003) & $-$0.000 (0.032) \\ 
  Duration: 6 Rounds & $-$1.817 (1.172) & $-$0.001 (0.005) & $-$0.057 (0.177) \\ 
  Duration: 9 Rounds & $-$0.627 (1.250) & $-$0.001 (0.005) & $-$0.028 (0.165) \\ 
  Strategic Prompt & 0.053 (0.292) & 0.001 (0.003) & 0.000 (0.113) \\ 
  Medium (t3s110) & 2.200 (2.635) & $-$0.000 (0.003) & 0.038 (1.808) \\ 
  Hard (t3s111) & 5.327$^{*}$ (2.317) & 0.286$^{**}$ (0.106) & 12.470$^{.}$ (7.451) \\ 
  Very Hard (t3s111o2ye2) & 6.658$^{*}$ (2.974) & 0.700$^{***}$ (0.084) & 21.771$^{***}$ (4.744) \\ 
  Initial Error: Low & 0.014 (0.120) & 0.000 (0.004) & 0.000 (0.069) \\ 
  Initial Error: Medium & 0.225 (0.186) & $-$0.000 (0.003) & $-$0.000 (0.063) \\ 
  Average Intelligence & $-$0.221$^{***}$ (0.059) & $-$0.0004 (0.0004) & $-$0.069$^{**}$ (0.025) \\ 
  Intelligence SD & 0.044 (0.227) & $-$0.0003 (0.0003) & $-$0.052$^{*}$ (0.026) \\ 
  6 Rounds x Strategic & $-$0.067 (0.261) & $-$0.001 (0.004) & 0.000 (0.128) \\ 
  9 Rounds x Strategic & 0.076 (0.385) & $-$0.001 (0.004) & 0.028 (0.161) \\ 
  Medium Struct x Initial Error (Low) & 1.309 (1.279) & $-$0.000 (0.005) & 0.019 (3.595) \\ 
  Hard Struct x Initial Error (Low) & 0.504 (0.682) & 0.032 (0.124) & 19.889 (12.267) \\ 
  Very Hard x Initial Error (Low) & 2.861$^{.}$ (1.623) & 0.013 (0.167) & 10.425$^{*}$ (5.186) \\ 
  Medium Struct x Initial Error (Medium) & $-$0.725$^{.}$ (0.436) & 0.000 (0.004) & 0.019 (1.810) \\ 
  Hard Struct x Initial Error (Medium) & $-$0.236 (0.715) & 0.077 (0.149) & $-$9.945 (12.141) \\ 
  Very Hard x Initial Error (Medium) & $-$0.337 (1.021) & $-$0.007 (0.083) & 10.080 (9.342) \\ 
 \hline \\[-1.8ex] 
Observations & 1,772 & 1,772 & 1,772 \\ 
R$^{2}$ & 0.115 &  &  \\ 
\end{tabular}%
}
\parbox{0.95\linewidth}{\textit{Notes:} $^{.}$p$<$0.1; $^{*}$p$<$0.05; $^{**}$p$<$0.01; $^{***}$p$<$0.001. Q50 and Q80 are quantile regressions. Q50 models the median log error, whereas Q80 models the upper 20\% tail of the log error distribution (the worst-performing markets). OLS model uses CR2 cluster-robust standard errors clustered by exact AI team composition, featuring small-sample degrees of freedom adjustments. Quantile models report standard bootstrapped standard errors due to small-cluster convergence constraints. All three models use a mixed-contrast specification. Rounds and Structure use treatment contrasts with Round 3 and Structure t3s111y2 as baselines. All other controls (Initial Error, Strategic, Comments) use sum contrasts, where coefficients represent deviations from the grand mean.}
\end{table}


In our OLS specification, we observe that as the structure becomes highly difficult, the average log error increases. The intercept (7.548) is evaluated at zero group intelligence, which lies outside the observed range (7 to 46), and should not be read directly as the average Easy baseline. Evaluated at the sample mean intelligence (23.6), the predicted baseline log error is approximately 2.5, consistent with the raw Easy three-round cell in Supplementary Appendix Table \ref{tab:treatment_outcomes_both_waves}: even the average Easy market exhibits noticeable mispricing, driven by a small number of crashed markets. Using this baseline, the Medium structure shows no statistically significant increase in error ($\beta=2.200,p>0.1$). While the Hard ($\beta=5.327,p<0.05$) and Very Hard ($\beta=6.658,p<0.05$) structures generate large increases in log error relative to the baseline, post-estimation Wald tests using CR2 small-sample adjustments reveal that the means of the adjacent structures are statistically indistinguishable. Accounting for finite-cluster penalties, we fail to reject the equality of coefficients between Medium and Hard ($F_{1,11}=0.62,p=0.448$), as well as between Hard and Very Hard ($F_{1,11}=0.095,p=0.764$).

This extreme variance suggests that the mean effect is driven by crash markets rather than a uniform shift in behavior. To disentangle typical market behavior from these outliers, we analyze the median $(\tau=0.5)$. The intercept of the median regression ($\approx 0.018$; the intelligence terms are negligible at the median, so the zero-intelligence evaluation is immaterial here) implies that the typical baseline market converges close to the truth (implied probability $\approx 98\%$). The Medium structure performs similarly well. However, at the Hard structure, typical success drops (implied probability $\approx 75\%, p<0.01$). At the Very Hard structure, the median error penalty expands to 0.700 ($p<0.001$), dropping the implied probability to $\approx 50\%$. While linear restrictions confirm that the mean effects are statistically indistinguishable due to extreme variance, bootstrapped Wald tests on the median distributions confirm the strict tiered deterioration. The median deterioration from Medium to Hard is highly significant ($\chi^2=7.26, p=0.007$), as is the subsequent deterioration from Hard to Very Hard ($\chi^2=9.58, p=0.002$). Thus, while the Very Hard structure produces extreme errors in the average market, it causes the typical market to be no better than a coin toss.

\begin{figure}[htbp]
    \centering
    \includegraphics[width=0.9\textwidth]{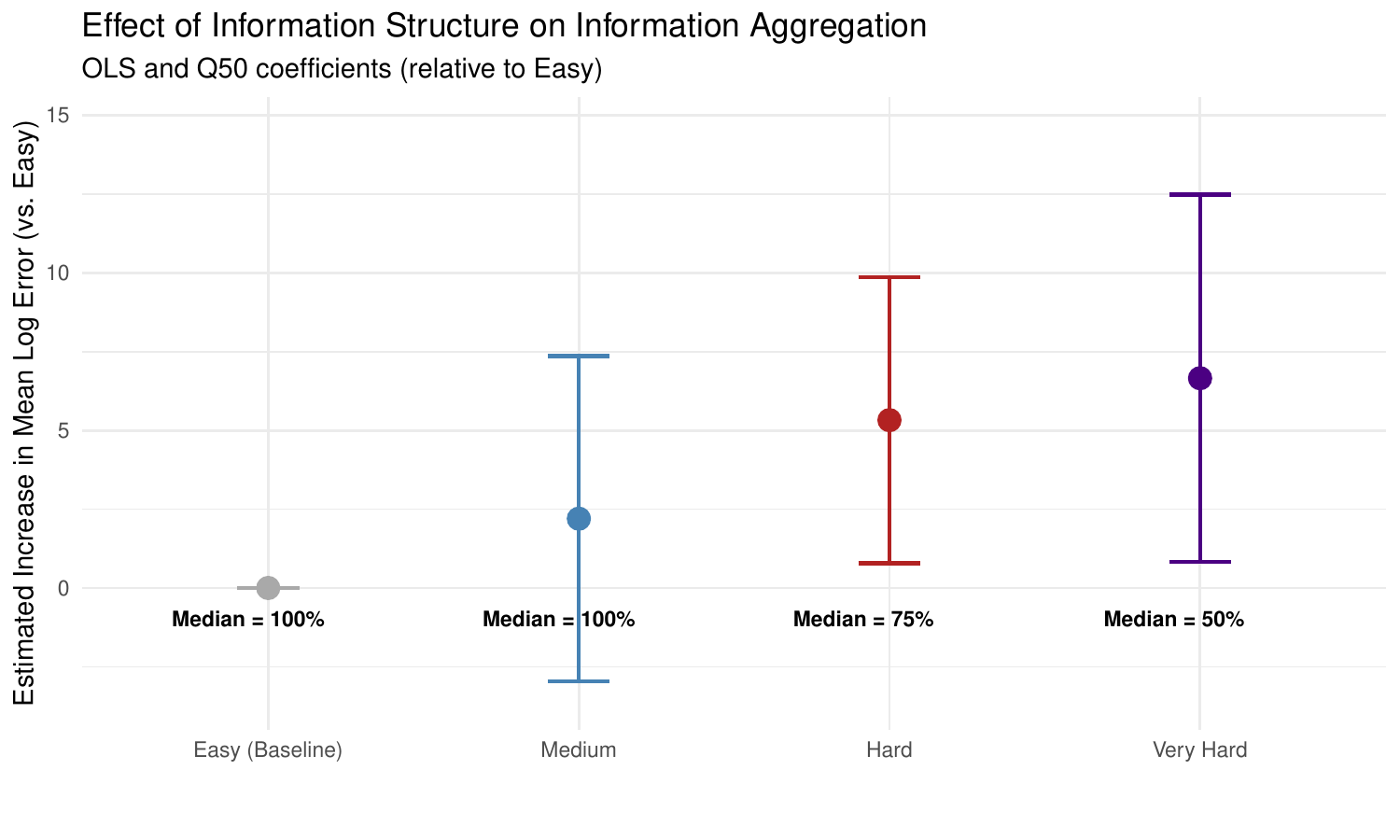}
    \caption{\textbf{The Complexity Effect.} Point estimates and 95\% CR2 confidence intervals represent the OLS mean marginal effect of the structure. Text labels indicate the Q50 median marginal effects, displayed as implied probabilities ($p = e^{\text{-log error}}$).}
    \vspace{0.2cm}
    \begin{singlespace}
        \setlength{\leftskip}{0pt}\setlength{\rightskip}{0pt}\parfillskip=0pt plus 1fil
    \end{singlespace}
    \label{fig:complexity_effect}
\end{figure}

In summary, we reject the hypothesis that information aggregates across all market structures. Figure \ref{fig:complexity_effect} shows the mean and median log errors across the four structures.


\begin{Result}
\label{res:info deteriorates}
Information aggregation deteriorates strictly at higher levels of complexity. While extreme outliers obscure the mean effects, the typical (median) market follows a tiered deterioration: Easy $\approx$ Medium  < Hard  < Very Hard.
\end{Result}

This failure threshold can be benchmarked against the experimental evidence on the depth of interactive reasoning in humans. The Very Hard structure is a version of the dirty faces puzzle, whose laboratory analogue is studied by \cite{weber2001}. He finds that although almost all subjects behave consistently with the theoretical prediction in the simplest case, compliance declines considerably as more steps of iterated rationality are required, in line with earlier evidence that subjects rarely perform more than two steps. Using ring games that identify orders of rationality directly, \cite{kneeland2015} estimates that 93\% of subjects are rational, but only 44\% are third-order and 22\% fourth-order rational. The level-k and cognitive hierarchy literatures reach a similar conclusion, placing the typical depth of human strategic reasoning at one to two levels \citep{nagel1995, costaCrawford2006, camererEtAl2004}. Read against this benchmark, the tiered deterioration of Result \ref{res:info deteriorates} mirrors this human pattern: the median market aggregates information when each trader needs to reason only about his own signal, or about the signal of only one other trader, but breaks down when he needs to reason about the two signals of two other agents, hence requiring more than two levels of interactive reasoning.  
In this respect, AI agents do not appear to relax the cognitive constraint that limits interactive reasoning in human experiments.


\begin{figure}[htbp]
    \centering
    \includegraphics[width=1\textwidth]{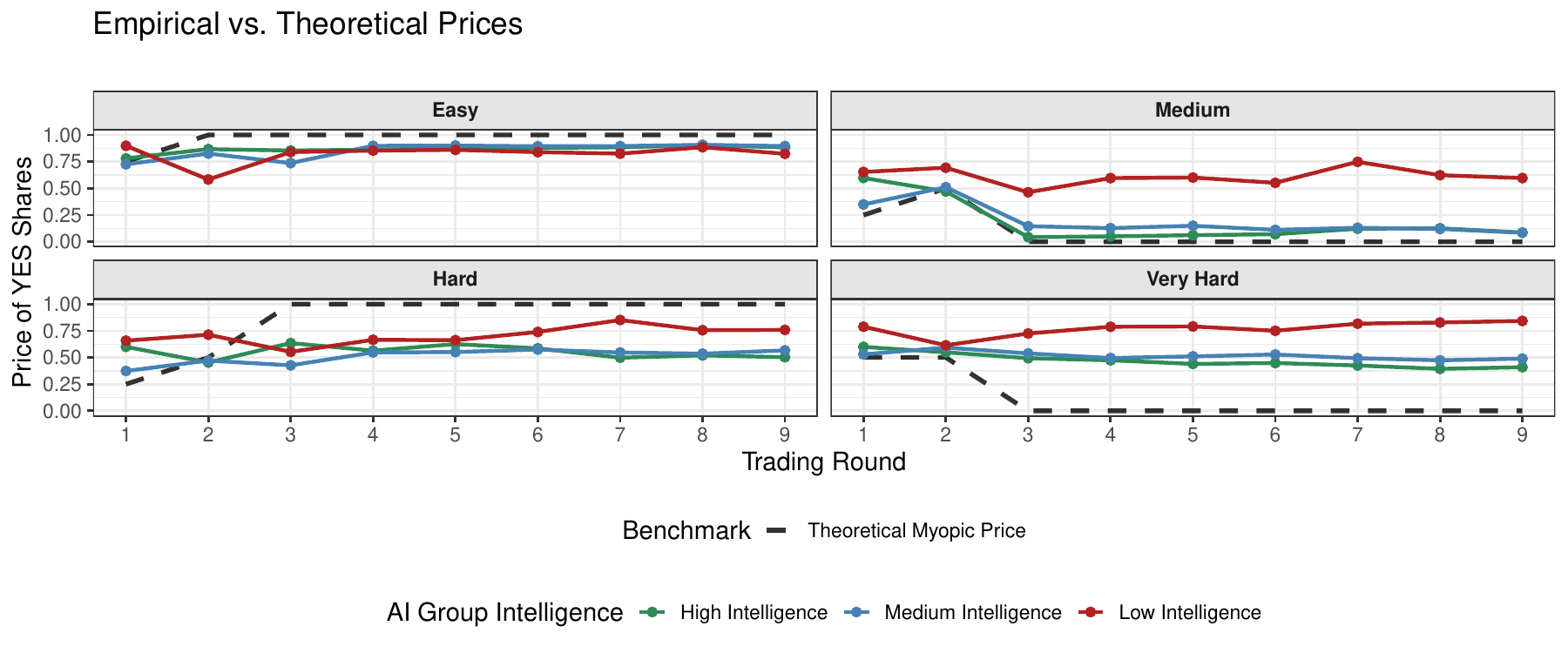}
    \caption{\textbf{Myopically Optimal and Actual Prices Across Structures.} Smarter markets tightly track the myopically optimal prices in the Easy and Medium structures, but revert to 0.5 in the Hard and Very Hard. Low-intelligence markets perform better in the Hard structure but this seems more an artifact of noisy trading rather than sophisticated interactive reasoning.}
    \vspace{0.2cm}
    \begin{singlespace}
        \setlength{\leftskip}{0pt}\setlength{\rightskip}{0pt}\parfillskip=0pt plus 1fil
    \end{singlespace}
    \label{fig:graph_ext_theory_trajectory}
\end{figure}

While Figure \ref{fig:complexity_effect}  demonstrates the magnitude of the final aggregation error, Figure \ref{fig:graph_ext_theory_trajectory} illustrates the dynamic mechanism behind this failure. By plotting the empirical price paths against the theoretical myopic equilibrium derived in Section \ref{sec:myopic prices}, we observe a clear divergence from the myopic benchmark. In the Easy and Medium structures, the High-Intelligence agents tightly track the myopically optimal prices. However, in the Hard and Very Hard structures, this tracking deteriorates. Rather than following the theoretical price to 1 or 0, the smartest agents anchor their prices near 0.5, a trajectory consistent with difficulty in higher-order reasoning. Conversely, the Low-Intelligence (red) agents have a similar price trajectory that converges to 0.75 in both Hard and Very Hard, even though the true value is 1 in the former and 0 in the latter. This suggests that any apparent `success' in complex structures reflects noisy trading rather than sophisticated interactive reasoning.

Our results confirm the hypothesis that neither communication nor strategic prompting significantly influences information aggregation. The main effect of Comments Allowed is statistically insignificant in the OLS model ($\beta=-0.191,p>0.47$), as well as in the median and tail-risk quantile regressions. Similarly, the Strategic Prompt coefficient is statistically indistinguishable from zero ($\beta=0.053,p>0.8$). This lack of significance holds even when controlling for interaction effects with duration, suggesting that the core aggregation mechanism is robust to these variations.

\begin{Result} \label{result: strategic}
Information aggregation is unaffected by prompting AI agents to be strategic or myopic.
\end{Result}

\begin{Result} \label{result: comments}
Information aggregation is unaffected by allowing AI agents to post public comments.
\end{Result}

Communication and strategic prompting may fail to influence information aggregation for at least two reasons. The first is that highly sophisticated AI agents behave according to the theoretical predictions. The second is that they are not very sophisticated, so they fail to understand the prompt, or they cannot plan ahead and implement a strategic plan. 
To understand their thought process better, we conduct a text analysis of their public comments and private justifications in Section \ref{sec:public private comments}.

The initial log error is the log error between the initial price and the true value of $X$. The low initial log error is when the initial price is 0.7 and the true value of $X$ is 1, or the initial price is 0.3 and the true value is 0, whereas the medium initial log error is when the initial price is 0.5. We observe no significant main effect of the low initial log error on market performance ($\beta = 0.01, p > 0.9$), and similarly for the medium initial log error. This suggests that under average conditions, information aggregation is not strongly related to the initial price.

This average result does not hold in the most complex environment. The interaction between the Very Hard structure and the low initial log error (Very Hard $\times$ Initial Error (Low)) generates a large, positive, and statistically significant coefficient in both the OLS mean regression ($\beta = 2.86,p<0.1$) and the 80th percentile tail regression ($\beta = 10.42, p<0.05$). This positive coefficient indicates that, in the most complex environment, markets that begin with a more accurate price signal (lower initial error) end with worse performance than those starting with higher error. The magnitude of this effect in the tail model ($\beta = 10.42$) is very high, suggesting that for fragile markets, a low initial error does not act as an anchor for truth. This effect is absent in the median regression ($\beta  = 0.01, p>0.9$), suggesting that the `typical' market ignores the initial price, which is consistent with the theory. It is interesting to note  that, when human participants are ambiguity averse, \cite{galanisEtAl24} show theoretically and experimentally that the initial price can also degrade information aggregation.

\begin{Result} \label{result:initial price}
On average, information aggregation is unaffected by manipulating the initial price of the market, but the initial price matters in the Very Hard structure.
\end{Result}


Our results support the hypothesis that information aggregation does not deteriorate when the number of rounds increases, which is consistent with the theory. 
Furthermore, the quantile regressions (Models 2 and 3) show no effect of duration on the median or 80th percentile.

\begin{Result} \label{result: duration}
Information aggregation is unaffected by the duration of the market.
\end{Result}



\begin{figure}[htbp]
    \centering
    \includegraphics[width=0.9\textwidth]{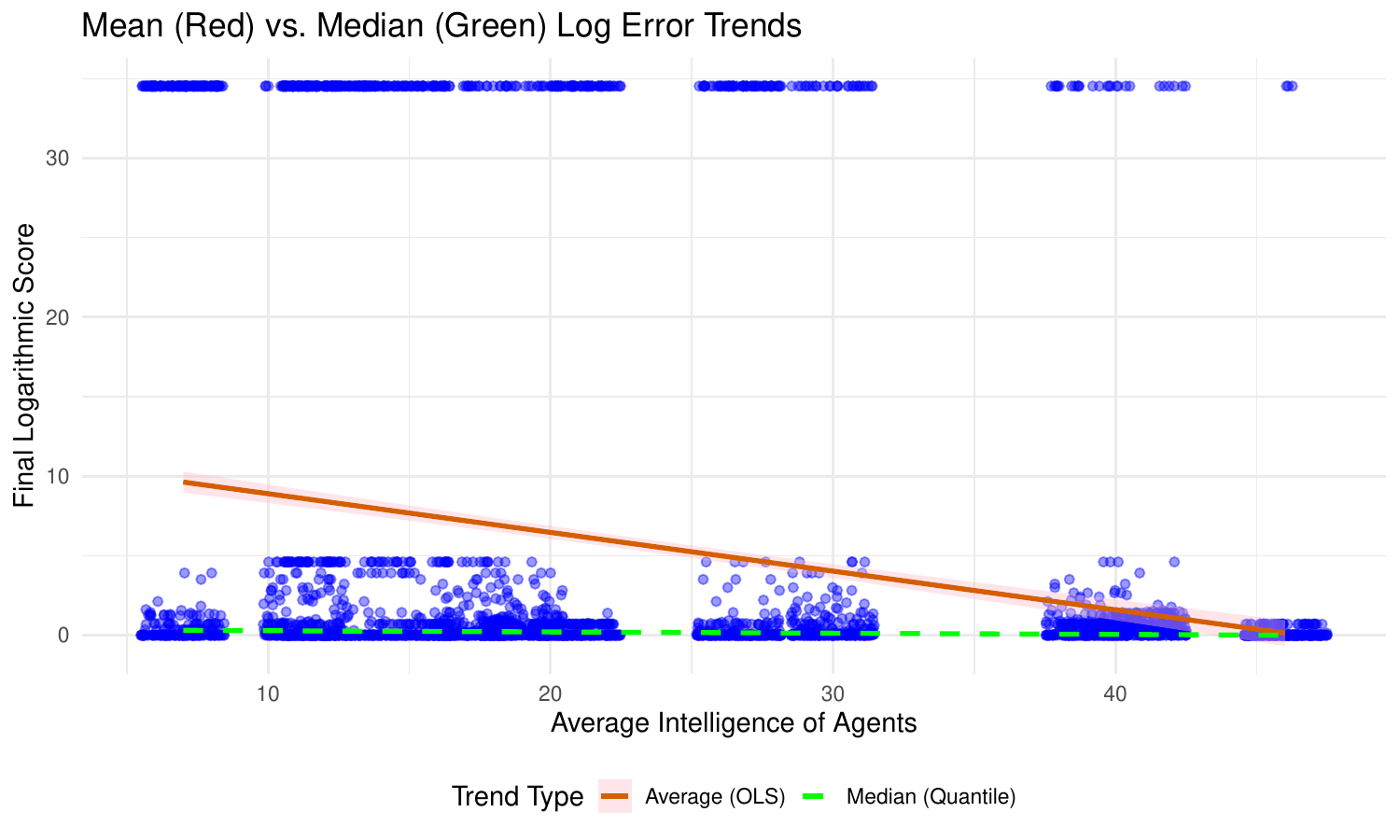}
    \caption{\textbf{The Intelligence Effect.} As the AI agents become smarter, information aggregation (mean log error) improves, but the median log error is unaffected. Points are jittered horizontally to show density of markets with the same intelligence.}
    \vspace{0.2cm}
    \begin{singlespace}
        \setlength{\leftskip}{0pt}\setlength{\rightskip}{0pt}\parfillskip=0pt plus 1fil
    \end{singlespace}
    \label{fig:intelligence_effect}
\end{figure}

Our analysis strongly supports the hypothesis that average intelligence improves information aggregation (Figure \ref{fig:intelligence_effect}). In the OLS specification, Average Intelligence is a highly significant predictor of market success ($\beta=-0.221,p<0.001$), with smarter teams achieving consistently lower log errors. This means that an increase from the lower intelligence scores (7) teams to the highest (46) reduces the logarithmic error penalty by 8.6 points. Because logarithmic scoring scales exponentially, an 8.6 reduction in error implies that the smartest agents assign a probability to the true outcome that is roughly $e^{8.6}$ (over 5,000 times) larger than the probability assigned by the least intelligent agents. Part of this multiplier is mechanical, reflecting the exponential scale of the logarithmic error; the more informative reading is that intelligence acts primarily as a safeguard against extreme mispricing, preventing low-intelligence agents from confidently driving the security price in the opposite direction.

\begin{figure}[htbp]
    \centering
    \includegraphics[width=0.9\textwidth]{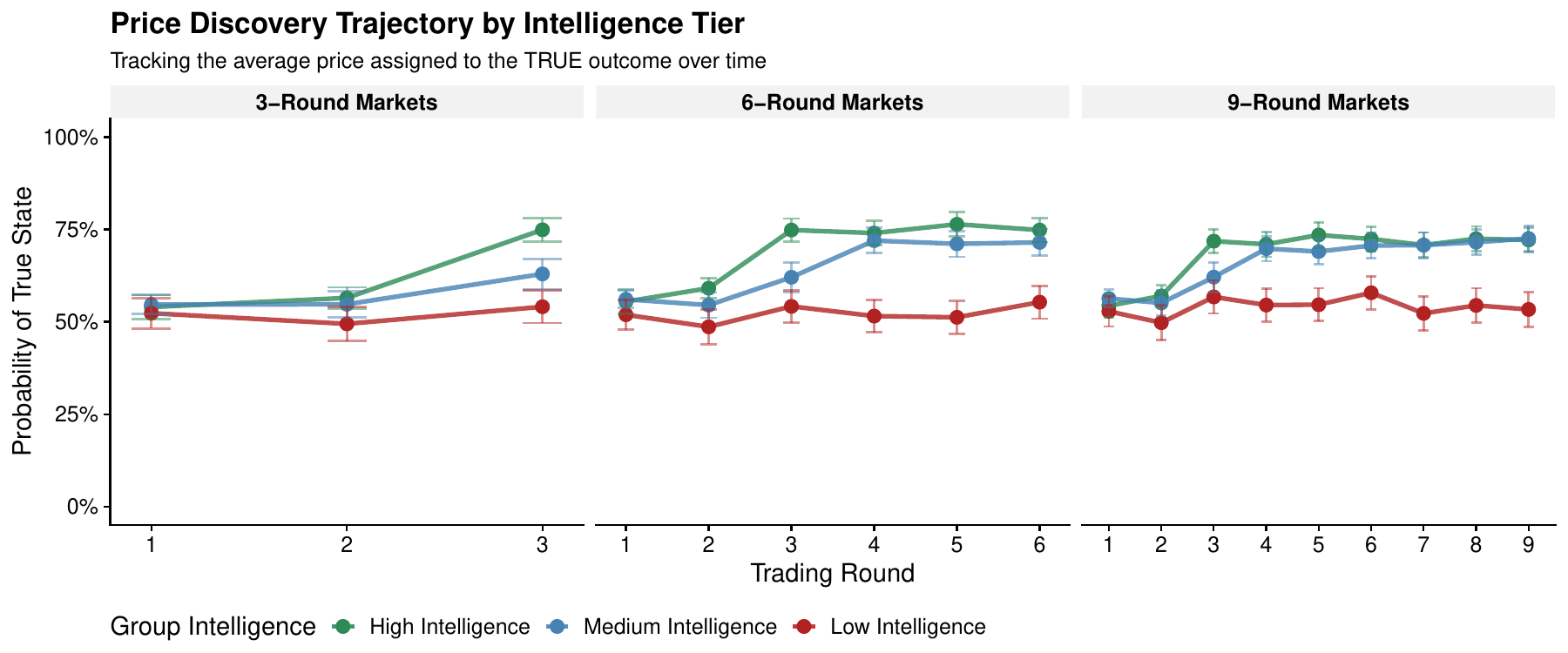}
    \caption{\textbf{The Price Discovery Trajectory.} The vertical axis is the market price of the security that pays 1 in the realized state, and the horizontal axis is the trading round. The three panels correspond to 3-, 6-, and 9-round markets. Lines group markets by team intelligence cohort; higher-intelligence teams are closer to the true value across horizons.}
    \vspace{0.2cm}
    \begin{singlespace}
        \setlength{\leftskip}{0pt}\setlength{\rightskip}{0pt}\parfillskip=0pt plus 1fil
    \end{singlespace}
    \label{fig:graph_price_trajectory}
\end{figure}

Figure \ref{fig:graph_price_trajectory} maps how AI agents, grouped in three intelligence tiers, discover the true value of the security. The probability of the true state is the price of Yes if the answer is Yes and the price of No if the answer is No. The High Intelligence cohort (green line) exhibits rapid informational efficiency, sharply converging toward the true state and subsequently stabilizing. In contrast, the Low Intelligence cohort (red line) exhibits a `sawtooth' trajectory, indicating noisy, inefficient belief updating. 

While higher intelligence improves performance on average, the quantile regressions reveal that this is driven primarily by tail risk mitigation rather than universal optimisation. We find no significant effect of intelligence on the median market $(\beta \approx 0,p>0.5$), suggesting that typical markets reach an accuracy ceiling regardless of marginal gains in cognition. However, intelligence plays a critical role in the tail of the distribution (Q80), where it significantly reduces error ($\beta=-0.069,p<0.01$). This implies that average intelligence acts as a form of crash protection: improving the bottom markets rather than the median. The Intelligence SD (Standard Deviation) measures whether increasing the diversity of the group improves information aggregation. We found an effect only in the tail of the distribution (Q80), where it significantly reduces error ($\beta=-0.052,p<0.05$).

\begin{Result} \label{result: average intelligence}
Information aggregation improves as the average intelligence increases. 
\end{Result}

%
%

\subsection{Trading Volume and Profits}
\label{sec:trading volume profits}

In Table \ref{tab:trading volume and profits robust}, we report two regressions, on trading volume and individual profits, using the same independent variables as in the previous models on information aggregation. We first find that trading volume across structures broadly aligns with myopic theoretical predictions, though with notable statistical caveats. First, while point estimates suggest volume increases from the Easy baseline to the Very Hard structure, this difference is completely statistically insignificant (p>0.1). Second, both the Hard and Medium structures generate marginally significantly more volume than the Easy baseline (p<0.1). Consistent with Hypothesis \ref{hypothesis: volume},  a post-estimation Wald test using the CR2 cluster-robust variance-covariance matrix fails to reject equality between the Hard and Medium coefficients ($p=0.332$), indicating that trading volume between the Hard and Medium structures is statistically indistinguishable. The Very Hard coefficient is too imprecisely estimated to be statistically distinguished from either the Easy baseline or the Hard and Medium structures, so the placement of Very Hard in the ordering below reflects point estimates rather than pairwise tests.

\begin{Result} 
Trading volume is ordered, from lowest to highest, as follows: Easy $\approx$ Very Hard  < Hard  $\approx$ Medium.
\end{Result}

Second, we reject Hypothesis \ref{hypothesis: positive profits} that individual profits are positive. The average realized profit per trader is $-46$, and the covariate-adjusted position-specific estimates reported below are significantly negative for the first two trading positions, so that on average, participation in the market generates financial losses. See Figure \ref{fig:graph_profits_by_model} for a breakdown of profits by AI model, where the only model with positive profits is Gemini 3 Flash. However, an analysis of the trading sequence reveals a clear pattern by trading position. At first glance, the coefficients for trading second or third appear statistically indistinguishable from the baseline first-mover ($p>0.1$). Yet, the point estimates diverge: trading second carries a $-42.1$ penalty, while trading third yields a $77.4$ premium. A post-estimation Wald test, using CR2 errors clustered at the unique market level (the level at which the three competing positions are correlated; see the note to Figure \ref{fig:graph_profit_position_theory_wald}), rejects equality between the second and third trading positions ($p<0.001$). Under the more conservative team-composition clustering with small-sample corrections, the position-level losses reported below remain significant (trader 1: $p \approx 0.01$; trader 2: $p<0.001$), while the second-versus-third contrast weakens to marginal significance ($p=0.076$); the late-mover advantage is therefore robust in sign and magnitude, but its statistical significance depends on the assumed correlation structure, consistent with our discussion in Section \ref{sec:methodological}.
 
 To test our theoretical prediction that all market participants should generate weakly positive profits, we calculated the estimated marginal means for profit across all three trading positions. Applying one-sided tests using the market-level CR2 variance-covariance matrix ($H_a$: Profit<0), we reject the theoretical baseline for agents who trade first and second. The first mover generates a highly significant loss (Estimated Profit $=-57.7, p < 0.001$), as does the second mover (Estimated Profit $=-99.8, p < 0.001$). The third trader is the only participant to extract positive value from the market (Estimated Profit $= 19.7$). See Figure \ref{fig:graph_profit_position_theory_wald} for an illustration.



\begin{Result}
Market participation generates losses for the first two traders and profits for the third. Trading third yields a significant financial advantage over trading second.
\end{Result}

We also reject Hypothesis \ref{hypothesis:profits correlated} that average profits are uncorrelated with the difficulty of the structure. As shown in Table \ref{tab:trading volume and profits robust}, the negative coefficient observed in the Medium ($\beta=-30.5, p>0.1$) treatment is not statistically significant relative to the Easy baseline. However, as complexity increases further, profits decline. Post-estimation Wald tests using the market-level CR2 matrix reject equality between the Medium and Hard structures ($p<0.001$), as well as between the Hard and Very Hard structures ($p=0.015$). This suggests that once complexity crosses a certain threshold, average profits decrease. This is theoretically consistent with Result \ref{res:info deteriorates}: worse information aggregation equates to an increasing divergence between the final market price and the true value of the security, mechanically resulting in lower average profits.

\begin{Result}
Individual profits decrease as the structure becomes more complex: Easy $\approx$ Medium > Hard > Very Hard.
\end{Result}

%

The analysis of individual agent profitability (Table \ref{tab:trading volume and profits robust}, Column 2) reveals that financial success in these markets is primarily driven by relative rather than absolute cognitive advantage. We find a strong positive effect of individual intelligence on profit ($\beta =14.08,p<0.001$), coupled with a nearly equal and opposite negative effect from the average intelligence of the market ($\beta =-11.02,p<0.001$). This confirms Hypothesis \ref{hypothesis: profits correlated with intelligence} that agents benefit significantly from their own cognitive capacity, but suffer from the intelligence of their competitors.

\begin{Result}
Individual profitability is driven by relative cognitive advantage: an agent's profits increase with their own intelligence, but decrease as the average intelligence of their competitors rises.
\end{Result}

%



\begin{table}[!htbp] \centering 
  \caption{Trading Volume and Profits Without Disclosure} 
  \label{tab:trading volume and profits robust} 
\small 
\begin{tabular}{@{\extracolsep{5pt}}lcc} 
\\[-1.8ex]\hline 
\hline \\[-1.8ex] 
 & \multicolumn{2}{c}{\textit{Dependent variable:}} \\ 
\cline{2-3} 
 \\[-1.8ex] & Trading Volume & Individual Profits \\ 
\\[-1.8ex] & (1) & (2)\\ 
\hline \\[-1.8ex] 
 Constant & 1,795.988$^{.}$ (1,064.073) & $-$80.719$^{*}$ (31.683) \\ 
  Comments Allowed & 85.077$^{*}$ (42.516) & 2.723 (4.048) \\ 
  Duration: 6 Rounds & 543.398$^{***}$ (108.871) & 2.447 (20.007) \\ 
  Duration: 9 Rounds & 1,037.835$^{***}$ (192.429) & $-$5.396 (17.292) \\ 
  Strategic Prompt & $-$65.396$^{.}$ (36.642) & $-$0.491 (2.876) \\ 
  Medium (t3s110) & 308.391$^{.}$ (174.854) & $-$30.548 (33.429) \\ 
  Hard (t3s111) & 128.183$^{.}$ (74.893) & $-$69.955$^{*}$ (32.035) \\ 
  Very Hard (t3s111o2ye2) & 55.372 (266.568) & $-$98.904$^{*}$ (42.381) \\ 
  Initial Error: Low & $-$180.641$^{**}$ (68.906) & $-$10.767$^{**}$ (3.809) \\ 
  Initial Error: Medium & $-$14.704 (34.607) & $-$4.386 (2.691) \\ 
  Average Intelligence & $-$27.925 (24.667) & $-$11.019$^{***}$ (2.252) \\ 
  Individual Intelligence &  & 14.079$^{***}$ (2.150) \\ 
  Intelligence SD & 22.347 (41.617) & 0.606 (2.473) \\ 
  Trades Second &  & $-$42.127 (28.229) \\ 
  Trades Third &  & 77.390 (55.932) \\ 
  6 Rounds x Strategic & 9.810 (55.547) & $-$1.725 (4.474) \\ 
  9 Rounds x Strategic & 87.983 (62.172) & $-$3.680 (4.496) \\ 
  Medium Struct x Initial Error (Low) & 273.357$^{.}$ (158.440) & $-$21.173 (19.587) \\ 
  Hard Struct x Initial Error (Low) & 87.880 (70.669) & $-$13.015 (10.054) \\ 
  Very Hard x Initial Error (Low) & 414.759 (264.776) & $-$32.672 (23.473) \\ 
  Medium Struct x Initial Error (Med) & 41.568 (67.534) & 4.394 (4.099) \\ 
  Hard Struct x Initial Error (Med) & 62.984 (55.047) & 11.491 (8.982) \\ 
  Very Hard x Initial Error (Med) & $-$238.720 (401.800) & $-$1.778 (11.232) \\ 
 \hline \\[-1.8ex] 
Observations & 1,772 & 5,316 \\ 
R$^{2}$ & 0.104 & 0.060 \\ 
\end{tabular} 
\parbox{0.95\linewidth}{\textit{Notes:} $^{.}$p$<$0.1; $^{*}$p$<$0.05; $^{**}$p$<$0.01; $^{***}$p$<$0.001. Column 1 is a market-level regression where the dependent variable is total trading volume, measured as the absolute quantity of shares traded in the market. Column 2 is a trader-level regression where the dependent variable is the trader's realized LMSR profit. Both models use CR2 cluster-robust standard errors clustered by exact AI team composition, featuring small-sample degrees of freedom adjustments. Rounds, Information Structure, and Trader Position (Model 2 only) use treatment contrasts, with `Round 3', `t3s111y2' (Easy), and `Trades First' serving as their respective baselines. All other controls (Initial Error, Strategic, Comments) use sum contrasts, where coefficients represent deviations from the grand mean.}
\end{table}

\subsection{Information Provision}
\label{sec:information provision}

To test Hypotheses \ref{hypothesis: info on accuracy} and \ref{hypothesis: info on profits}, we introduce an information provision treatment (e.g., \cite{bergEtAl95},  \cite{caiEtAl2009}). Specifically, we use the empirical results from our initial baseline sample (1,772 markets) to construct a disclosure prompt. This summary information is then provided to an independent, non-overlapping later-wave sample of 1,728 new markets. By pooling the baseline and treatment samples, we estimate the association between this disclosure prompt and market outcomes using a treatment indicator.

Sequential later-wave comparisons can be confounded by secular model or platform changes, prompt-context changes, or other implementation differences. We reduce this concern by using identical, version-pinned LLM models across both the baseline and treatment samples via a stateless API, so the models have no endogenous memory of the baseline markets. Nevertheless, because the disclosure was not randomized within a single contemporaneous wave, we interpret the estimates as later-wave disclosure associations rather than as strictly equivalent to within-wave random assignment.

%
%

In Supplementary Appendix \ref{sa:sec:information provision} we report the Experiment Disclosure prompt, which was dynamically generated based on which model was trading. Supplementary Appendix Table \ref{tab:disclosure_quantile_regressions_robust} presents the mixed-contrast regression models for information aggregation across the full, pooled sample of 3,500 markets, introducing a dummy variable for the Experiment Disclosure treatment. We find no support for Hypothesis \ref{hypothesis: info on accuracy}. The provision of historical market data yields no statistically significant improvement in market accuracy across the mean ($\beta=2.075, p>0.1$), the median ($\beta=-0.00004, p>0.1$), or the tail-risk distributions. Furthermore, the interaction term between Disclosure and Average Intelligence is  null. Within the limits of this later-wave comparison, even the most highly capable LLMs in our sample did not detectably improve their probability estimations in response to the explicit strategic feedback.

\begin{Result} \label{result: information provision}
Information provision has a null effect on information aggregation.
\end{Result}

Supplementary Appendix Table \ref{tab:disclosure trading volume and profits robust} replicates this analysis for market activity and individual profitability. Consistent with Hypothesis \ref{hypothesis: info on profits}, the disclosure treatment has no statistically observable impact on either trading volume ($\beta=171.9, p>0.1$) or individual profits ($\beta=-21.05, p>0.1$). Providing all participants with the exact same strategic heuristics did not generate a measurable advantage for any individual trader.

\begin{Result}
Information provision has a null effect on individual profits.
\end{Result}

Finally, analyzing the pooled sample of 3,500 markets confirms that all primary findings from the first wave of experiments are highly robust. The baseline market mechanics remain unchanged: strategic prompting, communication, initial price, and market duration have no significant impact on information aggregation. Conversely, the structural complexity of the market continues to significantly hinder aggregation, while higher average intelligence strictly improves it.

\subsection{Public and Private Comments}
\label{sec:public private comments}

Our finding that strategic prompting does not influence information aggregation is consistent with our theoretical prediction that being myopic or strategic does not have an impact. However, this null result may also be driven by the inability of LLMs to execute dynamic plans \citep{kambhampatiEtAl2024}. In our experiment, a strategic agent could coordinate an inter-temporal plan by leaving instructions for its future self in the private comments. Alternatively, he could manipulate competitors by deceiving them in public comments, either by lying outright or by hoarding information (revealing less in public than in private).

%

To investigate these behavioral mechanics, we construct three measures of communication strategy. First, we compute the {\bf cosine similarity} between an agent's private and public messages, a standard natural language processing metric for semantic alignment ranging from 0 (divergent) to 1 (identical)  \citep{manning1999}.%
\footnote{Each text is transformed into a vector, where a dimension is a unique word and its value is the number of times it appears in the text. The cosine of the angle between the two vectors measures the lexical overlap between the two texts, a standard proxy for their semantic similarity, ranging from 0 (very different) to 1 (very similar).}

Second, we measure {\bf information hoarding} via the word gap (the number of words in the private minus the public message). In markets where public messages are allowed, private messages average 83.5 words while public messages average only 40; private messages are longer on average in 99\% of markets, and strictly longer in 94\% of individual messages (Figure \ref{fig:graph_density_length_messages}). 

Finally, we measure explicit {\bf deception}. Using an LLM-as-a-judge methodology, we task our most capable model in the first wave, Gemini 3 Flash, with evaluating, in every round of every comments-allowed market, whether the public comment truthfully reveals the agent's private signal (Truthful Revelation), avoids disclosing the signal (Information Withholding), or states the opposite of the true signal (Lying).%
\footnote{Note that for structure t3s111o2ye2 we asked twice, as there are two signals for each trader.}
We estimate an ordered logit model across these three hierarchical categories, where positive coefficients indicate a shift toward withholding or lying.

Table \ref{tab:combined_communication_models} presents the pooled models for cosine similarity (OLS, Column 1), information hoarding (OLS, Column 2), and deception (Ordered Logit, Column 3). The strategic prompt has a statistically insignificant effect on cosine similarity and only marginally increases information hoarding by less than one word ($\beta=0.788, p<0.1$). In the Supplementary Appendix Tables \ref{tab:split_sim_models}, \ref{tab:split_duration_models}, and \ref{tab:split_ord_models}, we report the same regressions when splitting the sample by market duration (3, 6, and 9 rounds). We find that even this weak, marginal effect on the word gap fails to survive consistently when splitting the sample by market duration; it retains the same sign in all three subsamples but it is statistically significant only in the 9-round markets. Thus, we do not find robust evidence that the strategic prompt changes the communication metrics, and the design cannot distinguish whether agents are unable or unwilling to implement more complex deceptive strategies. Individual intelligence similarly provides mixed results, correlating with less explicit lying but marginally more information hoarding. Moreover, information hoarding and deception increase as the structure becomes more complex.

The Experiment Disclosure treatment does, however, change communication behavior. Disclosing the past history widens the gap between private and public messages ($14.8$ words, p<0.001) and increases deception ($0.86$, $p<0.001$). Recall that  one of the qualitative results we disclosed was that public comments have no effect on information aggregation. 

Finally, by modelling the current round as a categorical variable, the regression uncovers a pattern which is consistent with end-of-horizon disclosure, but also with simpler round-number heuristics. Rather than exhibiting a simple linear decay in deception and information hoarding over time, there is a ``sawtooth'' pattern. Agents are most deceptive and guarded in Round 1. However, the probability of truthful revelation increases significantly, indicated by large negative coefficients, at Rounds 3, 6, and 9 (p < 0.001). These increases coincide with the scheduled final rounds of the short, medium, and long market treatments. In the intervening rounds (e.g., Rounds 4, 5, and 7), concealment rates rise back toward the Round 1 baseline. This suggests that AI agents understand that in the final round revealing the truth will not have an adverse effect on their profits.

Because the trading order is fixed, this pattern admits a sharper test. Each trader has his \emph{own} final trading opportunity, and these differ by position: in a six-round market, trader 1 trades for the last time in Round 4 and trader 2 in Round 5; in a nine-round market, in Rounds 7 and 8 respectively. Once a trader has no further trades, revealing his private signal cannot reduce his profits, irrespective of what he knows or believes about the other signals: his LMSR payoff depends only on his existing holdings and the realised outcome. A trader who reasons about his own horizon should therefore reveal at his final move, whatever his position. This prediction fails for traders 1 and 2. In the deception models split by duration (Table \ref{tab:split_ord_models}), the revelation response at trader 1's final move is weak (Round 4: $-0.278$ in six-round markets) or statistically null (Round 7: $-0.183$ in nine-round markets), and null at trader 2's (Round 5: $-0.145$; Round 8: $-0.062$). By contrast, Round 6 spikes strongly ($-0.615$, $p<0.001$) even in nine-round markets, where it is terminal for nobody: the trader acting in Round 6 trades again in Round 9, so protecting his information rents dictates concealment there. Revelation thus tracks the salient round numbers 3, 6, and 9, which are the possible market durations stated in the prompt, rather than any trader's actual horizon, individual or aggregate.%
\footnote{Rounds 3, 6, and 9 are always trader 3's turns, so a round-number heuristic cannot be fully separated from behavior specific to the third trading position. The distinction is immaterial for our conclusion: under either interpretation, no trader conditions his revelation on his own remaining trading opportunities, which is the hallmark of backward induction. The failure of traders 1 and 2 to reveal at their own final moves is identified from their own turns and is unaffected by this equivalence.}
Agents behave as if rounds that are multiples of three could be terminal, even when the horizon is known to be nine rounds, and fail to exploit the costless revelation available at their own final moves, which provides evidence against backward induction over their individual trading horizons.

\begin{table}[!htbp] \centering 
\small
  \caption{Agent Communication Strategy: Text Similarity, Word Gap, and Deception} 
  \label{tab:combined_communication_models} 
\resizebox{\linewidth}{!}{%
\begin{tabular}{@{\extracolsep{5pt}}lccc} 
\\[-1.8ex]\hline 
\hline \\[-1.8ex] 
 & \multicolumn{3}{c}{\textit{Dependent variable:}} \\ 
\cline{2-4} 
 & Cosine Similarity (OLS) & Word Gap (OLS) & Deception (Ordered Logit) \\ 
\\[-1.8ex] & (1) & (2) & (3)\\ 
\hline \\[-1.8ex] 
 Constant & 0.324$^{***}$ (0.043) & 33.880$^{*}$ (15.423) &  \\ 
  Strategic Prompt & $-$0.003 (0.002) & 0.788$^{.}$ (0.416) & 0.006 (0.036) \\ 
  Round 2 & 0.030$^{*}$ (0.014) & $-$13.194$^{.}$ (7.874) & $-$0.398$^{***}$ (0.074) \\ 
  Round 3 & 0.004 (0.013) & $-$25.105$^{*}$ (9.969) & $-$0.665$^{***}$ (0.076) \\ 
  Round 4 & 0.030$^{*}$ (0.012) & $-$12.247$^{**}$ (3.723) & $-$0.205$^{**}$ (0.079) \\ 
  Round 5 & 0.040$^{**}$ (0.015) & $-$23.619$^{**}$ (7.966) & $-$0.160$^{.}$ (0.082) \\ 
  Round 6 & 0.005 (0.016) & $-$30.972$^{**}$ (9.860) & $-$0.560$^{***}$ (0.091) \\ 
  Round 7 & 0.038$^{.}$ (0.020) & $-$18.893$^{***}$ (5.068) & $-$0.201$^{*}$ (0.101) \\ 
  Round 8 & 0.047$^{**}$ (0.015) & $-$26.708$^{**}$ (8.346) & $-$0.081 (0.099) \\ 
  Round 9 & 0.033$^{.}$ (0.019) & $-$32.644$^{***}$ (9.907) & $-$0.487$^{***}$ (0.112) \\ 
  Duration: 6 Rounds & 0.002 (0.005) & $-$0.468 (0.906) & $-$0.036 (0.093) \\ 
  Duration: 9 Rounds & $-$0.002 (0.005) & $-$1.072 (0.866) & $-$0.017 (0.094) \\ 
  Medium (t3s110) & $-$0.022$^{**}$ (0.007) & $-$1.213 (1.063) & 1.198$^{***}$ (0.108) \\ 
  Hard (t3s111) & 0.002 (0.006) & 2.417$^{*}$ (1.099) & 0.907$^{***}$ (0.124) \\ 
  Very Hard (t3s111o2ye2) & 0.001 (0.008) & 5.666$^{***}$ (1.423) & 1.369$^{***}$ (0.103) \\ 
  Experiment Disclosure & 0.010 (0.018) & 14.858$^{***}$ (3.174) & 0.861$^{***}$ (0.074) \\ 
  Initial Error & 0.002 (0.012) & 2.867 (2.227) & 0.268 (0.208) \\ 
  Individual Intelligence & 0.001 (0.002) & 0.844$^{.}$ (0.484) & $-$0.004$^{.}$ (0.002) \\ 
 \hline \\[-1.8ex] 
Observations & 10,620 & 10,620 & 13,189 \\ 
R$^{2}$ & 0.027 & 0.156 &  \\ 
\end{tabular}%
}
\parbox{0.95\linewidth}{\textit{Notes:} $^{.}$p$<$0.1; $^{*}$p$<$0.05; $^{**}$p$<$0.01; $^{***}$p$<$0.001. Column 1 reports cosine similarity between private and public messages, ranging from 0 (semantically different) to 1 (identical). Column 2 reports the word gap, defined as private-message word count minus public-message word count. Column 3 is an ordered logit where the deception outcome is ordered Truthful Revelation $<$ Information Withholding $<$ Lying, so positive coefficients indicate a shift toward withholding or lying. Column 3 has more observations because the Very Hard structure contains two private signals per trader and is judged separately for each signal. All models use a mixed-contrast specification. Treatment contrast baselines are: Round 1 for Round, 3 Rounds for Duration, Easy (t3s111y2) for Structure, and No Disclosure for Experiment Disclosure. The Strategic Prompt variable uses sum contrasts, representing the deviation from the grand mean, whereas Initial Error is mean-centered. Columns 1 and 2 (OLS) report CR2 cluster-robust standard errors in parentheses, clustered by exact AI team composition with small-sample degrees of freedom adjustments. Column 3 (Ordered Logit) reports CR1 cluster-robust standard errors clustered at the market level to account for intra-market correlation across competing agents and repeated communication rounds.}
\end{table}

\subsection{Robustness with Frontier AI Models}
\label{sec:robustness frontier}

The rapid pace of innovation in AI is evidenced by the frequent release of new models every few months.  To test whether the failure of information aggregation in complex environments (Result \ref{res:info deteriorates}) is a transient artifact of older AI capabilities, we conducted in April 2026 an out-of-sample robustness check using a cohort of frontier LLMs. This subsample comprises 576 markets played across all four structures (Easy, Medium, Hard, and Very Hard). We employ four distinct teams: three homogeneous markets populated by GPT-5.4, Claude Opus 4.6, and Gemini 3.1 Pro, and a heterogeneous team containing one agent of each model. Rather than comparing average log errors across the two cohorts, which is sensitive to the small number of markets that crash, we describe both cohorts by classifying each market according to how close to the truth its final price ends up.

\begin{figure}[htbp]
    \centering
    \includegraphics[width=1\textwidth]{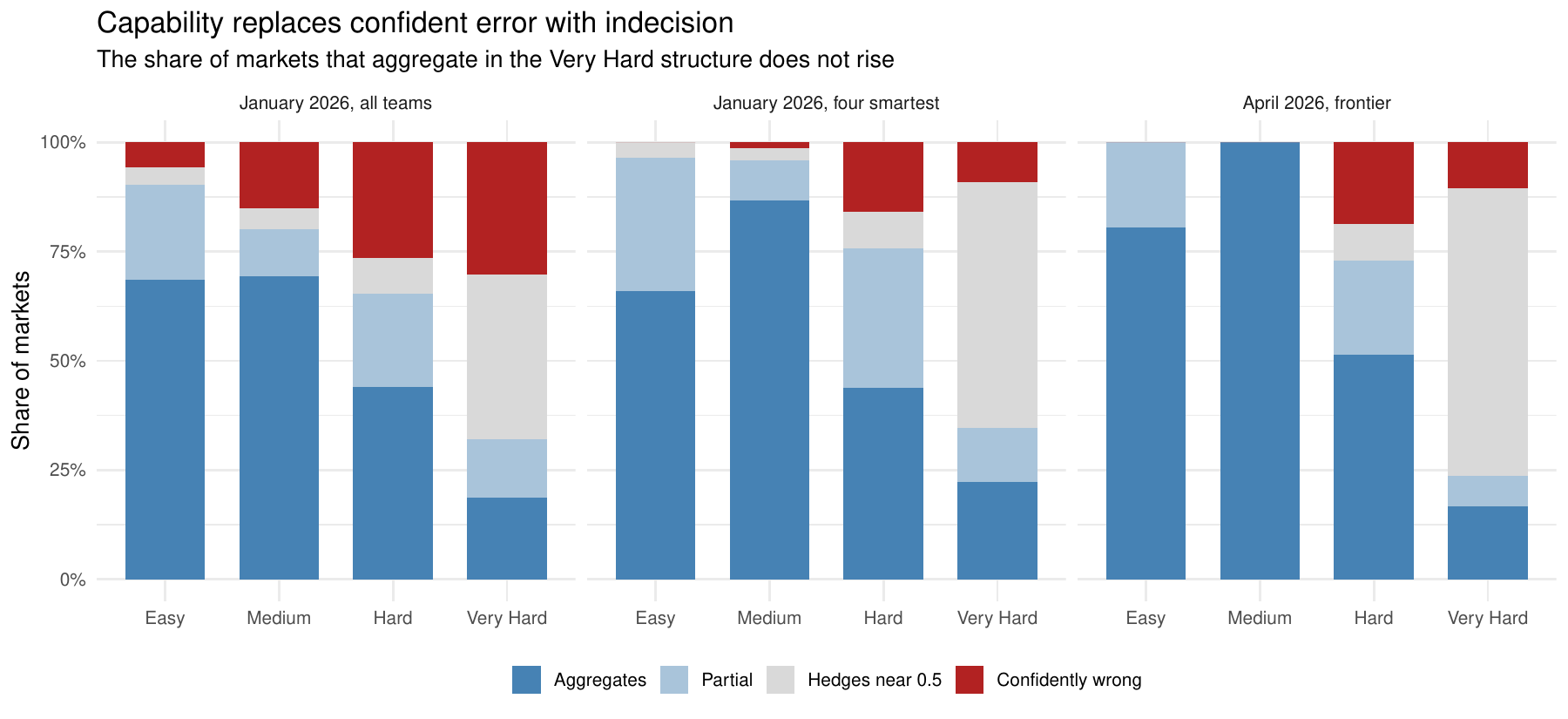}
    \caption{\textbf{January vs. April Cohorts.} Each market is classified by the distance between its final price and the true value of the security. As capability rises, from all twelve teams of the first wave to its four smartest teams and then to the frontier cohort, confidently wrong markets are replaced by markets that hedge near 0.5, while the share of markets that aggregate in the Very Hard structure does not rise.}
    \vspace{0.2cm}
    \begin{singlespace}
        \setlength{\leftskip}{0pt}\setlength{\rightskip}{0pt}\parfillskip=0pt plus 1fil
    \end{singlespace}
    \label{fig:frontier_outcome_shape}
\end{figure}

We define four mutually exclusive and exhaustive categories, using the final price $p$ and the true value $v$ of the security. A market \textit{aggregates} if $|p-v|<0.1$, it is \textit{confidently wrong} if $|p-v|>0.9$, it \textit{hedges} if $|p-v|$ lies within 0.1 of 0.5, and it is \textit{partial} otherwise.%
\footnote{In the Very Hard structure, tightening the tolerance to 0.05 gives aggregation shares of 17.3\%, 20.8\% and 15.3\% for the three groups below, and loosening it to 0.2 gives 19.6\%, 25.0\% and 17.4\%, against 18.7\%, 22.2\% and 16.7\% at the tolerance we use, so the difference is small.}
We report three groups of markets: all twelve teams of the first wave, the four \textit{smartest} teams of the first wave, ranked by the average Artificial Analysis intelligence index accessed in January 2026, and the four frontier teams of April 2026.%
\footnote{These are: 3x Gemini 3 Flash, 3x GPT-5 mini, the mixed team of Claude Haiku 4.5, Gemini 3 Flash and GPT-5 mini, and 3x Claude Haiku 4.5, with index values of 46, 41, 39 and 30, as shown in Table \ref{tab:all ai models}. The index is used only within the first wave and only at that single access date, since the scores of the April models were accessed later and are not comparable. Selecting instead the four teams with the lowest realized mean log error gives 19.4\% of markets aggregating in the Very Hard structure, so the picture does not depend on the selection rule.} %
Table \ref{tab:frontier_outcome_shape} reports the shares and Figure \ref{fig:frontier_outcome_shape} displays them.%
\footnote{Figures \ref{fig:graph_mse_by_model_structure} and \ref{fig:graph_frontier_mse_by_model_structure} in the Supplementary Appendix report the mean squared error by structure for each team of the two cohorts.}

The frontier models are more capable in the first three structures than those of the first wave. In the Easy structure the share of markets that aggregate rises from 68.5\% to 80.6\%, and no frontier market is confidently wrong, against 5.8\% in the first wave. In the Medium structure every one of the 144 frontier markets aggregates, against 69.3\%, and the 15.2\% of first-wave markets that were confidently wrong disappear entirely. The frontier cohort also outperforms the four smartest teams of the first wave in the Easy, Medium and Hard structures, at 80.6\% against 66.0\%, 100\% against 86.8\%, and 51.4\% against 43.8\%. 

\begin{figure}[htbp]
    \centering
    \includegraphics[width=0.9\textwidth]{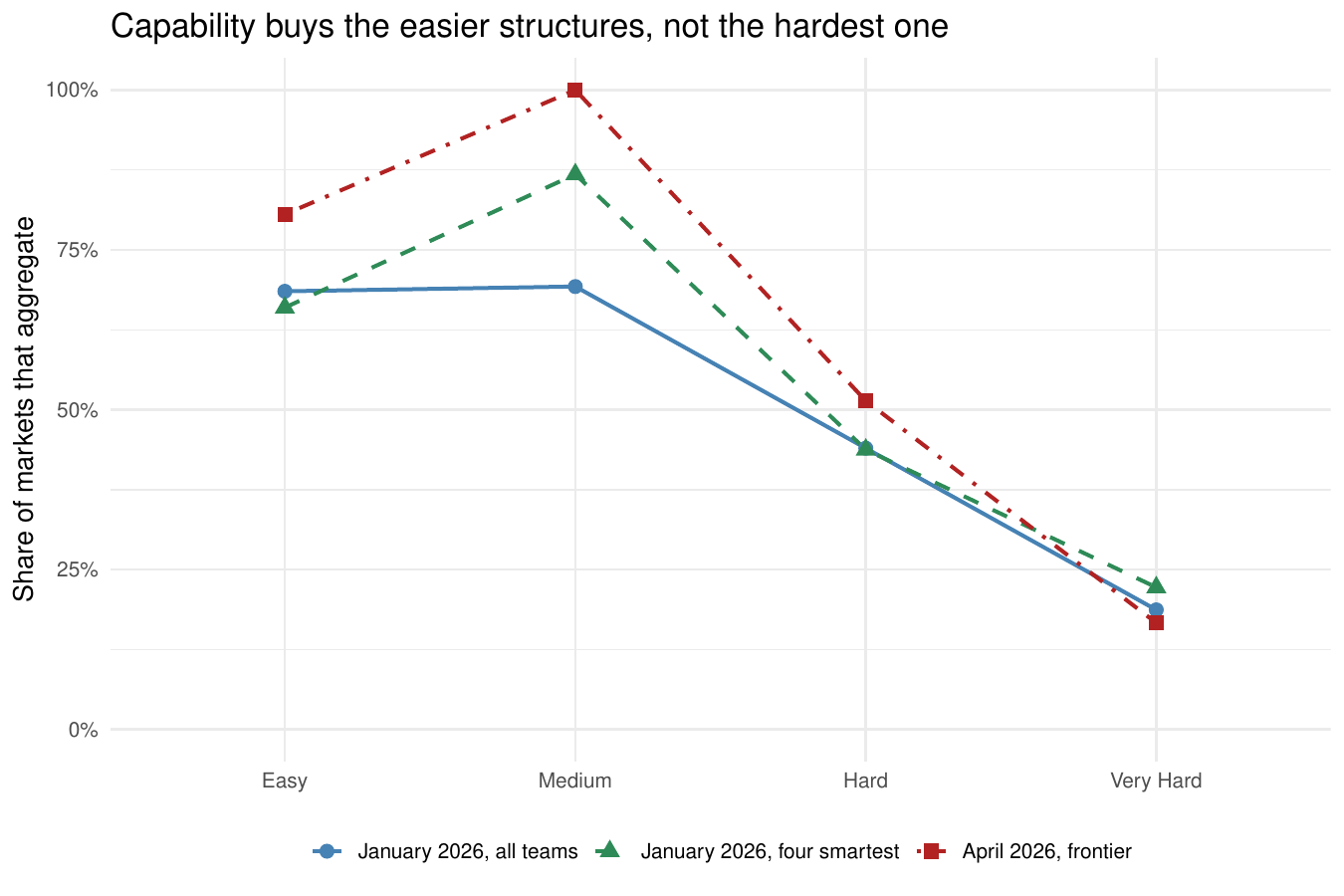}
    \caption{\textbf{Aggregation by Structure and Group.} The share of markets whose final price lies within 0.1 of the true value of the security. The gain from capability is concentrated in the Easy and Medium structures and disappears in the Very Hard structure.}
    \vspace{0.2cm}
    \begin{singlespace}
        \setlength{\leftskip}{0pt}\setlength{\rightskip}{0pt}\parfillskip=0pt plus 1fil
    \end{singlespace}
    \label{fig:frontier_aggregation_by_structure}
\end{figure}

In the Very Hard structure, however, the frontier group does not perform better. The share of markets that aggregate is 18.7\% for the first wave, 22.2\% for its four smartest teams, and 16.7\% for the frontier cohort. What changes with capability is not the frequency of aggregation but the composition of the failures. The share of markets that are confidently wrong falls from 30.3\% to 9.0\% and 10.4\%, while the share that hedge near 0.5 rises from 37.6\% to 56.2\% and 66.0\%. Two thirds of frontier markets in the hardest structure therefore end at a price that conveys no information about the state, and the gain from capability consists of avoiding the confident mispricing of the weaker models rather than of pricing the security correctly more often. Figure \ref{fig:frontier_aggregation_by_structure} makes the same point in terms of the aggregation share alone: the three groups separate in the Easy and Medium structures and get very close in the Very Hard structure.

\begin{Result}
\label{res:frontier shape}
Capability replaces confident error with indecision rather than with information aggregation. Moving from the first wave to its smartest teams and then to the frontier cohort, the share of Very Hard markets that are confidently wrong falls from 30.3\% to 9.0\% and 10.4\% and the share that hedge near 0.5 rises from 37.6\% to 56.2\% and 66.0\%, while the share that aggregate remains between 16.7\% and 22.2\%. Two thirds of the frontier markets in the Very Hard structure end at a price close to 0.5.
\end{Result}

Read together with Result \ref{res:info deteriorates}, this suggests that the deterioration we document is not an artifact of the models available when the first wave was run. Capability gains are spent where the required depth of interactive reasoning is low, and where that depth rises the additional capability shows up as a reluctance to move the price away from 0.5 rather than as an ability to infer the private information of others from their orders. The market becomes less confidently wrong without becoming more informative, which is a benign failure for a trader and an unhelpful one for anyone who reads prices to learn about the state.

\subsection{Statistical Power and Minimum Detectable Effects}
\label{sec:power}

Several of our central findings are null results: communication, strategic prompting, the initial price, and market duration do not significantly improve information aggregation. Because a failure to reject is informative only to the extent that the design could have detected an effect had one been present, we compute the minimum detectable effect (MDE) for each treatment null by Monte Carlo simulation operating through the exact CR2/Satterthwaite estimator used in our main specifications. On the mean, the design is well powered for communication and duration: it detects a comments effect of $0.83$ log-error units (about 7\% of the outcome standard deviation) and duration effects of roughly $2.0$ units (about 17\%), so these nulls are informative. For strategic prompting, the joint MDE for the main effect and its duration interactions is $1.09$ log-error units, so we can rule out large strategic effects but not small ones, and the initial-price contrasts are not detectable on the mean. The quantile specifications are better powered, because they are robust to the small number of ``crash'' markets that dominate the sampling variance of the mean. In the Very Hard structure the median MDE for strategic prompting is $0.36$ against a baseline median log error of $0.69$, so that null is informative and not an artifact of a floor; in the Hard structure the median MDE ($0.38$) exceeds the baseline error ($0.28$), so the null is uninformative there and we make no tail claims. Supplementary Appendix \ref{sec:power_ia} reports the simulation design and the full set of MDEs, by specification and by information structure.

\section{Conclusion}
\label{conclusion}

Understanding whether AI agents can reason interactively and aggregate information is important for several reasons. First, the informational efficiency of prices is a basic property of financial markets, and AI agents are increasingly participating in them, as well as being deployed in firms. Second, an essential ability of AI agents is to make predictions, hence it is natural to examine whether this ability can be improved by making them trade in a prediction market. Finally, the capacity to reason about the private information of others by observing their actions is a fundamental human quality. While it is tempting to anthropomorphize highly capable AI models and assume they inherently possess these higher-order interactive reasoning skills, empirical evidence remains scarce.

This paper provides experimental evidence on information aggregation and interactive reasoning among autonomous AI agents within a prediction market framework. Consistent with economic theory, we find little evidence that cheap talk communication, market duration, or strategic prompting materially improve market accuracy; initial price has little average effect but matters in the Very Hard structure. Markets perform worse in the harder information and payoff structures. While the median AI team successfully solves the Easy and Medium structures, performance deteriorates sharply in the Hard and Very Hard environments. This suggests that current LLMs do not yet demonstrate the hyper-rational interactive reasoning capabilities often assumed in economic theory. Instead, they exhibit interactive reasoning capabilities comparable to human subjects in similar experiments.

Within our main sample, the intelligence index is positively associated with information aggregation. A robustness check with a cohort of frontier models released in April 2026 shows that they aggregate information more often than the first-wave models in the three easier structures, but not in the Very Hard structure. There, higher capability raises the share of markets that hedge near 0.5 and lowers the share that are confidently wrong, without raising the share that aggregate (Result \ref{res:frontier shape}).


At the individual level, an agent's profitability is more closely related to relative than absolute cognitive advantage. Higher individual intelligence is associated with higher profits, while higher average team intelligence is associated with lower individual profits. Furthermore, the order of trading matters: traders who move third earn higher profits than those who move second, consistent with a late-mover advantage in this design.

Our later-wave disclosure exercise also reveals limitations in how LLMs process feedback from past play in this setting. Contrary to recent literature suggesting LLMs can self-improve via iterative prompting, we find that providing agents with information about past play does not improve market accuracy. In the message-based outcomes, the disclosure condition is associated with more information hoarding and deceptive communication.

Finally, text analysis of the agents' private and public messages shows that they struggle to act strategically and show no sign of designing or executing a dynamic plan. They reveal more information as the game approaches the end, indicating that they have some understanding about its dynamic nature. However, they also exhibit a sawtooth pattern of revelation that spikes in rounds 3, 6, and 9, which are always trader 3's turns, even when the game ends in 9 rounds, whereas traders 1 and 2 do not reveal more when it is their last period. This suggests that their reasoning may rely on token-matching heuristics instead of backward induction.

Ultimately, our findings suggest a nuanced frontier for trading with AI agents. While they demonstrate competence in basic market mechanics and, within our main sample, perform better when the team has a higher intelligence index, their difficulty with strategic communication and multi-step interactive reasoning indicates a significant distance until they resemble the sophisticated agents usually assumed in economic theory.

\appendix


\bibliographystyle{apalike}


\bibliography{master}

\clearpage

\setcounter{section}{0}
\setcounter{table}{0}
\setcounter{figure}{0}
\setcounter{equation}{0}
\renewcommand{\thesection}{SA.\arabic{section}}
\renewcommand{\thetable}{SA.\arabic{table}}
\renewcommand{\thefigure}{SA.\arabic{figure}}
\renewcommand{\theequation}{SA.\arabic{equation}}
\renewcommand{\theHsection}{SA.\arabic{section}}
\renewcommand{\theHtable}{SA.\arabic{table}}
\renewcommand{\theHfigure}{SA.\arabic{figure}}

\setcounter{page}{1}
\renewcommand{\thepage}{\arabic{page}}

\begin{center}
{\Large \textbf{Supplementary Appendix}}\\[0.35cm]
{\large Information Aggregation with AI Agents}\\[0.35cm]
\end{center}

%

\section{Separability}
\label{separability}

In this section we define the notion of separability and show that the securities are separable in all four structures of the experiment. We consider a finite state space $\Omega$ and a set of traders $I$ with $n=|I|$.  Trader \is initial private information is represented by partition $\Pi_{i}$ of $\Omega$. Let $\Pi_i(\omega)$ be a partition element of $\Pi_{i}$ that contains $\omega$, so that $\omega \in \Pi_i(\omega) \in \Pi_{i}$. When the true state is $\omega \in \Omega$, Trader $i$ considers all states in $\Pi_i(\omega) \sq \Omega$ to be possible. We assume that the join (the coarsest common refinement) of partitions $\Pi = \{\Pi_1, \ldots \Pi_n\}$ consists of singleton sets so that $\underset{i \in I}{\bigcap} \Pi_i(\omega) = \omega$ for all $\omega \in \Omega$, which means that the traders' pooled information always reveals the true state. Note that this assumption is satisfied in all four structures of the experiment. Let $E_{\mu}[X|\Pi_{i}(\omega)]$ be the expected value of security $X$, conditional on private information $\Pi_i(\omega)$ and prior $\mu$. Let $Supp(\mu)$ be the support of $\mu$.

\begin{Definition}
\label{separable}
A security $X$ is called non-separable under information structure $\Pi$ if there exists probability distribution $\mu$ and value $v\in \mathbb{R}$ such that:
\begin{itemize}
\item[$(i)$] $X(\omega) \neq v$ for some $\omega \in Supp(\mu)$,
\item[$(ii)$] $E_{\mu}[X|\Pi_{i}(\omega)]=v$ for all $i=1,...,n$ and $\omega \in Supp(\mu)$. 
\end{itemize}
Otherwise, it is called separable.
\end{Definition}

For structures t3s111 and t3s110, the security $X$ is Arrow-Debreu, paying 1 in one state and 0 otherwise. \cite{ostrovsky12} shows that Arrow-Debreu securities are separable irrespective of the information structure. For the other two structures, we will use the following characterization of separable securities by \cite{ostrovsky12}. It specifies that $X$ is separable if and only if, for any $v$,  we can find numbers $\lambda_i(\Pi_i(\omega))$, for each $i$ and $\omega$, such that the sum over all traders has the same sign as the difference of $X(\omega) -v$. Intuitively, for any $v$ and at each $\omega$, all traders ``vote'' and the sign of the sum of the votes has to agree with the sign of the difference between the value of the security and $v$.

\begin{Proposition}[\cite{ostrovsky12}] \label{ostrovsky char thm}
Security $X$ is separable under partition structure $\Pi$ if and only if, for every $v \in {\mathbb R}$, there exist functions $\lambda_i: \Pi_i \rightarrow {\mathbb R}$ for $i = 1, \ldots, n$ such that, for every state $\omega$ with $X(\omega) \neq v$, 
\[(X(\omega) - v)\underset{i \in I}{\sum} \lambda_i(\Pi_i(\omega)) > 0.\]
\end{Proposition} 

Suppose that $v$ is strictly between $0$ and $1$. For structure t3s111y2, assign $\lambda_1 = 1$ to partition cells $\{a,b,c,d\}, \{a,b,e,f\}, \{a,c,e,g\}$ and $\lambda_2 = -1$ to all other cells. For structure t3s111o2ye2, assign $\lambda_1 = -1$ to cells $\{a,e\}, \{a,c\}, \{a,b\}$, $\lambda_2 = -10$ to cells $\{d,h\}, \{f,h\}, \{g,h\}$, and $\lambda_3 = 4$ to all other cells. Then, it is straightforward to check that the inequality $(X(\omega) - v)\underset{i \in I}{\sum} \lambda_i(\Pi_i(\omega)) > 0$ is satisfied for both structures and all states. If $v \geq 1$ or $v \leq 0$, then $(X(\omega) - v)$ has the same sign for all $\omega$ with $X(\omega) \neq v$, so finding appropriate $\lambda$ is trivial.

\section{Statistical Power and Minimum Detectable Effects}
\label{sec:power_ia}

Several of our central findings are null results: communication, strategic prompting, the initial price, and market duration do not significantly improve information aggregation. Because a failure to reject is informative only to the extent that the design could have detected an effect had one been present, we report an ex-ante power analysis that quantifies the smallest effect each test can detect. We deliberately avoid closed-form power formulas (e.g., those based on the non-central $F$ distribution), because they assume independent and identically distributed errors and would substantially overstate power by ignoring both the intra-team correlation and the small number of clusters ($G= 12$) that govern our CR2/Satterthwaite inference. Instead, we compute power by Monte Carlo simulation operating through the exact estimator and inference used in our main specifications. 

For each treatment coefficient we hold the design matrix fixed, inject a known effect of magnitude $\delta$, and regenerate the outcome by resampling residuals in two components, a cluster-level shock and a within-cluster deviation, so that both the intra-team correlation and the pronounced right-skew of the log-error distribution are preserved. We then re-estimate the model, apply the CR2/Satterthwaite test, and record whether it rejects at the 5\% level; repeating this $1{,}000$ times yields the rejection probability (power) at each $\delta$. The minimum detectable effect (MDE) is the value of $\delta$ at which power reaches 80\%. Under the null ($\delta = 0$) the simulated rejection rate is close to the nominal level (between 0.04 and 0.07 across coefficients), confirming that the CR2 procedure is correctly sized rather than mechanically conservative.

Table \ref{tab:power_mde} reports the MDEs for the mean (OLS), median (Q50), and tail (Q80) specifications. For the mean outcome, the design is well powered for communication and duration: it detects a comments effect of $0.83$ log-error units (about 7\% of the outcome standard deviation) and duration effects of roughly $2.0$ units (about 17\%). These nulls are therefore informative, as effects of the magnitude that would be economically meaningful would have been detected. For strategic prompting, the relevant test of the strategic-prompting hypothesis is the joint nullity of the main effect and its interactions with duration; the joint MDE is $1.09$ log-error units, so we can rule out strategic effects larger than roughly one log-error unit on the mean, though not smaller ones. The initial-price contrasts are the least precisely estimated on the mean: their effects are not detectable within the simulated range, reflecting that these contrasts enter an interaction with the information structure and are identified off limited within-cell variation. The common thread is that the mean outcome is the least precise of the three, because its sampling variance is dominated by a small number of ``crash'' markets; the quantile specifications, which are robust to these outliers, are correspondingly better powered.

\begin{table}[!htbp] \centering
\caption{Minimum Detectable Effects for the Treatment Nulls}
\label{tab:power_mde}
\small
\makeatletter
\begin{tabular}{lccccc}
\hline\hline \\[-1.8ex]
 & CR2 SE & MDE: Mean & (\% of SD) & MDE: Median & MDE: Tail \\
 & (OLS)  & (OLS, sim.) &          & (Q50)       & (Q80)      \\
\hline \\[-1.8ex]
Comments Allowed & 0.319 & 0.810 & 6.9\% & 0.009 & 0.107 \\
Duration: 6 Rounds & 1.172 & 2.031 & 17.3\% & 0.014 & 0.379 \\
Duration: 9 Rounds & 1.250 & 2.117 & 18.0\% & 0.014 & 0.298 \\
Strategic Prompt & 0.292 & 1.376 & 11.7\% & 0.008 & 0.215 \\
Initial Error (Low) & 0.120 & $>$1.2 & --- & 0.012 & 0.165 \\
Initial Error (Medium) & 0.186 & $>$1.9 & --- & 0.009 & 0.190 \\
\hline\hline
\end{tabular}
\makeatother

\vspace{0.5ex}
\begin{minipage}{0.95\linewidth}
\footnotesize \textit{Notes}: The MDE is the smallest true effect detectable with 80\% power at $\alpha=0.05$, expressed in log-error units. The Mean (OLS) column is from Monte Carlo simulation ($1{,}000$ replications) operating through the CR2/Satterthwaite estimator; the outcome standard deviation is $11.8$. For the Strategic Prompt, the table reports the single-coefficient MDE for comparability across rows; the joint MDE for the main effect and its duration interactions is $1.09$. The initial-error effects are not detectable within the simulated grid ($\pm 10$ CR2 standard errors), so no value is reported. The Low and Medium rows are the two estimated sum-contrast deviations of the initial log error; the High deviation is the omitted level, equal to minus the sum of the other two. The Median (Q50) and Tail (Q80) columns are not simulated: they report the analytic MDE, equal to $2.802$ times the bootstrapped standard error ($200$ replications) of the corresponding quantile regression. The $1{,}000$ Monte Carlo replications apply only to the OLS mean column; the $200$ replications refer to the quantile-regression bootstrap and match the bootstrapped inference used elsewhere in the paper.
\end{minipage}
\end{table}

The median (Q50) specification is well powered throughout, with pooled MDEs near $0.01$ log-error units. This precision is partly mechanical, however: in the Easy and Medium structures the typical market already aggregates, so the median log error sits at zero and there is little room for any treatment to move it. To separate this floor from genuine precision, Table \ref{tab:power_mde_bystructure} re-estimates the quantile MDEs separately within each structure. 
In the Hard and Very Hard structures, the median log error sits well above zero (0.28 and 0.69), creating room for treatments to matter. However, the strategic MDE at the median in the Hard structure is 0.38, which exceeds the baseline error of 0.28, indicating that the design lacks the power to reliably detect even a complete correction to perfect accuracy in this specific setting. By contrast, in the Very Hard structure, the MDE of 0.36 is well below the median outcome of 0.69. The null for strategic prompting is therefore highly informative in the most complex environment: strategic prompting does not meaningfully improve the typical Very Hard market, and this failure is not an artifact of a statistical floor. By contrast, the tail (Q80) is well powered only in the Easy structure; in the harder structures the crash-driven variance inflates the bootstrapped standard errors, so tail MDEs become large (up to $19$ log-error units in the Hard structure). We accordingly refrain from making tail claims about strategic prompting or initial price in the Hard structure, where crash-driven variance inflates the tail MDEs. However, in the Very Hard structure, the tail MDE for the low initial price drops to 3.48, providing adequate power to detect the large mispricings ($\beta=10.42$) reported in Result \ref{result:initial price}.


%

\begin{table}[!htbp] \centering
\caption{Minimum Detectable Effects by Information Structure}
\label{tab:power_mde_bystructure}
\small
\makeatletter
\resizebox{\linewidth}{!}{%
\begin{tabular}{lccccccc}
\hline\hline \\[-1.8ex]
          & Median   & \multicolumn{3}{c}{Median (Q50) MDE} & \multicolumn{3}{c}{Tail (Q80) MDE} \\
\cline{3-5}\cline{6-8} \\[-1.8ex]
Structure & Log Err. & Strategic & Init.\ (Low) & Init.\ (Med.) & Strategic & Init.\ (Low) & Init.\ (Med.) \\
\hline \\[-1.8ex]
Easy & 0.00 & 0.04 & 0.01 & 0.00 & 0.21 & 0.15 & 0.15 \\
Medium & 0.00 & 0.01 & 0.02 & 0.02 & 3.07 & 4.79 & 2.50 \\
Hard & 0.28 & 0.38 & 0.16 & 0.17 & 19.26 & 8.48 & 6.45 \\
Very Hard & 0.69 & 0.36 & 0.43 & 0.40 & 1.67 & 3.48 & 2.66 \\
\hline\hline
\end{tabular}}
\makeatother

\vspace{0.5ex}
\begin{minipage}{0.95\linewidth}
\footnotesize \textit{Notes}: Analytic MDE $= 2.802 \times$ the bootstrapped standard error ($200$ replications) from quantile regressions estimated separately within each structure, in log-error units. A median log error near zero indicates that the typical market already aggregates (a floor), so the corresponding MDE is uninformative; a median log error above zero indicates room for treatments to move outcomes, so a small MDE represents an informative null.
\end{minipage}
\end{table}


Taken together, the power analysis disciplines the interpretation of our nulls. 
The nulls for communication and duration are informative on the mean: effects larger than roughly $0.8$ and $2.0$ log-error units, respectively, would have been detected. The null for strategic prompting is informative at the typical market specifically in the Very Hard structure, where the design detects effects as small as 0.36 log-error units (well below the baseline error of 0.69). In the Hard structure, however, the median MDE (0.38) exceeds the baseline error (0.28), rendering the null uninformative for the typical market. On the mean, we can exclude only large strategic effects, and we make no tail claims in either complex structure. 
The null for the initial price is the weakest overall: on the pooled mean, its average effect is not detectable. However, the structure-specific divergence identified in Result \ref{result:initial price} is adequately powered. Specifically, the design is well powered to detect an effect on the typical (median) Very Hard market, as its MDE is small ($0.40 - 0.43$). Simultaneously, the tail is adequately powered (Very Hard Q80 MDE of 3.48) to detect the large mispricings ($\beta =10.42$) associated with the initial price in the worst-performing markets. This pattern is consistent with our broader finding that the informational content of these markets resides in the typical outcome and the tail behaves idiosyncratically.

\section{Robustness Checks by Market Duration}
\label{sec:duration_robustness}

In this section we report the text-analysis regressions of Table \ref{tab:combined_communication_models} estimated separately for markets lasting 3, 6, and 9 rounds.

%

\begin{table}[!htbp] \centering 
\small
  \caption{Robustness Check: Cosine Similarity by Market Duration (CR2 Standard Errors)} 
  \label{tab:split_sim_models} 
\resizebox{\linewidth}{!}{%
\begin{tabular}{@{\extracolsep{5pt}}lccc} 
\\[-1.8ex]\hline 
\hline \\[-1.8ex] 
 & \multicolumn{3}{c}{\textit{Dependent variable:}} \\ 
\cline{2-4} 
\\[-1.8ex] & \multicolumn{3}{c}{Cosine Similarity  (Public, Private)} \\ 
 & 3-Round & 6-Round & 9-Round \\ 
\\[-1.8ex] & (1) & (2) & (3)\\ 
\hline \\[-1.8ex] 
 Constant & 0.331$^{***}$ (0.058) & 0.311$^{***}$ (0.044) & 0.329$^{***}$ (0.042) \\ 
  Strategic Prompt & $-$0.004 (0.004) & $-$0.003 (0.004) & $-$0.003 (0.004) \\ 
  Round 2 & 0.025$^{.}$ (0.014) & 0.033$^{*}$ (0.015) & 0.032$^{*}$ (0.013) \\ 
  Round 3 & 0.005 (0.016) & 0.005 (0.014) & 0.003 (0.014) \\ 
  Round 4 &  & 0.029$^{.}$ (0.015) & 0.032$^{**}$ (0.011) \\ 
  Round 5 &  & 0.042$^{**}$ (0.016) & 0.039$^{**}$ (0.014) \\ 
  Round 6 &  & 0.009 (0.017) & 0.003 (0.018) \\ 
  Round 7 &  &  & 0.039$^{*}$ (0.018) \\ 
  Round 8 &  &  & 0.047$^{***}$ (0.013) \\ 
  Round 9 &  &  & 0.032$^{.}$ (0.019) \\ 
  Medium (t3s110) & $-$0.017 (0.014) & $-$0.015 (0.010) & $-$0.029$^{***}$ (0.008) \\ 
  Hard (t3s111) & $-$0.006 (0.010) & 0.012 (0.010) & $-$0.001 (0.007) \\ 
  Very Hard (t3s111o2ye2) & 0.001 (0.015) & 0.005 (0.010) & $-$0.001 (0.007) \\ 
  Experiment Disclosure & 0.004 (0.019) & 0.011 (0.019) & 0.011 (0.018) \\ 
  Initial Error & 0.009 (0.026) & 0.001 (0.016) & 0.001 (0.012) \\ 
  Individual Intelligence & 0.001 (0.002) & 0.001 (0.002) & 0.001 (0.002) \\ 
 \hline \\[-1.8ex] 
Observations & 1,776 & 3,552 & 5,292 \\ 
R$^{2}$ & 0.015 & 0.034 & 0.029 \\ 
\end{tabular}%
}
\parbox{0.95\linewidth}{\textit{Notes:} $^{.}$p$<$0.1; $^{*}$p$<$0.05; $^{**}$p$<$0.01; $^{***}$p$<$0.001. CR2 cluster-robust standard errors, clustered by AI team composition, are reported in parentheses. The model uses a mixed-contrast specification. Treatment contrast baselines are: Round 1 for Round, Easy (t3s111y2) for Structure, and No Disclosure for Experiment Disclosure. The Strategic Prompt variable uses sum contrasts, representing the deviation from the grand mean, whereas Initial Error is mean-centered.}
\end{table} 

\begin{table}[!htbp] \centering 
\small
  \caption{Robustness Check: Word Gap by Market Duration (CR2 Standard Errors)} 
  \label{tab:split_duration_models} 
\resizebox{\linewidth}{!}{%
\begin{tabular}{@{\extracolsep{5pt}}lccc} 
\\[-1.8ex]\hline 
\hline \\[-1.8ex] 
 & \multicolumn{3}{c}{\textit{Dependent variable:}} \\ 
\cline{2-4} 
\\[-1.8ex] & \multicolumn{3}{c}{Word Gap} \\ 
 & 3-Round Markets & 6-Round Markets & 9-Round Markets \\ 
\\[-1.8ex] & (1) & (2) & (3)\\ 
\hline \\[-1.8ex] 
 Constant & 31.127$^{.}$ (17.734) & 35.468$^{*}$ (15.756) & 32.393$^{*}$ (14.326) \\ 
  Strategic Prompt & 0.682 (0.596) & 0.114 (0.480) & 1.282$^{*}$ (0.645) \\ 
  Round 2 & $-$13.385 (8.313) & $-$12.328 (8.341) & $-$13.802$^{.}$ (7.461) \\ 
  Round 3 & $-$24.149$^{*}$ (10.091) & $-$25.950$^{*}$ (10.454) & $-$24.945$^{*}$ (9.743) \\ 
  Round 4 &  & $-$12.730$^{**}$ (3.883) & $-$11.833$^{**}$ (4.002) \\ 
  Round 5 &  & $-$23.211$^{**}$ (8.601) & $-$24.144$^{**}$ (7.531) \\ 
  Round 6 &  & $-$31.712$^{**}$ (10.433) & $-$30.467$^{**}$ (9.379) \\ 
  Round 7 &  &  & $-$18.861$^{***}$ (5.144) \\ 
  Round 8 &  &  & $-$26.700$^{***}$ (8.090) \\ 
  Round 9 &  &  & $-$32.712$^{***}$ (9.620) \\ 
  Medium (t3s110) & $-$2.959 (2.270) & $-$3.839$^{*}$ (1.759) & 1.157 (1.207) \\ 
  Hard (t3s111) & 0.958 (2.769) & $-$0.100 (1.270) & 4.582$^{**}$ (1.554) \\ 
  Very Hard (t3s111o2ye2) & 3.865 (2.826) & 4.459$^{***}$ (1.302) & 7.073$^{***}$ (1.788) \\ 
  Experiment Disclosure & 12.640$^{***}$ (2.191) & 15.753$^{***}$ (3.654) & 14.981$^{***}$ (3.707) \\ 
  Initial Error & 2.330 (2.853) & 2.005 (4.383) & 3.618$^{*}$ (1.491) \\ 
  Individual Intelligence & 1.050$^{.}$ (0.578) & 0.813$^{.}$ (0.486) & 0.796$^{.}$ (0.452) \\ 
 \hline \\[-1.8ex] 
Observations & 1,776 & 3,552 & 5,292 \\ 
R$^{2}$ & 0.143 & 0.152 & 0.159 \\ 
\end{tabular}%
}
\parbox{0.95\linewidth}{\textit{Notes:} $^{.}$p$<$0.1; $^{*}$p$<$0.05; $^{**}$p$<$0.01; $^{***}$p$<$0.001. CR2 cluster-robust standard errors, clustered by AI team composition, are reported in parentheses. The model uses a mixed-contrast specification. Treatment contrast baselines are: Round 1 for Round, Easy (t3s111y2) for Structure, and No Disclosure for Experiment Disclosure. The Strategic Prompt variable uses sum contrasts, representing the deviation from the grand mean, whereas Initial Error is mean-centered.}
\end{table}

\begin{table}[!htbp] \centering 
\small
  \caption{Robustness Check: Propensity to Deceive by Market Duration} 
  \label{tab:split_ord_models} 
\resizebox{\linewidth}{!}{%
\begin{tabular}{@{\extracolsep{5pt}}lccc} 
\\[-1.8ex]\hline 
\hline \\[-1.8ex] 
 & \multicolumn{3}{c}{\textit{Dependent variable:}} \\ 
\cline{2-4} 
\\[-1.8ex] & \multicolumn{3}{c}{Deception} \\ 
 & 3-Round & 6-Round & 9-Round \\ 
\\[-1.8ex] & (1) & (2) & (3)\\ 
\hline \\[-1.8ex] 
 Strategic Prompt & $-$0.039 (0.066) & 0.006 (0.056) & 0.020 (0.056) \\ 
  Round 2 & $-$0.426$^{***}$ (0.126) & $-$0.364$^{**}$ (0.130) & $-$0.388$^{**}$ (0.126) \\ 
  Round 3 & $-$0.628$^{***}$ (0.130) & $-$0.696$^{***}$ (0.134) & $-$0.645$^{***}$ (0.131) \\ 
  Round 4 &  & $-$0.278$^{*}$ (0.120) & $-$0.114 (0.115) \\ 
  Round 5 &  & $-$0.145 (0.125) & $-$0.161 (0.124) \\ 
  Round 6 &  & $-$0.510$^{***}$ (0.137) & $-$0.615$^{***}$ (0.140) \\ 
  Round 7 &  &  & $-$0.183 (0.125) \\ 
  Round 8 &  &  & $-$0.062 (0.124) \\ 
  Round 9 &  &  & $-$0.482$^{***}$ (0.140) \\ 
  Medium (t3s110) & 1.703$^{***}$ (0.242) & 0.917$^{***}$ (0.172) & 1.270$^{***}$ (0.163) \\ 
  Hard (t3s111) & 1.650$^{***}$ (0.268) & 0.693$^{***}$ (0.199) & 0.853$^{***}$ (0.190) \\ 
  Very Hard (t3s111o2ye2) & 1.994$^{***}$ (0.230) & 1.093$^{***}$ (0.160) & 1.401$^{***}$ (0.157) \\ 
  Experiment Disclosure & 0.693$^{***}$ (0.139) & 0.790$^{***}$ (0.117) & 0.971$^{***}$ (0.118) \\ 
  Initial Error & 0.552 (0.401) & 0.113 (0.323) & 0.284 (0.332) \\ 
  Individual Intelligence & 0.013$^{**}$ (0.004) & $-$0.009$^{*}$ (0.004) & $-$0.007$^{*}$ (0.004) \\ 
 \hline \\[-1.8ex] 
Observations & 2,203 & 4,410 & 6,576 \\ 
\end{tabular}%
}
\parbox{0.95\linewidth}{\textit{Notes:} $^{.}$p$<$0.1; $^{*}$p$<$0.05; $^{**}$p$<$0.01; $^{***}$p$<$0.001. Ordered logit estimates predicting deception, ordered as Truth $<$ Withhold $<$ Lie. Positive coefficients indicate a shift toward higher propensities of withholding or lying. All specifications employ a mixed-contrast framework: treatment contrasts are used for Round, Structure, and Experiment Disclosure, with Round 1, Easy (t3s111y2), and No Disclosure serving as the respective omitted baselines. The Strategic Prompt indicator uses sum contrasts (representing deviations from the grand mean), while the Initial Error continuous variable is mean-centered. Standard errors reported in parentheses are CR1 cluster-robust standard errors clustered at the unique market level (`slug') to account for intra-market correlation across competing agents and repeated communication rounds.}
\end{table}

\clearpage
\section{Information Provision}
\label{sa:sec:information provision}

This section reports the two regression tables for the information provision treatment discussed in Section \ref{sec:information provision}, estimated on the pooled sample of 3,500 markets. First, we display the text appended to the prompt in the Experiment Disclosure treatment, which was dynamically generated based on which model was trading.%
\footnote{The qualitative summaries provided in the information treatment were constructed using provisional analysis (employing HC1 robust standard errors) immediately following the first wave of the experiment. While our final, more conservative CR2 clustered specifications render some adjacent structural boundaries statistically indistinguishable (e.g., the effect of Easy and Medium on market accuracy), the directional point estimates for the treatment parameters remain identical. Consequently, the information provided represents a valid, directionally accurate heuristic regarding the historical market environment. In particular, the prompt's Result 7 states an unqualified null for the initial price, reflecting the provisional average estimate; our final analysis (Result \ref{result:initial price}) qualifies this with a Very Hard-specific interaction that the provisional analysis had not identified. The provisional analysis is reported in Section 2 of the Supplementary Appendix.}
We provided qualitative historical guidance in the prompt, without specifying a trading strategy:

\begin{Verbatim}[breaklines=true]

=== Experimental Findings & Strategic Context ===
Before trading, all traders are informed about the following qualitative 
results from a study of over 1,700 similar prediction markets involving LLM 
agents. Use them to guide your decisions.
Definition of market accuracy: denotes how close is the last price of the 
Yes shares to the true value of the Yes shares, and similarly for the No
shares.
Definition of Intelligence: Agents are scored on the "Artificial Analysis 
Intelligence Index" (reasoning, math, coding). The observed range in our 
study is 7 (Low) to 46 (High).

1. Intelligence
* Your Intelligence (trader_2): 46
* Intelligence of trader_1: 30
* Intelligence of trader_3: 41
* Average Group Intelligence: 39.00
Result 1 shows that higher individual intelligence directly correlates 
with higher profits.

2. Market Complexity
* We ranked the market structures by complexity of reasoning: Level 1 
(Easiest) < Level 2 < Level 3 < Level 4 (Hardest).
* Current Status: You are trading in a Level 4 market.
Result 2 shows that as complexity rises, trader profits decrease and 
the market becomes less accurate.

3. Market Design
* Result 3: Trading order significantly impacts profitability. The most 
profitable position is 3rd, followed by 1st, with 2nd being the least 
profitable (3rd > 1st > 2nd).
* Result 4: Higher average group intelligence leads to a more accurate 
market but lower individual profits.
* Neutral Factors: The following factors have no statistically significant 
effect on market accuracy:
* Result 5: Posting public comments has no effect on market accuracy.
* Result 6: Being "myopic" (maximize profits on current round only) vs. 
"strategic" (maximize profits for current and all future rounds) has no 
effect on market accuracy.
* Result 7: The initial price of the market has no effect on market 
accuracy.
* Result 8: Increasing the duration of the market (9 rounds vs 3) has no 
effect on market accuracy.

\end{Verbatim}

\begin{table}[!htbp] \centering 
  \caption{Information Aggregation With Disclosure} 
  \label{tab:disclosure_quantile_regressions_robust} 
\small 
\resizebox{\linewidth}{!}{%
\begin{tabular}{@{\extracolsep{5pt}}lccc} 
\\[-1.8ex]\hline 
\hline \\[-1.8ex] 
 & \multicolumn{3}{c}{\textit{Dependent variable:}} \\ 
\cline{2-4} 
\\[-1.8ex] & \multicolumn{3}{c}{Log Error} \\ 
 & Mean (OLS) & Median (Q50) & Tail Risk (Q80) \\ 
\\[-1.8ex] & (1) & (2) & (3)\\ 
\hline \\[-1.8ex] 
 Constant & 7.196$^{***}$ (2.061) & 0.018$^{*}$ (0.008) & 5.638$^{***}$ (0.954) \\ 
  Comments Allowed & $-$0.318 (0.258) & $-$0.001 (0.001) & $-$0.014 (0.038) \\ 
  Duration: 6 Rounds & $-$1.550 (1.118) & $-$0.001 (0.001) & $-$0.107 (0.112) \\ 
  Duration: 9 Rounds & $-$0.328 (1.183) & $-$0.001 (0.001) & $-$0.019 (0.086) \\ 
  Strategic Prompt & 0.007 (0.231) & 0.001 (0.001) & 0.007 (0.064) \\ 
  Medium (t3s110) & 2.772 (3.002) & $-$0.00001 (0.001) & 0.075 (0.971) \\ 
  Hard (t3s111) & 5.222$^{*}$ (2.225) & 0.206$^{*}$ (0.093) & 11.809$^{.}$ (6.577) \\ 
  Very Hard (t3s111o2ye2) & 7.192$^{*}$ (3.332) & 0.697$^{***}$ (0.024) & 20.126$^{***}$ (5.325) \\ 
  Initial Error: Low & 0.123 (0.198) & 0.00000 (0.001) & 0.008 (0.046) \\ 
  Initial Error: Medium & 0.234 (0.236) & $-$0.00001 (0.001) & $-$0.003 (0.048) \\ 
  Experiment Disclosure & 2.075 (1.963) & $-$0.00004 (0.008) & 0.980 (4.638) \\ 
  Average Intelligence & $-$0.218$^{***}$ (0.058) & $-$0.0004$^{*}$ (0.0002) & $-$0.123$^{***}$ (0.021) \\ 
  Intelligence SD & $-$0.012 (0.223) & $-$0.0003$^{**}$ (0.0001) & $-$0.102$^{***}$ (0.020) \\ 
  6 Rounds x Strategic & $-$0.083 (0.294) & $-$0.001 (0.001) & $-$0.021 (0.093) \\ 
  9 Rounds x Strategic & 0.182 (0.286) & $-$0.001 (0.001) & 0.014 (0.084) \\ 
  Medium Struct x Init. Error (Low) & 0.657 (0.786) & $-$0.00000 (0.001) & 0.003 (1.897) \\ 
  Hard Struct x Init. Error (Low) & 0.120 (0.474) & 0.045 (0.097) & $-$8.149 (11.199) \\ 
  Very Hard x Init. Error (Low) & 1.665 (1.222) & 0.008 (0.047) & 9.638$^{*}$ (4.721) \\ 
  Medium Struct x Init. Error (Med) & $-$0.399 (0.378) & 0.00001 (0.001) & 0.006 (0.957) \\ 
  Hard Struct x Init. Error (Med) & 0.187 (0.594) & $-$0.077 (0.130) & 18.273 (11.427) \\ 
  Very Hard x Init. Error (Med) & $-$0.321 (0.975) & $-$0.004 (0.024) & 8.923 (6.687) \\ 
  Disclosure x Av. Intelligence & $-$0.053 (0.061) & 0.00000 (0.0002) & $-$0.023 (0.109) \\ 
 \hline \\[-1.8ex] 
Observations & 3,500 & 3,500 & 3,500 \\ 
R$^{2}$ & 0.119 &  &  \\ 
\end{tabular}%
}
\parbox{0.95\linewidth}{\textit{Notes:} $^{.}$p$<$0.1; $^{*}$p$<$0.05; $^{**}$p$<$0.01; $^{***}$p$<$0.001. Q50 and Q80 are quantile regressions. Q50 models the median log error, whereas  Q80 models the upper 20\% tail of the log error distribution (the worst-performing markets). OLS model uses CR2 cluster-robust standard errors clustered by exact AI team composition, featuring small-sample degrees of freedom adjustments. Quantile models use bootstrapped standard errors. All three models use a mixed-contrast specification. Rounds, Structure, and Experiment Disclosure use treatment contrasts with Round 3, Structure t3s111y2, and No Disclosure as baselines. All other controls (Initial Error, Strategic, Comments) use sum contrasts, where coefficients represent deviations from the grand mean.}
\end{table}

\begin{table}[!htbp] \centering 
  \caption{Trading Volume and Agent Profitability With Disclosure} 
  \label{tab:disclosure trading volume and profits robust} 
\small 
\begin{tabular}{@{\extracolsep{5pt}}lcc} 
\\[-1.8ex]\hline 
\hline \\[-1.8ex] 
 & \multicolumn{2}{c}{\textit{Dependent variable:}} \\ 
\cline{2-3} 
 \\[-1.8ex] & Trading Volume & Individual Profits \\ 
\\[-1.8ex] & (1) & (2)\\ 
\hline \\[-1.8ex] 
 Constant & 1,757.638$^{.}$ (1,057.684) & $-$77.335$^{*}$ (30.791) \\ 
  Comments Allowed & 36.269 (29.026) & 3.552 (2.747) \\ 
  Duration: 6 Rounds & 563.399$^{***}$ (104.689) & 3.359 (17.416) \\ 
  Duration: 9 Rounds & 1,080.922$^{***}$ (209.860) & $-$12.727 (19.250) \\ 
  Strategic Prompt & $-$37.028$^{.}$ (21.233) & $-$0.345 (2.338) \\ 
  Medium (t3s110) & 316.783$^{*}$ (156.405) & $-$35.852 (38.603) \\ 
  Hard (t3s111) & 212.977$^{**}$ (67.246) & $-$71.160$^{*}$ (32.793) \\ 
  Very Hard (t3s111o2ye2) & 43.866 (267.233) & $-$101.644$^{*}$ (45.614) \\ 
  Initial Error: Low & $-$145.678$^{*}$ (66.395) & $-$12.168$^{***}$ (3.042) \\ 
  Initial Error: Medium & 44.294 (56.922) & $-$5.193 (3.645) \\ 
  Experiment Disclosure & 171.986 (137.059) & $-$21.052 (14.316) \\ 
  Average Intelligence & $-$28.182 (24.757) & $-$11.057$^{***}$ (1.855) \\ 
  Individual Intelligence &  & 14.091$^{***}$ (1.787) \\ 
  Intelligence SD & 24.293 (40.089) & 1.154 (2.467) \\ 
  Trades Second &  & $-$39.316 (32.790) \\ 
  Trades Third &  & 74.604 (53.060) \\ 
  6 Rounds x Strategic & $-$26.283 (34.055) & $-$0.524 (3.928) \\ 
  9 Rounds x Strategic & 45.471 (41.863) & $-$5.343 (3.828) \\ 
  Medium Struct x Initial Error (Low) & 166.553 (110.724) & $-$9.165 (10.672) \\ 
  Hard Struct x Initial Error (Low) & 69.433 (64.362) & $-$6.278 (7.583) \\ 
  Very Hard x Initial Error (Low) & 349.973 (261.708) & $-$18.174 (15.738) \\ 
  Medium Struct x Initial Error (Med) & $-$28.194 (77.289) & 2.954 (4.152) \\ 
  Hard Struct x Initial Error (Med) & 93.975 (73.295) & 2.619 (8.980) \\ 
  Very Hard x Initial Error (Med) & $-$252.898 (441.683) & 1.676 (11.363) \\ 
  Disclosure x Average Intelligence & $-$3.394 (5.256) & 0.493 (0.445) \\ 
 \hline \\[-1.8ex] 
Observations & 3,500 & 10,500 \\ 
R$^{2}$ & 0.110 & 0.059 \\ 
\end{tabular} 
\parbox{0.95\linewidth}{\textit{Notes:} $^{.}$p$<$0.1; $^{*}$p$<$0.05; $^{**}$p$<$0.01; $^{***}$p$<$0.001. Both models use CR2 cluster-robust standard errors clustered by exact AI team composition, featuring small-sample degrees of freedom adjustments. Rounds, Information Structure, Experiment Disclosure, and Trader Position use treatment contrasts, with `Round 3', `t3s111y2', No Disclosure, and `Trades First' serving as their respective baselines. All other controls (Initial Error, Strategic, Comments) use sum contrasts, where coefficients represent deviations from the grand mean.}
\end{table}

\clearpage
\section{Theoretical Myopic Profits and Volume}
\label{sec:profit_calculations}

In this section, we derive the theoretical profits and trading volume for all four structures, if all agents are myopic and this is common knowledge. Let $p_{start}$ denote the price of Yes in the previous round (or the initial price we are in the first round) and $p_{end}$ denote the price after the AI agents conducts his trades. The $\eta$ parameter represents the liquidity sensitivity parameter, which is set to 0.01 \citep{rajtmajerEtAl2022, galanisEtAl24}.

Under the Logarithmic Market Scoring Rule (LMSR), the price of a binary contract is strictly a function of the difference in outstanding shares. The quantity of Yes shares, $\Delta q$, required to shift the market probability from $p_{start}$ to $p_{end}$ is given by
\[
\Delta q = \frac{1}{\eta} \ln\left(\frac{p_{end}(1 - p_{start})}{p_{start}(1 - p_{end})}\right)
\]

The cost of buying these Yes shares is
\[
\text{Cost} = \frac{1}{\eta} \ln\left(\frac{1 - p_{start}}{1 - p_{end}}\right)
\]

If the market resolves to Yes (yielding a payout of 1 per share), the trader's net profit is the total payout minus the initial cost. If it resolves to No, the Yes shares are worthless and the net profit is simply the negative cost:
\[
\text{Profit}_{\text{Yes}} = \Delta q - \text{Cost} = \frac{1}{\eta} \ln\left(\frac{p_{end}}{p_{start}}\right)
\]
\[
\text{Profit}_{\text{Yes}} = -\text{Cost} = \frac{1}{\eta} \ln\left(\frac{1 - p_{end}}{1 - p_{start}}\right)
\]

The calculations for buying No shares are similar. If a trader wishes to reduce the price of Yes from $p_{start}$ to $p_{end}$ (where $p_{end} < p_{start}$) and he has no Yes shares to sell,  he must acquire No shares. The quantity of No shares, $\Delta q_{no}$, required to decrease the Yes price to $p_{end}$ is given by:
\[
\Delta q_{no} = \frac{1}{\eta} \ln\left(\frac{p_{start}(1 - p_{end})}{p_{end}(1 - p_{start})}\right)
\]

The capital required to execute this transaction depends entirely on the ratio of the starting and ending ``Yes'' probabilities, mirroring the profit function of the opposing side:
\[
\text{Cost}_{no} = \frac{1}{\eta} \ln\left(\frac{p_{start}}{p_{end}}\right)
\]

If the market resolves to No' (yielding a payout of 1 per share), the trader's net profit is the total payout minus the initial cost. If the market resolves to Yes, then the profit from this transaction is the negative cost:
\[
\text{Profit}_{\text{No}} = \Delta q_{no} - \text{Cost}_{no} = \frac{1}{\eta} \ln\left(\frac{1 - p_{end}}{1 - p_{start}}\right)
\]

\[
\text{Profit}_{\text{No}} = - \text{Cost}_{no} = -\frac{1}{\eta} \ln\left(\frac{p_{start}}{p_{end}}\right)
\]

Using these formulas, we calculate the average profits and trading volume for each structure and trader, which we report in Table \ref{tab:myopic prices}.

\begin{table}[!htbp] \centering 
  \caption{Average Volume and Profits} 
  \label{tab:myopic_volume_profits_average} 
\begin{tabular}{@{\extracolsep{5pt}}lccccc} 
\\[-1.8ex]\hline 
\hline \\[-1.8ex] 
 \textbf{Average over} & \textbf{Easy} & \textbf{Medium} & \textbf{Hard} & \textbf{Very Hard} & \textbf{Average over} \\ 
 \textbf{initial prices} & \textbf{t3s111y2} & \textbf{t3s110} & \textbf{t3s111} & \textbf{t3s111o2ye2} & \textbf{structures} \\
\hline \\[-1.8ex] 
 Volume & 990 & 1210 & 1210 & 1046 & 1115 \\ 
 Trader 1 Volume & 110 & 110 & 110 & 56 & 97 \\ 
 Trader 2 Volume & 880 & 110 & 110 & 0 & 275 \\ 
 Trader 3 Volume & 0 & 990 & 990 & 990 & 743 \\ 
 Trader 1 Profits & 46 & 46 & -64 & 6 & 9 \\ 
 Trader 2 Profits & 29 & -41 & 69 & 0 & 14 \\ 
 Trader 3 Profits & 0 & 69 & 69 & 69 & 52 \\ 
Average Profits & 25 & 25 & 25 & 25 & 25 \\

\hline 
\hline \\[-1.8ex] 
\end{tabular} 
\parbox{0.95\linewidth}{\textit{Notes:} Entries are deterministic myopic benchmark calculations from the theoretical LMSR price path, averaged over the three initial prices used in the experiment. Volume is the absolute quantity of shares traded. Profits are realized LMSR profits in experimental payoff units.}
\end{table}

\clearpage
\section{Supplementary Graphs and Tables}
\label{sec:supplementary graphs}

\begin{figure}[htbp]
    \centering
    \includegraphics[width=0.9\textwidth]{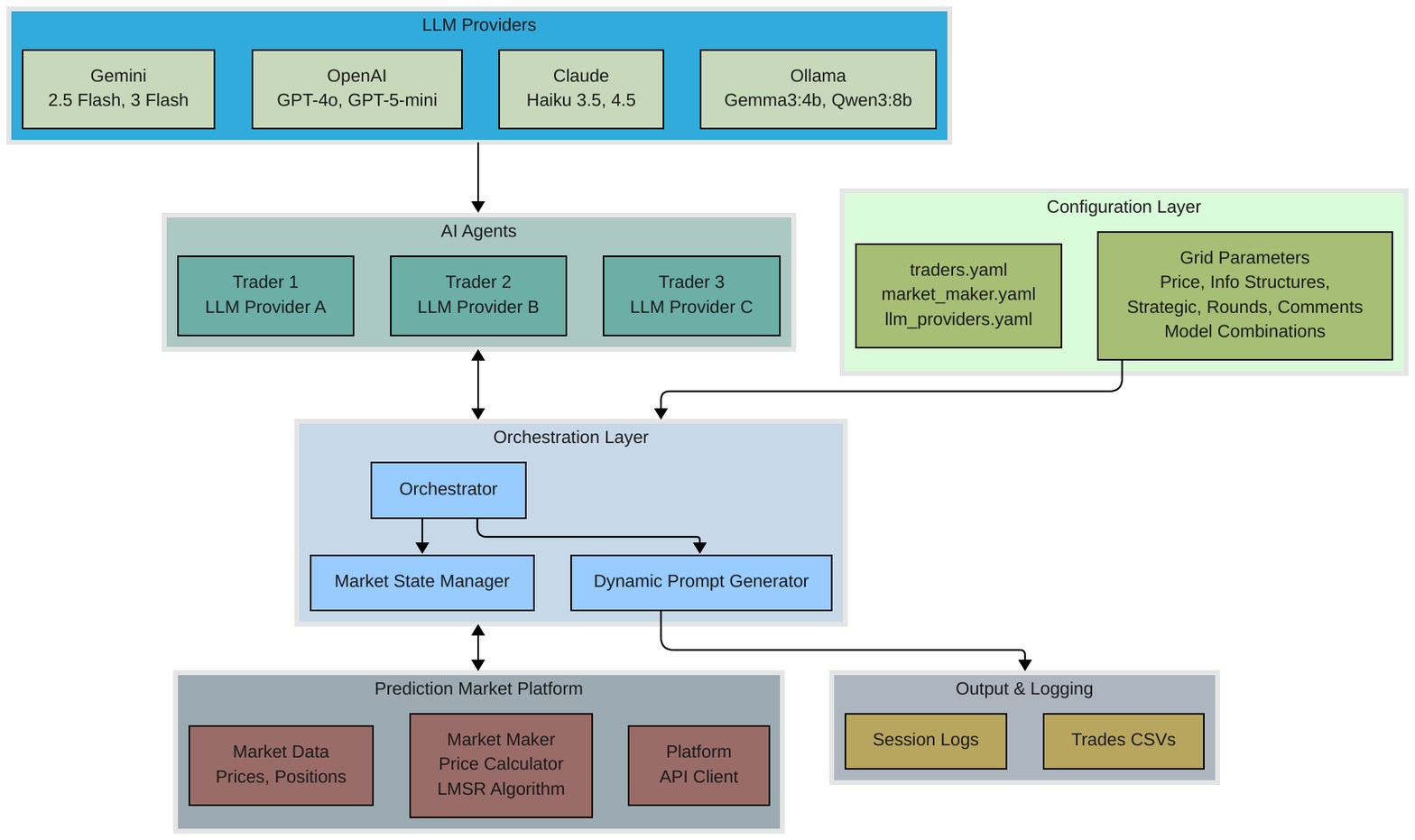}
    \caption{\textbf{Prediction market platform.} The figure summarizes the automated workflow. The experimental grid creates a market for each team, structure, duration, initial price, strategic-prompt condition, and communication condition; the agent orchestrator calls the relevant LLM API, constructs the trader-specific prompt from private signals, public history, private memory, portfolio information, and hypothetical price impacts, submits trades and comments through the API, and exports text and trade logs for analysis.}
    \vspace{0.2cm}
    \begin{singlespace}
        \setlength{\leftskip}{0pt}\setlength{\rightskip}{0pt}\parfillskip=0pt plus 1fil
    \end{singlespace}
    \label{fig:flow_chart}
\end{figure}

\begin{figure}[htbp]
    \centering
    \includegraphics[width=0.9\textwidth]{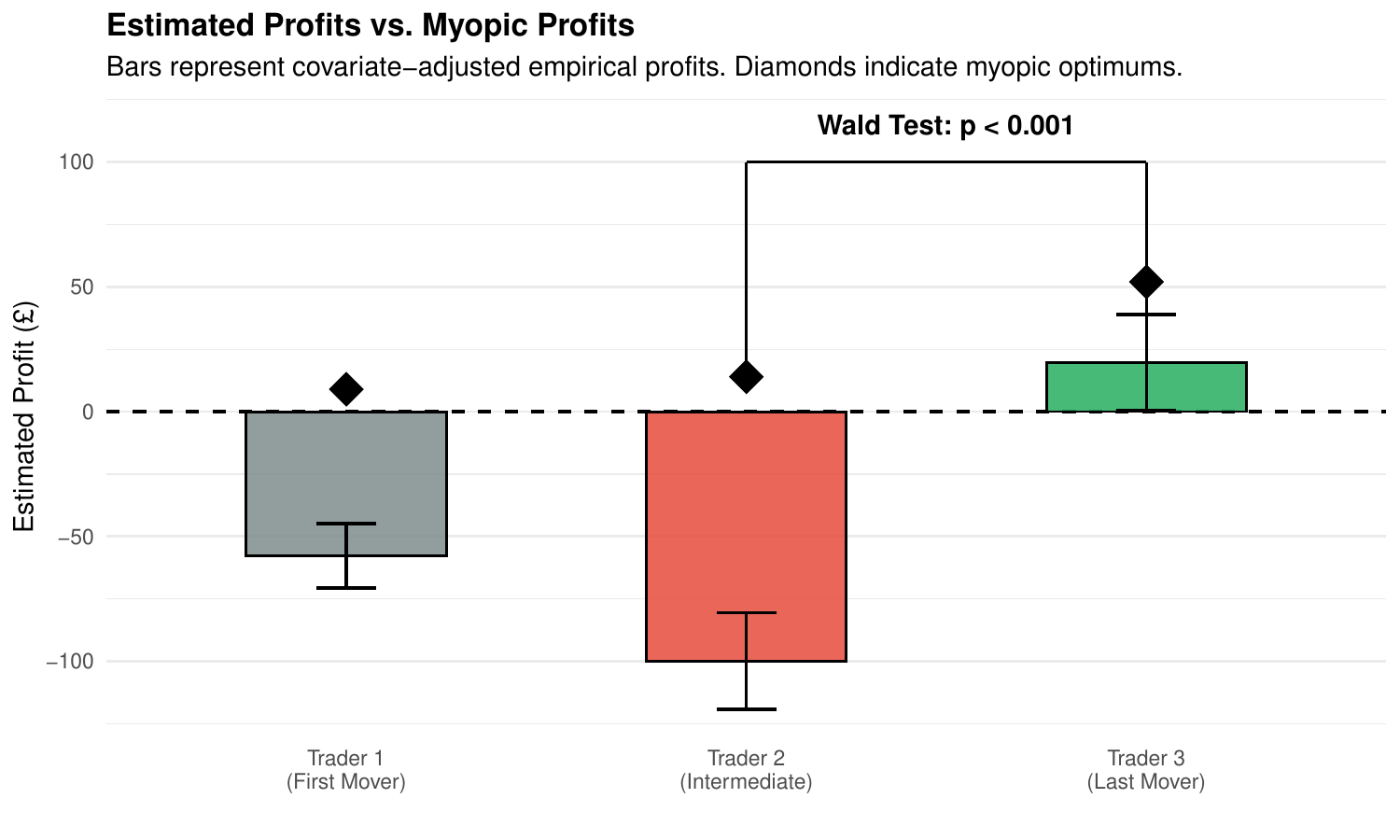}
    \caption{\textbf{The Trading Position Effect.} Bars represent covariate-adjusted estimated marginal means (EMMs) of profit for each trading position, derived from the primary OLS specification. Error bars indicate 95\% confidence intervals constructed using CR2 cluster-robust standard errors clustered at the unique market level. Unlike the team-composition clustering of Table \ref{tab:trading volume and profits robust}, we cluster the position contrasts at the market level because they are within-market comparisons among the three competing traders, and team-composition clustering would leave only 12 clusters for trader-level marginal-mean inference. Black diamonds denote the theoretically optimal myopic profit for each respective position. The annotated bracket reports the result of a post-estimation Wald test (p<0.001), confirming a highly significant late-mover advantage for the third trader relative to the second. }
    \vspace{0.2cm}
    \begin{singlespace}
        \setlength{\leftskip}{0pt}\setlength{\rightskip}{0pt}\parfillskip=0pt plus 1fil
    \end{singlespace}
    \label{fig:graph_profit_position_theory_wald}
\end{figure}

\begin{figure}[htbp]
    \centering
    \includegraphics[width=0.9\textwidth]{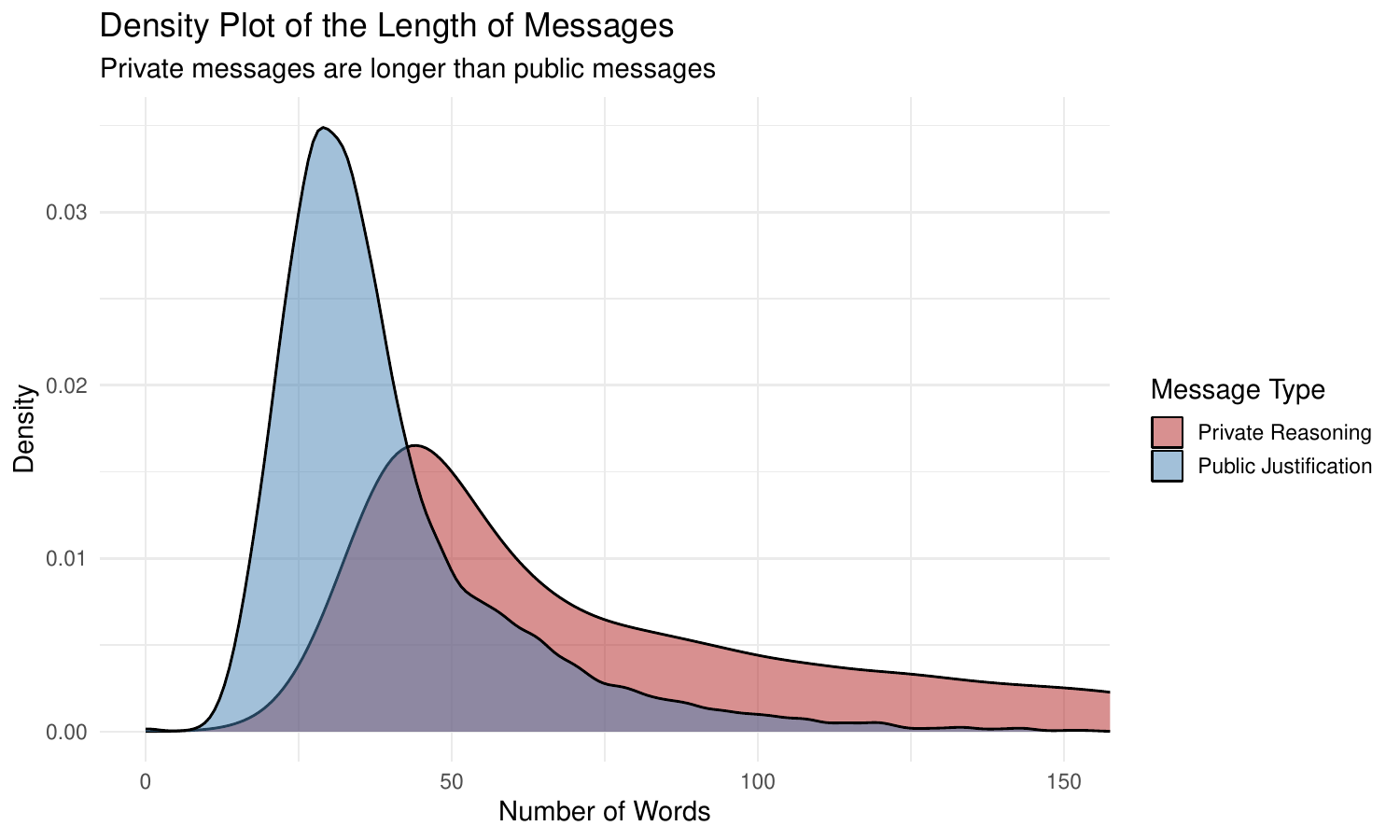}
    \caption{\textbf{Length of Private and Public Messages.} The figure plots kernel densities of message length, measured by word count on the horizontal axis, separately for private reasoning messages and public comments. The distribution shows that private messages are typically longer than public comments.}
    \vspace{0.2cm}
    \begin{singlespace}
        \setlength{\leftskip}{0pt}\setlength{\rightskip}{0pt}\parfillskip=0pt plus 1fil
    \end{singlespace}
    \label{fig:graph_density_length_messages}
\end{figure}

\begin{table}[!htbp] \centering
  \caption{Information Aggregation, by Group and Structure}
  \label{tab:frontier_outcome_shape}
\small
\makeatletter
\begin{tabular}{@{\extracolsep{5pt}}llccccc}
\\[-1.8ex]\hline
\hline \\[-1.8ex]
Structure & Group & N & Aggregates & Partial & Hedges & Confidently wrong \\
\hline \\[-1.8ex]
Easy & January 2026, all teams & 432 & 68.5\% & 21.8\% & 3.9\% & 5.8\% \\
Easy & January 2026, four smartest & 144 & 66.0\% & 30.6\% & 3.5\% & 0.0\% \\
Easy & April 2026, frontier & 144 & 80.6\% & 19.4\% & 0.0\% & 0.0\% \\
Medium & January 2026, all teams & 475 & 69.3\% & 10.9\% & 4.6\% & 15.2\% \\
Medium & January 2026, four smartest & 144 & 86.8\% & 9.0\% & 2.8\% & 1.4\% \\
Medium & April 2026, frontier & 144 & 100.0\% & 0.0\% & 0.0\% & 0.0\% \\
Hard & January 2026, all teams & 432 & 44.0\% & 21.3\% & 8.3\% & 26.4\% \\
Hard & January 2026, four smartest & 144 & 43.8\% & 31.9\% & 8.3\% & 16.0\% \\
Hard & April 2026, frontier & 144 & 51.4\% & 21.5\% & 8.3\% & 18.8\% \\
Very Hard & January 2026, all teams & 433 & 18.7\% & 13.4\% & 37.6\% & 30.3\% \\
Very Hard & January 2026, four smartest & 144 & 22.2\% & 12.5\% & 56.2\% & 9.0\% \\
Very Hard & April 2026, frontier & 144 & 16.7\% & 6.9\% & 66.0\% & 10.4\% \\
\hline \\[-1.8ex]
\end{tabular}
\makeatother
\parbox{0.95\linewidth}{\textit{Notes:} A market aggregates if the absolute distance between the final price and the true value of the security is below 0.1, it is confidently wrong if that distance exceeds 0.9, it hedges if the distance lies within 0.1 of 0.5, and it is partial otherwise. The first group comprises all twelve teams of the first wave, excluding the information provision treatment, and the second group is the subset of four teams with the highest intelligence index in January 2026, so their markets are counted in both rows. The third group comprises the four frontier teams of April 2026.}
\end{table}

\begin{figure}[htbp]
    \centering
    \includegraphics[width=0.9\textwidth]{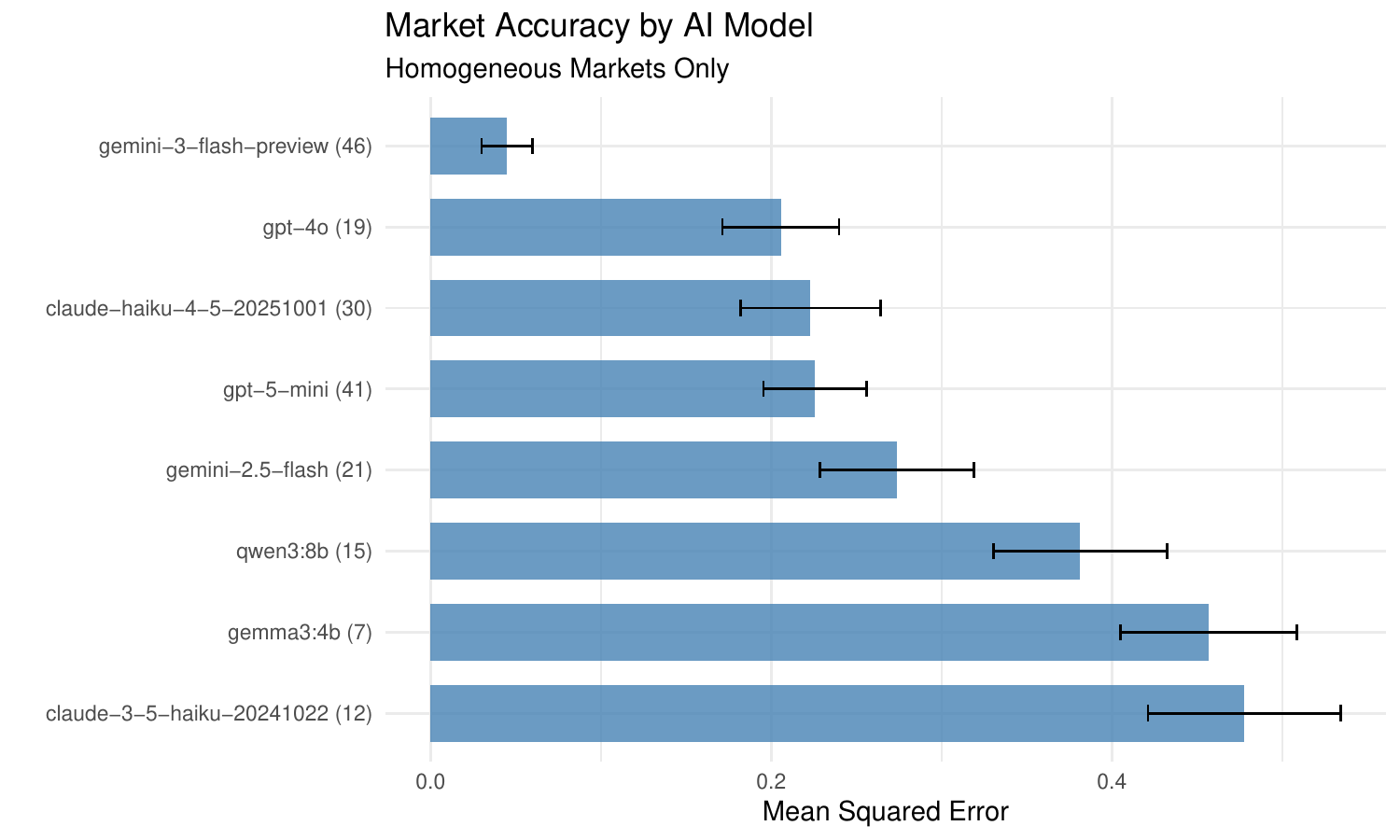}
    \caption{Mean Squared Error by AI Model}
    \vspace{0.2cm}
    \begin{singlespace}
        \setlength{\leftskip}{0pt}\setlength{\rightskip}{0pt}\parfillskip=0pt plus 1fil
    \end{singlespace}
    \label{fig:graph_mse_by_model}
\end{figure}

\begin{figure}[htbp]
    \centering
    \includegraphics[width=0.9\textwidth]{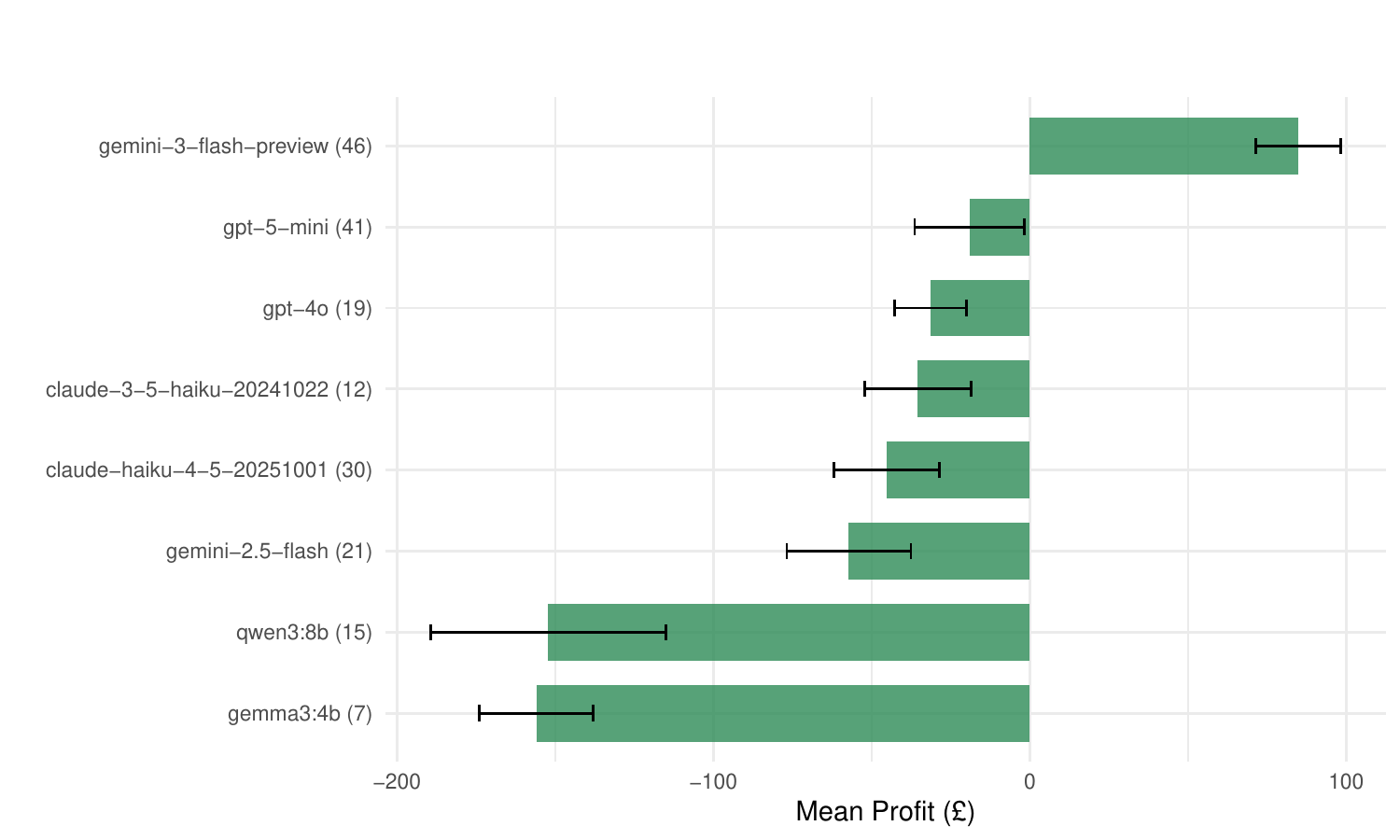}
    \caption{Average Profits by AI Model}
    \vspace{0.2cm}
    \begin{singlespace}
        \setlength{\leftskip}{0pt}\setlength{\rightskip}{0pt}\parfillskip=0pt plus 1fil
    \end{singlespace}
    \label{fig:graph_profits_by_model}
\end{figure}

%

\begin{figure}[htbp]
    \centering
    \includegraphics[width=0.9\textwidth]{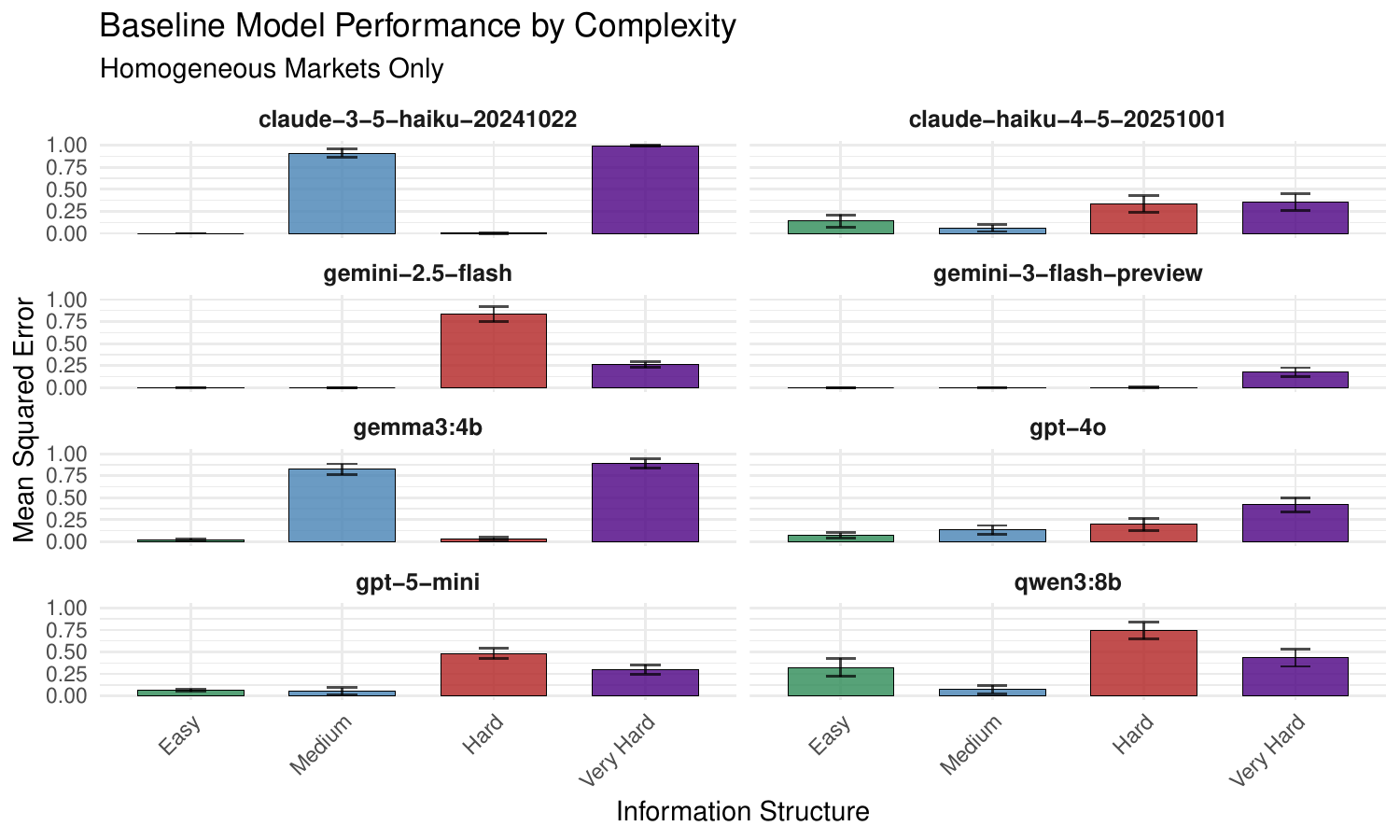}
    \caption{Mean Squared Error by Structure and Model}
    \vspace{0.2cm}
    \begin{singlespace}
        \setlength{\leftskip}{0pt}\setlength{\rightskip}{0pt}\parfillskip=0pt plus 1fil
    \end{singlespace}
    \label{fig:graph_mse_by_model_structure}
\end{figure}

\begin{figure}[htbp]
    \centering
    \includegraphics[width=0.9\textwidth]{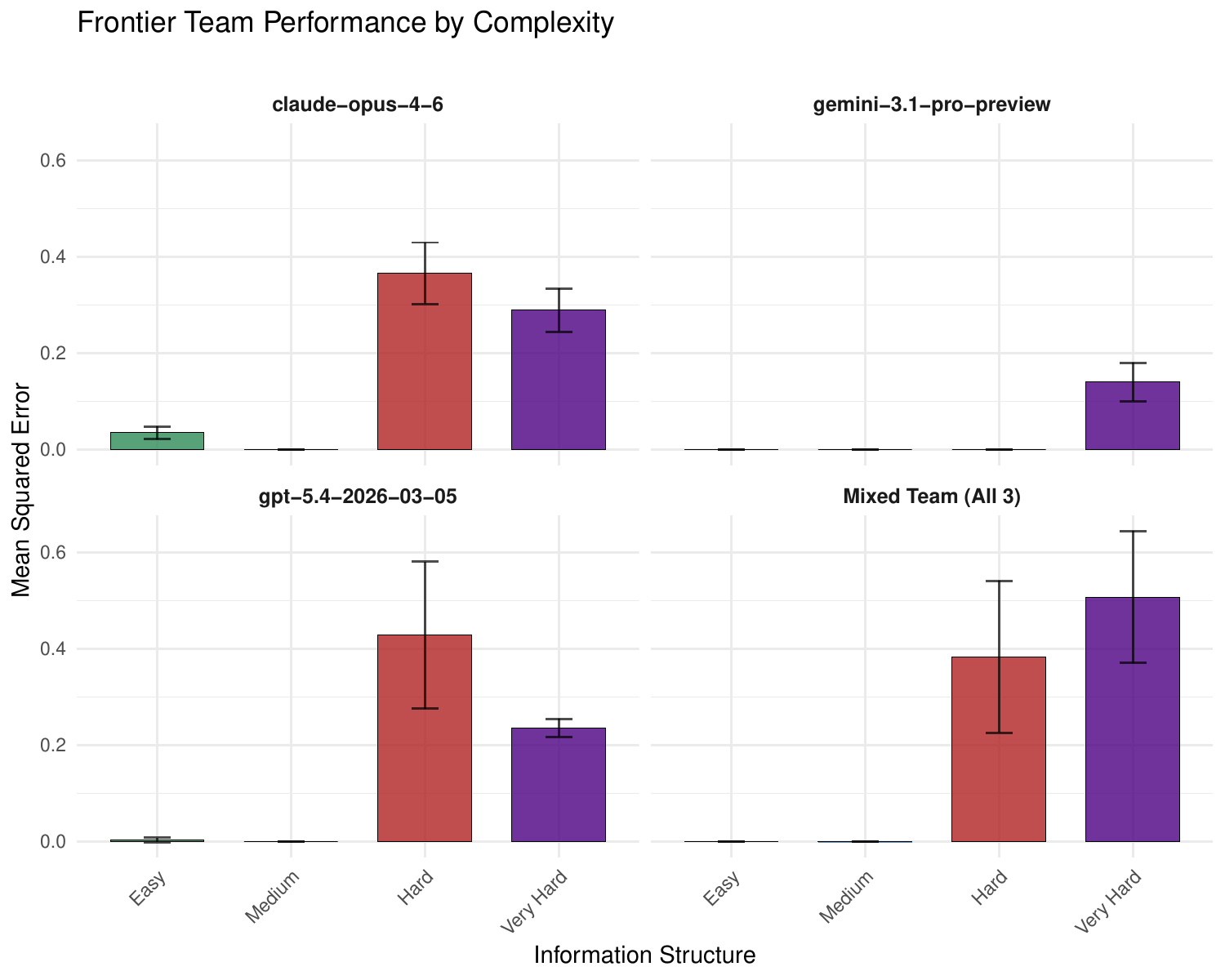}
    \caption{Mean Squared Error by Structure and Frontier Model}
    \vspace{0.2cm}
    \begin{singlespace}
        \setlength{\leftskip}{0pt}\setlength{\rightskip}{0pt}\parfillskip=0pt plus 1fil
    \end{singlespace}
    \label{fig:graph_frontier_mse_by_model_structure}
\end{figure}

\clearpage
\section{Market Outcomes by Treatment Cell}
\label{sa:sec:treatment outcomes}

This section reports the market outcomes by information structure and market duration discussed in Section \ref{sec:data}.

\begin{table}[!htbp] \centering
  \caption{Market Outcomes by Information Structure and Duration}
  \label{tab:treatment_outcomes_both_waves}
\small
\makeatletter
\resizebox{\linewidth}{!}{%
\begin{tabular}{@{\extracolsep{1pt}} cccccccccc}
\\[-1.8ex]\hline
\hline \\[-1.8ex]
Structure & Rounds & N & MSE & SE (Sq Er) & Mean Log & SE (Log) & Crash Rate & Avg Volume & Avg IQ \\
\hline \\[-1.8ex]
First Wave \\
\hline \\[-1.8ex]
Easy & 3 & 144 & 0.108 & 0.023 & 2.56 & 0.73 & 6.9\% & 1,255 & 23.8 \\ 
 & 6 & 144 & 0.070 & 0.018 & 0.69 & 0.34 & 1.4\% & 1,808 & 23.8 \\ 
 & 9 & 144 & 0.078 & 0.018 & 1.58 & 0.58 & 4.2\% & 2,102 & 23.8 \\ 
Medium & 3 & 159 & 0.183 & 0.027 & 3.56 & 0.80 & 9.4\% & 1,508 & 23.0 \\ 
 & 6 & 160 & 0.200 & 0.029 & 3.80 & 0.82 & 10.0\% & 2,065 & 22.9 \\ 
 & 9 & 156 & 0.202 & 0.030 & 4.69 & 0.92 & 12.8\% & 2,639 & 23.1 \\ 
Hard & 3 & 144 & 0.387 & 0.036 & 9.30 & 1.24 & 25.7\% & 1,328 & 23.8 \\ 
 & 6 & 144 & 0.319 & 0.034 & 4.80 & 0.92 & 11.8\% & 1,865 & 23.8 \\ 
 & 9 & 144 & 0.332 & 0.035 & 6.70 & 1.10 & 18.1\% & 2,356 & 23.8 \\ 
Very Hard & 3 & 145 & 0.464 & 0.032 & 8.82 & 1.19 & 23.4\% & 1,229 & 23.7 \\ 
 & 6 & 144 & 0.433 & 0.032 & 7.46 & 1.12 & 19.4\% & 1,752 & 23.8 \\ 
 & 9 & 144 & 0.436 & 0.032 & 8.56 & 1.19 & 22.9\% & 2,354 & 23.8 \\ 
\hline \\[-1.8ex]
Both Waves \\
\hline \\[-1.8ex]
Easy & 3 & 288 & 0.131 & 0.018 & 2.76 & 0.53 & 7.3\% & 1,264 & 23.8 \\ 
 & 6 & 288 & 0.081 & 0.013 & 1.15 & 0.34 & 2.8\% & 1,844 & 23.8 \\ 
 & 9 & 288 & 0.081 & 0.014 & 1.39 & 0.37 & 3.5\% & 2,140 & 23.8 \\ 
Medium & 3 & 303 & 0.173 & 0.020 & 3.46 & 0.57 & 9.2\% & 1,521 & 23.4 \\ 
 & 6 & 304 & 0.212 & 0.021 & 4.62 & 0.65 & 12.5\% & 2,072 & 23.3 \\ 
 & 9 & 300 & 0.223 & 0.022 & 5.75 & 0.73 & 16.0\% & 2,650 & 23.4 \\ 
Hard & 3 & 288 & 0.407 & 0.026 & 9.50 & 0.88 & 26.0\% & 1,377 & 23.8 \\ 
 & 6 & 288 & 0.307 & 0.024 & 4.52 & 0.63 & 11.1\% & 1,995 & 23.8 \\ 
 & 9 & 288 & 0.339 & 0.025 & 6.95 & 0.78 & 18.8\% & 2,515 & 23.8 \\ 
Very Hard & 3 & 289 & 0.469 & 0.023 & 9.21 & 0.85 & 24.6\% & 1,234 & 23.7 \\ 
 & 6 & 288 & 0.450 & 0.023 & 8.30 & 0.82 & 21.9\% & 1,739 & 23.8 \\ 
 & 9 & 288 & 0.456 & 0.023 & 9.39 & 0.87 & 25.3\% & 2,408 & 23.8 \\ 
\hline \\[-1.8ex]
Third Wave (Frontier Models, April 2026) \\
\hline \\[-1.8ex]
Easy & 3 & 48 & 0.016 & 0.005 & 0.07 & 0.02 & 0.0\% & 920 & 55.7 \\ 
 & 6 & 48 & 0.010 & 0.003 & 0.05 & 0.01 & 0.0\% & 1,226 & 55.7 \\ 
 & 9 & 48 & 0.003 & 0.001 & 0.03 & 0.01 & 0.0\% & 1,281 & 55.7 \\ 
Medium & 3 & 48 & 0.000 & 0.000 & 0.00 & 0.00 & 0.0\% & 1,121 & 55.7 \\ 
 & 6 & 48 & 0.000 & 0.000 & 0.00 & 0.00 & 0.0\% & 1,475 & 55.7 \\ 
 & 9 & 48 & 0.000 & 0.000 & 0.00 & 0.00 & 0.0\% & 1,532 & 55.7 \\ 
Hard & 3 & 48 & 0.263 & 0.053 & 5.36 & 1.76 & 14.6\% & 795 & 55.7 \\ 
 & 6 & 48 & 0.306 & 0.057 & 6.17 & 1.85 & 16.7\% & 1,274 & 55.7 \\ 
 & 9 & 48 & 0.314 & 0.058 & 5.52 & 1.75 & 14.6\% & 1,575 & 55.7 \\ 
Very Hard & 3 & 48 & 0.358 & 0.045 & 4.34 & 1.51 & 10.4\% & 379 & 55.7 \\ 
 & 6 & 48 & 0.267 & 0.037 & 3.39 & 1.37 & 8.3\% & 891 & 55.7 \\ 
 & 9 & 48 & 0.255 & 0.031 & 2.69 & 1.20 & 6.2\% & 794 & 55.7 \\ 
\hline
\end{tabular}
}
\makeatother
\textit{Notes}: The crash rate measures the percentage of markets that have a log error above 20. The Avg IQ column in the third-wave panel uses the April 2026 vintage of the Artificial Analysis Intelligence Index, which is not comparable to the January 2026 scores used in the first two panels.

\end{table}

\clearpage
\section{Randomization Balance Checks}

To ensure that differences in market performance were driven by the structure and not by exogenous starting conditions, we tested for balance in initial pricing errors across treatment groups. The results are reported in Table \ref{sa:tab:balance_check}. An OLS regression of initial squared error on information structure reveals no significant systematic differences ($F(3,3496)=0.231, p=0.875$). The regression explains only 0.02\% of the variance in initial conditions ($R^2=0.0002$) and we find no evidence of imbalances in the randomization procedure.

\begin{table}[!htbp] \centering 
  \caption{Randomization Balance Checks} 
  \label{sa:tab:balance_check} 
\begin{tabular}{@{\extracolsep{5pt}}lcc} 
\\[-1.8ex]\hline 
\hline \\[-1.8ex] 
 & \multicolumn{2}{c}{\textit{Dependent variable:}} \\ 
\cline{2-3} 
\\[-1.8ex] & Average Intelligence & Initial Squared Error \\ 
\\[-1.8ex] & (1) & (2)\\ 
\hline \\[-1.8ex] 
 Medium (t3s110) & $-$0.412 & $-$0.005 \\ 
  & (0.580) & (0.008) \\ 
  Hard (t3s111) & 0.000 & 0.000 \\ 
  & (0.589) & (0.008) \\ 
  Very Hard (t3s111o2ye2) & $-$0.010 & $-$0.0002 \\ 
  & (0.588) & (0.008) \\ 
  Constant & 23.778$^{***}$ & 0.277$^{***}$ \\ 
  & (0.416) & (0.006) \\ 
 \hline \\[-1.8ex] 
Robust F-Statistic & 0.252 (df = 3; 3496) & 0.231 (df = 3; 3496) \\ 
Observations & 3,500 & 3,500 \\ 
R$^{2}$ & 0.0002 & 0.0002 \\ 
\end{tabular} 
\parbox{0.8\linewidth}{\textit{Notes:} $^{.}$p$<$0.1; $^{*}$p$<$0.05; $^{**}$p$<$0.01; $^{***}$p$<$0.001. F-test checks joint significance of all treatments. Standard errors in parentheses are HC1 heteroskedasticity-consistent. Because experimental treatments were assigned via a perfectly balanced full-factorial grid across all agents, cluster-robust standard errors are not applied to baseline balance verifications to prevent artificial variance shrinkage and matrix rank-deficiency.}
\end{table}

%
%
%
%
%
%
%

\clearpage
\section{Robustness to Alternative Standard Error Specifications}
\label{sa:sec: robustness hc1}

This section replicates the primary regression analyses from the main text to assess whether the conclusions are sensitive to the conservative variance penalties imposed by our primary estimation strategy. In the main text, the OLS models utilize CR2 cluster-robust standard errors clustered by exact AI team composition to account for broad panel-level correlation. Here, we relax that assumption. For market-level outcomes (Volume and Information Aggregation), we apply unclustered HC1 robust standard errors, assuming independent error distributions across unique markets. For individual-level outcomes (Trader Profits, Cosine Similarity, and Word Gap),  we apply CR1 cluster-robust standard errors clustered at the unique market level. This relaxes the broad AI-team panel assumption while accounting for the mechanical, micro-level intra-market correlation inherent among competing agents.%
\footnote{We omit the Deception (Ordered Logit) model altogether because we already apply CR1 errors clustered at the market level in the main paper.}
Because these adjustments apply solely to the post-estimation variance-covariance matrix, the point estimates remain, by definition, mathematically identical to those reported in the main text.

The primary estimates for communication, strategic prompting, initial price, and extended duration (9 rounds) remain statistically insignificant even under these less restrictive inference specifications. Conversely, when the clustering penalty is removed, the 6-round duration, experiment disclosure, and the Medium information structure cross the threshold into statistical significance in the information aggregation OLS (Tables \ref{sa:tab:quantile_regressions_robust}, \ref{sa:tab:disclosure_quantile_regressions_robust}). Similarly, the order of trading is statistically significant for individual profits under market-level clustering (Tables  \ref{sa:tab:trading volume and profits robust}, \ref{sa:tab:disclosure trading volume and profits robust}). Because HC1 assumes strictly independent observations, ignoring the heavy intra-cluster correlation inherent in our multi-round, agent-based panel data, it produces smaller standard errors. We therefore interpret these results as evidence that several estimates are sensitive to the assumed correlation structure rather than as a basis for stronger causal claims.



\begin{table}[!htbp] \centering 
 \caption{Information Aggregation Without Disclosure (HC1 Standard Errors)} 
 \label{sa:tab:quantile_regressions_robust} 
\small 
\resizebox{\linewidth}{!}{%
\begin{tabular}{@{\extracolsep{5pt}}lccc} 
\\[-1.8ex]\hline 
\hline \\[-1.8ex] 
& \multicolumn{3}{c}{\textit{Dependent variable:}} \\ 
\cline{2-4} 
\\[-1.8ex] & \multicolumn{3}{c}{Log Error} \\ 
& Mean (OLS) & Median (Q50) & Tail Risk (Q80) \\ 
\\[-1.8ex] & (1) & (2) & (3)\\ 
\hline \\[-1.8ex]
 Constant & 7.548$^{***}$ (0.714) & 0.018 (0.019) & 3.170$^{**}$ (1.138) \\ 
  Comments Allowed & $-$0.191 (0.265) & $-$0.001 (0.003) & $-$0.000 (0.032) \\ 
  Duration: 6 Rounds & $-$1.817$^{**}$ (0.639) & $-$0.001 (0.005) & $-$0.057 (0.177) \\ 
  Duration: 9 Rounds & $-$0.627 (0.677) & $-$0.001 (0.005) & $-$0.028 (0.165) \\ 
  Strategic Prompt & 0.053 (0.487) & 0.001 (0.003) & 0.000 (0.113) \\ 
  Medium (t3s110) & 2.200$^{***}$ (0.573) & $-$0.000 (0.003) & 0.038 (1.808) \\ 
  Hard (t3s111) & 5.327$^{***}$ (0.730) & 0.286$^{**}$ (0.106) & 12.470$^{.}$ (7.451) \\ 
  Very Hard (t3s111o2ye2) & 6.658$^{***}$ (0.713) & 0.700$^{***}$ (0.084) & 21.771$^{***}$ (4.744) \\ 
  Initial Error: Low & 0.014 (0.493) & 0.000 (0.004) & 0.000 (0.069) \\ 
  Initial Error: Medium & 0.225 (0.508) & $-$0.000 (0.003) & $-$0.000 (0.063) \\ 
  Average Intelligence & $-$0.221$^{***}$ (0.018) & $-$0.0004 (0.0004) & $-$0.069$^{**}$ (0.025) \\ 
  Intelligence SD & 0.044 (0.060) & $-$0.0003 (0.0003) & $-$0.052$^{*}$ (0.026) \\ 
  6 Rounds x Strategic & $-$0.067 (0.639) & $-$0.001 (0.004) & 0.000 (0.128) \\ 
  9 Rounds x Strategic & 0.076 (0.676) & $-$0.001 (0.004) & 0.028 (0.161) \\ 
  Medium Struct x Initial Error (Low) & 1.309 (0.844) & $-$0.000 (0.005) & 0.019 (3.595) \\ 
  Hard Struct x Initial Error (Low) & 0.504 (1.039) & 0.032 (0.124) & 19.889 (12.267) \\ 
  Very Hard x Initial Error (Low) & 2.861$^{**}$ (1.050) & 0.013 (0.167) & 10.425$^{*}$ (5.186) \\ 
  Medium Struct x Initial Error (Medium) & $-$0.725 (0.803) & 0.000 (0.004) & 0.019 (1.810) \\ 
  Hard Struct x Initial Error (Medium) & $-$0.236 (1.047) & 0.077 (0.149) & $-$9.945 (12.141) \\ 
  Very Hard x Initial Error (Medium) & $-$0.337 (1.022) & $-$0.007 (0.083) & 10.080 (9.342) \\ 
 \hline \\[-1.8ex] 
Robust F-Statistic & 12.84$^{***}$ (df = 19; 1752) &  &  \\ 
Observations & 1,772 & 1,772 & 1,772 \\ 
R$^{2}$ & 0.115 &  &  \\ 
\end{tabular}%
}
\parbox{0.95\linewidth}{\textit{Notes:} $^{.}$p$<$0.1; $^{*}$p$<$0.05; $^{**}$p$<$0.01; $^{***}$p$<$0.001. Q50 and Q80 are quantile regressions. Q50 models the median log error, whereas Q80 models the upper 20\% tail of the log error distribution (the worst-performing markets). OLS model uses HC1 robust standard errors. Quantile models use bootstrapped standard errors. All three models use a mixed-contrast specification. Rounds and Structure use treatment contrasts with Round 3 and Structure t3s111y2 as baselines. All other controls (Initial Error, Strategic, Comments) use sum contrasts, where coefficients represent deviations from the grand mean.}
\end{table}

\begin{table}[!htbp] \centering 
  \caption{Trading Volume and Profits Without Disclosure (HC1 and CR1 Standard Errors)} 
  \label{sa:tab:trading volume and profits robust} 
\small 
\begin{tabular}{@{\extracolsep{5pt}}lcc} 
\\[-1.8ex]\hline 
\hline \\[-1.8ex] 
 & \multicolumn{2}{c}{\textit{Dependent variable:}} \\ 
\cline{2-3} 
 \\[-1.8ex] & Trading Volume & Individual Profits \\ 
\\[-1.8ex] & (1) & (2)\\ 
\hline \\[-1.8ex] 
 Constant & 1,795.988$^{***}$ (126.550) & $-$80.719$^{***}$ (10.174) \\ 
  Comments Allowed & 85.077$^{*}$ (41.093) & 2.723 (3.479) \\ 
  Duration: 6 Rounds & 543.398$^{***}$ (83.763) & 2.447 (8.182) \\ 
  Duration: 9 Rounds & 1,037.835$^{***}$ (106.092) & $-$5.396 (8.482) \\ 
  Strategic Prompt & $-$65.396 (53.805) & $-$0.491 (5.302) \\ 
  Medium (t3s110) & 308.391$^{**}$ (114.608) & $-$30.548$^{***}$ (7.264) \\ 
  Hard (t3s111) & 128.183 (119.400) & $-$69.955$^{***}$ (8.990) \\ 
  Very Hard (t3s111o2ye2) & 55.372 (118.947) & $-$98.904$^{***}$ (9.678) \\ 
  Initial Error: Low & $-$180.641 (115.824) & $-$10.767$^{*}$ (5.326) \\ 
  Initial Error: Medium & $-$14.704 (120.252) & $-$4.386 (5.892) \\ 
  Average Intelligence & $-$27.925$^{***}$ (2.657) & $-$11.019$^{***}$ (1.423) \\ 
  Individual Intelligence &  & 14.079$^{***}$ (1.402) \\ 
  Intelligence SD & 22.347$^{***}$ (6.159) & 0.606 (0.638) \\ 
  Trades Second &  & $-$42.127$^{**}$ (13.506) \\ 
  Trades Third &  & 77.390$^{***}$ (10.973) \\ 
  6 Rounds x Strategic & 9.810 (83.760) & $-$1.725 (8.181) \\ 
  9 Rounds x Strategic & 87.983 (106.100) & $-$3.680 (8.474) \\ 
  Medium Struct x Initial Error (Low) & 273.357$^{.}$ (159.892) & $-$21.173$^{.}$ (11.208) \\ 
  Hard Struct x Initial Error (Low) & 87.880 (165.985) & $-$13.015 (13.145) \\ 
  Very Hard x Initial Error (Low) & 414.759$^{*}$ (166.635) & $-$32.672$^{*}$ (13.956) \\ 
  Medium Struct x Initial Error (Med) & 41.568 (162.442) & 4.394 (10.453) \\ 
  Hard Struct x Initial Error (Med) & 62.984 (172.088) & 11.491 (12.328) \\ 
  Very Hard x Initial Error (Med) & $-$238.720 (161.232) & $-$1.778 (14.261) \\ 
 \hline \\[-1.8ex] 
Robust F-Statistic & 14.158$^{***}$ (df = 19; 1752) & 21.293$^{***}$ (df = 22; 5293) \\ 
Observations & 1,772 & 5,316 \\ 
R$^{2}$ & 0.104 & 0.060 \\ 
\end{tabular} 
\parbox{0.95\linewidth}{\textit{Notes:} $^{.}$p$<$0.1; $^{*}$p$<$0.05; $^{**}$p$<$0.01; $^{***}$p$<$0.001. Model 1 (market-level) utilizes HC1 heteroskedasticity-consistent robust standard errors. Model 2 (trader-level) utilizes CR1 cluster-robust standard errors clustered at the market level to account for intra-market profit correlation across competing agents. Rounds, Information Structure, and Trader Position (Model 2 only) use treatment contrasts, with `Round 3', `t3s111y2' (Easy), and `Trades First' serving as their respective baselines. All other controls use sum contrasts, representing deviations from the grand mean.}
\end{table}

%

\begin{table}[!htbp] \centering 
  \caption{Information Aggregation With Disclosure (HC1 Standard Errors)} 
  \label{sa:tab:disclosure_quantile_regressions_robust} 
\small 
\resizebox{\linewidth}{!}{%
\begin{tabular}{@{\extracolsep{5pt}}lccc} 
\\[-1.8ex]\hline 
\hline \\[-1.8ex] 
 & \multicolumn{3}{c}{\textit{Dependent variable:}} \\ 
\cline{2-4} 
\\[-1.8ex] & \multicolumn{3}{c}{Log Error} \\ 
 & Mean (OLS) & Median (Q50) & Tail Risk (Q80) \\ 
\\[-1.8ex] & (1) & (2) & (3)\\ 
\hline \\[-1.8ex] 
 Constant & 7.196$^{***}$ (0.678) & 0.018$^{*}$ (0.008) & 5.638$^{***}$ (0.954) \\ 
  Comments Allowed & $-$0.318 (0.194) & $-$0.001 (0.001) & $-$0.014 (0.038) \\ 
  Duration: 6 Rounds & $-$1.550$^{***}$ (0.469) & $-$0.001 (0.001) & $-$0.107 (0.112) \\ 
  Duration: 9 Rounds & $-$0.328 (0.492) & $-$0.001 (0.001) & $-$0.019 (0.086) \\ 
  Strategic Prompt & 0.007 (0.353) & 0.001 (0.001) & 0.007 (0.064) \\ 
  Medium (t3s110) & 2.772$^{***}$ (0.436) & $-$0.00001 (0.001) & 0.075 (0.971) \\ 
  Hard (t3s111) & 5.222$^{***}$ (0.528) & 0.206$^{*}$ (0.093) & 11.809$^{.}$ (6.577) \\ 
  Very Hard (t3s111o2ye2) & 7.192$^{***}$ (0.521) & 0.697$^{***}$ (0.024) & 20.126$^{***}$ (5.325) \\ 
  Initial Error: Low & 0.123 (0.379) & 0.00000 (0.001) & 0.008 (0.046) \\ 
  Initial Error: Medium & 0.234 (0.386) & $-$0.00001 (0.001) & $-$0.003 (0.048) \\ 
  Experiment Disclosure & 2.075$^{*}$ (0.929) & $-$0.00004 (0.008) & 0.980 (4.638) \\ 
  Average Intelligence & $-$0.218$^{***}$ (0.018) & $-$0.0004$^{*}$ (0.0002) & $-$0.123$^{***}$ (0.021) \\ 
  Intelligence SD & $-$0.012 (0.043) & $-$0.0003$^{**}$ (0.0001) & $-$0.102$^{***}$ (0.020) \\ 
  6 Rounds x Strategic & $-$0.083 (0.469) & $-$0.001 (0.001) & $-$0.021 (0.093) \\ 
  9 Rounds x Strategic & 0.182 (0.492) & $-$0.001 (0.001) & 0.014 (0.084) \\ 
  Medium Struct x Init. Error (Low) & 0.657 (0.631) & $-$0.00000 (0.001) & 0.003 (1.897) \\ 
  Hard Struct x Init. Error (Low) & 0.120 (0.751) & 0.045 (0.097) & $-$8.149 (11.199) \\ 
  Very Hard x Init. Error (Low) & 1.665$^{*}$ (0.757) & 0.008 (0.047) & 9.638$^{*}$ (4.721) \\ 
  Medium Struct x Init. Error (Med) & $-$0.399 (0.619) & 0.00001 (0.001) & 0.006 (0.957) \\ 
  Hard Struct x Init. Error (Med) & 0.187 (0.765) & $-$0.077 (0.130) & 18.273 (11.427) \\ 
  Very Hard x Init. Error (Med) & $-$0.321 (0.746) & $-$0.004 (0.024) & 8.923 (6.687) \\ 
  Disclosure x Av. Intelligence & $-$0.053$^{.}$ (0.028) & 0.00000 (0.0002) & $-$0.023 (0.109) \\ 
 \hline \\[-1.8ex] 
Robust F-Statistic & 23.324$^{***}$ (df = 21; 3478) &  &  \\ 
Observations & 3,500 & 3,500 & 3,500 \\ 
R$^{2}$ & 0.119 &  &  \\ 
\end{tabular}%
}
\parbox{0.95\linewidth}{\textit{Notes:} $^{.}$p$<$0.1; $^{*}$p$<$0.05; $^{**}$p$<$0.01; $^{***}$p$<$0.001. Q50 and Q80 are quantile regressions. Q50 models the median log error, whereas Q80 models the upper 20\% tail of the log error distribution (the worst-performing markets). OLS model uses HC1 robust standard errors. Quantile models use bootstrapped standard errors. All three models use a mixed-contrast specification. Rounds, Structure, and Experiment Disclosure use treatment contrasts with Round 3, Structure t3s111y2, and No Disclosure as baselines. All other controls (Initial Error, Strategic, Comments) use sum contrasts, where coefficients represent deviations from the grand mean.}
\end{table}

\begin{table}[!htbp] \centering 
  \caption{Trading Volume and Agent Profitability With Disclosure (HC1 and CR1 Standard Errors)} 
  \label{sa:tab:disclosure trading volume and profits robust} 
\small 
\begin{tabular}{@{\extracolsep{5pt}}lcc} 
\\[-1.8ex]\hline 
\hline \\[-1.8ex] 
 & \multicolumn{2}{c}{\textit{Dependent variable:}} \\ 
\cline{2-3} 
 \\[-1.8ex] & Trading Volume & Individual Profits \\ 
\\[-1.8ex] & (1) & (2)\\ 
\hline \\[-1.8ex] 
 Constant & 1,757.638$^{***}$ (111.920) & $-$77.335$^{***}$ (9.855) \\ 
  Comments Allowed & 36.269 (29.192) & 3.552 (2.501) \\ 
  Duration: 6 Rounds & 563.399$^{***}$ (59.237) & 3.359 (5.635) \\ 
  Duration: 9 Rounds & 1,080.922$^{***}$ (74.985) & $-$12.727$^{*}$ (6.256) \\ 
  Strategic Prompt & $-$37.028 (37.713) & $-$0.345 (3.769) \\ 
  Medium (t3s110) & 316.783$^{***}$ (80.555) & $-$35.852$^{***}$ (5.418) \\ 
  Hard (t3s111) & 212.977$^{*}$ (84.670) & $-$71.160$^{***}$ (6.670) \\ 
  Very Hard (t3s111o2ye2) & 43.866 (83.509) & $-$101.644$^{***}$ (6.698) \\ 
  Initial Error: Low & $-$145.678$^{.}$ (79.281) & $-$12.168$^{**}$ (4.122) \\ 
  Initial Error: Medium & 44.294 (87.228) & $-$5.193 (4.473) \\ 
  Experiment Disclosure & 171.986 (133.412) & $-$21.052 (13.080) \\ 
  Average Intelligence & $-$28.182$^{***}$ (2.715) & $-$11.057$^{***}$ (1.052) \\ 
  Individual Intelligence &  & 14.091$^{***}$ (1.019) \\ 
  Intelligence SD & 24.293$^{***}$ (4.342) & 1.154$^{*}$ (0.458) \\ 
  Trades Second &  & $-$39.316$^{***}$ (9.748) \\ 
  Trades Third &  & 74.604$^{***}$ (8.064) \\ 
  6 Rounds x Strategic & $-$26.283 (59.237) & $-$0.524 (5.634) \\ 
  9 Rounds x Strategic & 45.471 (74.986) & $-$5.343 (6.253) \\ 
  Medium Struct x Initial Error (Low) & 166.553 (111.310) & $-$9.165 (7.934) \\ 
  Hard Struct x Initial Error (Low) & 69.433 (116.184) & $-$6.278 (9.572) \\ 
  Very Hard x Initial Error (Low) & 349.973$^{**}$ (117.514) & $-$18.174$^{.}$ (9.513) \\ 
  Medium Struct x Initial Error (Med) & $-$28.194 (117.353) & 2.954 (7.820) \\ 
  Hard Struct x Initial Error (Med) & 93.975 (125.327) & 2.619 (9.608) \\ 
  Very Hard x Initial Error (Med) & $-$252.898$^{*}$ (116.905) & 1.676 (9.699) \\ 
  Disclosure x Average Intelligence & $-$3.394 (3.917) & 0.493 (0.394) \\ 
 \hline \\[-1.8ex] 
Robust F-Statistic & 27.855$^{***}$ (df = 21; 3478) & 38.592$^{***}$ (df = 24; 10475) \\ 
Observations & 3,500 & 10,500 \\ 
R$^{2}$ & 0.110 & 0.059 \\ 
\end{tabular} 
\parbox{0.95\linewidth}{\textit{Notes:} $^{.}$p$<$0.1; $^{*}$p$<$0.05; $^{**}$p$<$0.01; $^{***}$p$<$0.001. Model 1 (Market Level) reports HC1 heteroskedasticity-consistent robust standard errors. Model 2 (Individual Level) reports CR1 cluster-robust standard errors clustered at the market level to account for intra-market correlation among competing traders. Rounds, Information Structure, Experiment Disclosure, and Trader Position use treatment contrasts, with `Round 3', `t3s111y2', No Disclosure, and `Trades First' serving as their respective baselines. All other controls (Initial Error, Strategic, Comments) use sum contrasts, where coefficients represent deviations from the grand mean. Both specifications report adjusted robust Wald F-statistics.}
\end{table}


\begin{table}[!htbp] 
\centering 
\small
\caption{Agent Communication Strategy: Text Similarity and Word Gap (CR1 Standard Errors)} 
\label{sa:tab:combined_communication_models} 
\begin{tabular}{@{\extracolsep{5pt}}lcc} 
\\[-1.8ex]\hline 
\hline \\[-1.8ex] 
 & \multicolumn{2}{c}{\textit{Dependent variable:}} \\ 
\cline{2-3} 
 & Cosine Similarity (OLS) & Word Gap (OLS) \\ 
\\[-1.8ex] & (1) & (2)\\ 
\hline \\[-1.8ex] 
 Constant & 0.324$^{***}$ (0.009) & 33.880$^{***}$ (2.495) \\ 
  Strategic Prompt & $-$0.003 (0.003) & 0.788 (0.769) \\ 
  Round 2 & 0.030$^{***}$ (0.004) & $-$13.194$^{***}$ (1.129) \\ 
  Round 3 & 0.004 (0.004) & $-$25.105$^{***}$ (1.228) \\ 
  Round 4 & 0.030$^{***}$ (0.005) & $-$12.247$^{***}$ (1.284) \\ 
  Round 5 & 0.040$^{***}$ (0.005) & $-$23.619$^{***}$ (1.272) \\ 
  Round 6 & 0.005 (0.005) & $-$30.972$^{***}$ (1.318) \\ 
  Round 7 & 0.038$^{***}$ (0.006) & $-$18.893$^{***}$ (1.426) \\ 
  Round 8 & 0.047$^{***}$ (0.006) & $-$26.708$^{***}$ (1.455) \\ 
  Round 9 & 0.033$^{***}$ (0.006) & $-$32.644$^{***}$ (1.436) \\ 
  Duration: 6 Rounds & 0.002 (0.007) & $-$0.468 (2.072) \\ 
  Duration: 9 Rounds & $-$0.002 (0.007) & $-$1.072 (2.077) \\ 
  Medium (t3s110) & $-$0.022$^{**}$ (0.007) & $-$1.213 (2.021) \\ 
  Hard (t3s111) & 0.002 (0.007) & 2.417 (2.204) \\ 
  Very Hard (t3s111o2ye2) & 0.001 (0.007) & 5.666$^{*}$ (2.264) \\ 
  Experiment Disclosure & 0.010$^{.}$ (0.005) & 14.858$^{***}$ (1.543) \\ 
  Initial Error & 0.002 (0.016) & 2.867 (4.688) \\ 
  Individual Intelligence & 0.001$^{***}$ (0.0002) & 0.844$^{***}$ (0.043) \\ 
 \hline \\[-1.8ex] 
Observations & 10,620 & 10,620 \\ 
R$^{2}$ & 0.027 & 0.156 \\ 
F Statistic (df = 17; 10602) & 17.560$^{***}$ & 115.223$^{***}$ \\ 
\end{tabular}

\vspace{1ex} 
\parbox{0.95\linewidth}{\textit{Notes:} $^{.}$p$<$0.1; $^{*}$p$<$0.05; $^{**}$p$<$0.01; $^{***}$p$<$0.001. All models use a mixed-contrast specification. Treatment contrast baselines are: Round 1 for Round, 3 Rounds for Duration, Easy (t3s111y2) for Structure, and No Disclosure for Experiment Disclosure. The Strategic Prompt variable uses sum contrasts, representing the deviation from the grand mean, whereas Initial Error is mean-centered. Both models report CR1 cluster-robust standard errors in parentheses, clustered at the market level to account for intra-market correlation across competing agents and repeated communication rounds.}
\end{table}

\begin{table}[!htbp] \centering 
\small
  \caption{Robustness Check: Cosine Similarity by Market Duration (CR1 Standard Errors)} 
  \label{sa:tab:split_sim_models} 
\resizebox{\linewidth}{!}{%
\begin{tabular}{@{\extracolsep{5pt}}lccc} 
\\[-1.8ex]\hline 
\hline \\[-1.8ex] 
 & \multicolumn{3}{c}{\textit{Dependent variable:}} \\ 
\cline{2-4} 
\\[-1.8ex] & \multicolumn{3}{c}{Cosine Similarity  (Public, Private)} \\ 
 & 3-Round & 6-Round & 9-Round \\ 
\\[-1.8ex] & (1) & (2) & (3)\\ 
\hline \\[-1.8ex] 
 Constant & 0.331$^{***}$ (0.015) & 0.311$^{***}$ (0.012) & 0.329$^{***}$ (0.012) \\ 
  Strategic Prompt & $-$0.004 (0.005) & $-$0.003 (0.004) & $-$0.003 (0.004) \\ 
  Round 2 & 0.025$^{***}$ (0.006) & 0.033$^{***}$ (0.007) & 0.032$^{***}$ (0.007) \\ 
  Round 3 & 0.005 (0.007) & 0.005 (0.007) & 0.003 (0.007) \\ 
  Round 4 &  & 0.029$^{***}$ (0.007) & 0.032$^{***}$ (0.007) \\ 
  Round 5 &  & 0.042$^{***}$ (0.007) & 0.039$^{***}$ (0.007) \\ 
  Round 6 &  & 0.009 (0.007) & 0.003 (0.007) \\ 
  Round 7 &  &  & 0.039$^{***}$ (0.007) \\ 
  Round 8 &  &  & 0.047$^{***}$ (0.007) \\ 
  Round 9 &  &  & 0.032$^{***}$ (0.007) \\ 
  Medium (t3s110) & $-$0.017 (0.014) & $-$0.015 (0.012) & $-$0.029$^{**}$ (0.011) \\ 
  Hard (t3s111) & $-$0.006 (0.015) & 0.012 (0.012) & $-$0.001 (0.011) \\ 
  Very Hard (t3s111o2ye2) & 0.001 (0.015) & 0.005 (0.012) & $-$0.001 (0.011) \\ 
  Experiment Disclosure & 0.004 (0.010) & 0.011 (0.009) & 0.011 (0.008) \\ 
  Initial Error & 0.009 (0.031) & 0.001 (0.025) & 0.001 (0.024) \\ 
  Individual Intelligence & 0.001$^{*}$ (0.0004) & 0.001$^{***}$ (0.0003) & 0.001$^{**}$ (0.0003) \\ 
 \hline \\[-1.8ex] 
Observations & 1,776 & 3,552 & 5,292 \\ 
R$^{2}$ & 0.015 & 0.034 & 0.029 \\ 
Adjusted R$^{2}$ & 0.010 & 0.030 & 0.026 \\ 
F Statistic & 2.919$^{**}$ (df = 9; 1766) & 10.269$^{***}$ (df = 12; 3539) & 10.390$^{***}$ (df = 15; 5276) \\ 
\end{tabular}%
}
\parbox{0.95\linewidth}{\textit{Notes:} $^{.}$p$<$0.1; $^{*}$p$<$0.05; $^{**}$p$<$0.01; $^{***}$p$<$0.001. CR1 cluster-robust standard errors are reported in parentheses, clustered at the market level to account for intra-market correlation across multiple messaging rounds.  The model uses a mixed-contrast specification. Treatment contrast baselines are: Round 1 for Round, Easy (t3s111y2) for Structure, and No Disclosure for Experiment Disclosure. The Strategic Prompt variable uses sum contrasts, representing the deviation from the grand mean, whereas Initial Error is mean-centered.}
\end{table} 

\begin{table}[!htbp] \centering 
\small
  \caption{Robustness Check: Word Gap by Market Duration (CR1 Standard Errors)} 
  \label{sa:tab:split_duration_models} 
\resizebox{\linewidth}{!}{%
\begin{tabular}{@{\extracolsep{5pt}}lccc} 
\\[-1.8ex]\hline 
\hline \\[-1.8ex] 
 & \multicolumn{3}{c}{\textit{Dependent variable:}} \\ 
\cline{2-4} 
\\[-1.8ex] & \multicolumn{3}{c}{Word Gap} \\ 
 & 3-Round Markets & 6-Round Markets & 9-Round Markets \\ 
\\[-1.8ex] & (1) & (2) & (3)\\ 
\hline \\[-1.8ex] 
 Constant & 31.127$^{***}$ (4.020) & 35.468$^{***}$ (3.612) & 32.393$^{***}$ (3.234) \\ 
  Strategic Prompt & 0.682 (1.533) & 0.114 (1.282) & 1.282 (1.173) \\ 
  Round 2 & $-$13.385$^{***}$ (2.072) & $-$12.328$^{***}$ (1.942) & $-$13.802$^{***}$ (1.876) \\ 
  Round 3 & $-$24.149$^{***}$ (2.185) & $-$25.950$^{***}$ (2.128) & $-$24.945$^{***}$ (2.082) \\ 
  Round 4 &  & $-$12.730$^{***}$ (1.861) & $-$11.833$^{***}$ (1.884) \\ 
  Round 5 &  & $-$23.211$^{***}$ (2.012) & $-$24.144$^{***}$ (1.943) \\ 
  Round 6 &  & $-$31.712$^{***}$ (2.154) & $-$30.467$^{***}$ (1.984) \\ 
  Round 7 &  &  & $-$18.861$^{***}$ (1.731) \\ 
  Round 8 &  &  & $-$26.700$^{***}$ (1.945) \\ 
  Round 9 &  &  & $-$32.712$^{***}$ (1.966) \\ 
  Medium (t3s110) & $-$2.959 (4.020) & $-$3.839 (3.378) & 1.157 (3.080) \\ 
  Hard (t3s111) & 0.958 (4.548) & $-$0.100 (3.755) & 4.582 (3.296) \\ 
  Very Hard (t3s111o2ye2) & 3.865 (4.419) & 4.459 (3.765) & 7.073$^{*}$ (3.474) \\ 
  Experiment Disclosure & 12.640$^{***}$ (3.068) & 15.753$^{***}$ (2.582) & 14.981$^{***}$ (2.353) \\ 
  Initial Error & 2.330 (9.170) & 2.005 (7.834) & 3.618 (7.168) \\ 
  Individual Intelligence & 1.050$^{***}$ (0.085) & 0.813$^{***}$ (0.071) & 0.796$^{***}$ (0.066) \\ 
 \hline \\[-1.8ex] 
Observations & 1,776 & 3,552 & 5,292 \\ 
R$^{2}$ & 0.143 & 0.152 & 0.159 \\ 
Adjusted R$^{2}$ & 0.139 & 0.149 & 0.157 \\ 
F Statistic & 32.824$^{***}$ (df = 9; 1766) & 52.842$^{***}$ (df = 12; 3539) & 66.700$^{***}$ (df = 15; 5276) \\ 
\end{tabular}%
}
\parbox{0.95\linewidth}{\textit{Notes:} $^{.}$p$<$0.1; $^{*}$p$<$0.05; $^{**}$p$<$0.01; $^{***}$p$<$0.001. CR1 cluster-robust standard errors are reported in parentheses, clustered at the market level to account for intra-market correlation across multiple messaging rounds. The model uses a mixed-contrast specification. Treatment contrast baselines are: Round 1 for Round, Easy (t3s111y2) for Structure, and No Disclosure for Experiment Disclosure. The Strategic Prompt variable uses sum contrasts, representing the deviation from the grand mean, whereas Initial Error is mean-centered.}
\end{table}

\pagebreak 

\clearpage
\section{Prompts}
\label{sa:sec: prompts}

In this section, we provide excerpts from the prompts that are given to the AI agents. The following describes the public and private information that is provided to each agent in the four structures. In the Supplementary appendix we provide the full prompt for one market.

\subsection{t3s111y2} 

Public Information: "The prediction market question has two answers: Yes and No. There are three relevant dimensions to this prediction market question: dimension d\_a: Will sales in country A exceed 1 million? dimension d\_b: Will sales in country B exceed 1 million? dimension d\_c: Will sales in country C exceed 1 million? The answer to each dimension is either true, with probability 0.5, or false, with probability 0.5. Dimensions are independent, hence the probability of a dimension d resolving to true is independent of whether the other dimensions resolve to true or false. In summary, there are eight states of the world, depending on whether d\_a, d\_b, or d\_c are true or false. The answer to the prediction market question of whether the profits of Company X will exceed 1 million is Yes if at least two dimensions (d\_a,d\_b,d\_c) resolve to true. If more than one dimension resolves to false, then the answer to the question is No. There are three traders in the market. trader\_1 is privately informed whether d\_a is true or not, Trader\_2 is privately informed whether d\_b is true or not, and Trader\_3 is privately informed whether d\_c is true or not. All three traders assign the same prior probabilities to each dimension resolving to true or false."

Private Information:
    trader\_1: "A true state has now occurred. You (trader\_1) are now informed truthfully and privately that d\_a is true."
    trader\_2: "A true state has now occurred. You (trader\_2) are now informed truthfully and privately that d\_b is true." 
    trader\_3: "A true state has now occurred. You (trader\_3) are now informed truthfully and privately that d\_c is true."

\subsection{t3s111}

Public Information: The prediction market question has two answers: Yes and No. There are three relevant dimensions to this prediction market question: dimension d\_a: Will sales in country A exceed 1 million? dimension d\_b: Will sales in country B exceed 1 million? dimension d\_c: Will sales in country C exceed 1 million? The answer to each dimension is either true, with probability 0.5, or false, with probability 0.5. Dimensions are independent, hence the probability of a dimension d resolving to true is independent of whether the other dimensions resolve to true or false. In summary, there are eight states of the world, depending on whether d\_a, d\_b, or d\_c are true or false. The answer to the prediction market question of whether the profits of Company X will exceed 1 million is Yes if all three dimensions (d\_a,d\_b,d\_c) resolve to true. If at least one dimension resolves to false, then the answer to the question is No. There are three traders in the market. Trader\_1 is privately informed whether d\_a is true or not, Trader\_2 is privately informed whether d\_b is true or not, and Trader\_3 is privately informed whether d\_c is true or not. All three traders assign the same prior probabilities to each dimension resolving to true or false.

Private Information:
    trader\_1: "A true state has now occurred. You (trader\_1) are now informed truthfully and privately that d\_a is true."
    trader\_2: "A true state has now occurred. You (trader\_2) are now informed truthfully and privately that d\_b is true." 
    trader\_3: "A true state has now occurred. You (trader\_3) are now informed truthfully and privately that d\_c is true."
    
\subsection{t3s110}
    
Public Information: "The prediction market question has two answers: Yes and No. There are three relevant dimensions to this prediction market question: dimension d\_a: Will sales in country A exceed 1 million? dimension d\_b: Will sales in country B exceed 1 million? dimension d\_c: Will sales in country C exceed 1 million? The answer to each dimension is either true, with probability 0.5, or false, with probability 0.5. Dimensions are independent, hence the probability of a dimension d resolving to true is independent of whether the other dimensions resolve to true or false. In summary, there are eight states of the world, depending on whether d\_a, d\_b, or d\_c are true or false. The answer to the prediction market question of whether the profits of Company X will exceed 1 million is Yes if all three dimensions (d\_a,d\_b,d\_c) resolve to true. If at least one dimension resolves to false, then the answer to the question is No. There are three traders in the market. Trader\_1 is privately informed whether d\_a is true or not, Trader\_2 is privately informed whether d\_b is true or not, and Trader\_3 is privately informed whether d\_c is true or not. All three traders assign the same prior probabilities to each dimension resolving to true or false."

Private Information:
trader\_1: "A true state has now occurred. You (trader\_1) are now informed truthfully and privately that d\_a is true."
trader\_2: "A true state has now occurred. You (trader\_2) are now informed truthfully and privately that d\_b is true." 
trader\_3: "A true state has now occurred. You (trader\_3) are now informed truthfully and privately that d\_c is false."

\subsection{t3s111o2ye2}
    
      Public Information: "The prediction market question has two answers: Yes and No. There are three relevant dimensions to this prediction market question: dimension d\_a: Will sales in country A exceed 1 million? dimension d\_b: Will sales in country B exceed 1 million? dimension d\_c: Will sales in country C exceed 1 million? The answer to each dimension is either true, with probability 0.5, or false, with probability 0.5. Dimensions are independent, hence the probability of a dimension d resolving to true is independent of whether the other dimensions resolve to true or false. In summary, there are eight states of the world, depending on whether d\_a, d\_b, or d\_c are true or false. The answer to the prediction market question of whether the profits of Company X will exceed 1 million is Yes if exactly two of the three dimensions (d\_a,d\_b,d\_c) resolve to true. If all three dimensions resolve to true, or at least two dimensions resolve to false, then the answer to the question is No. There are three traders in the market. Trader\_1 is privately informed whether d\_b and d\_c are true or not, Trader\_2 is privately informed whether d\_a and d\_c are true or not, and Trader\_3 is privately informed whether d\_a and d\_b are true or not. All three traders assign the same prior probabilities to each dimension resolving to true or false."

Private Information:
    trader\_1: "A true state has now occurred. You (trader\_1) are now informed truthfully and privately that d\_b and d\_c are true."
    trader\_2: "A true state has now occurred. You (trader\_2) are now informed truthfully and privately that d\_a and d\_c are true." 
    trader\_3: "A true state has now occurred. You (trader\_3) are now informed truthfully and privately that d\_a and d\_b are true."

\subsection{Full prompt in a round}

The following is part of the report that was generated for market with slug `5gw55w', describing the prompt which is sent to the trader in Round 4, their actions and the execution of the trades. The complete report for this market, containing every prompt, every private and public message, and every trade, is included in the replication package.  

Note that the public and private information, as well as the explanation of what is a prediction market, are repeated in every round, as a  ``new'' LLM is invoked in every round. This is similar to the case where a user chats with an LLM.  Since the LLM has no memory, it reads the entire transcript of the conversation every time it is called to answer. The prompt also includes a history of prices and public messages, the private messages of the previous iterations of the same trader, a calculation of the price impact from various trades, and a report of the current portfolio of the trader.

\includepdf[pages={16-20}]{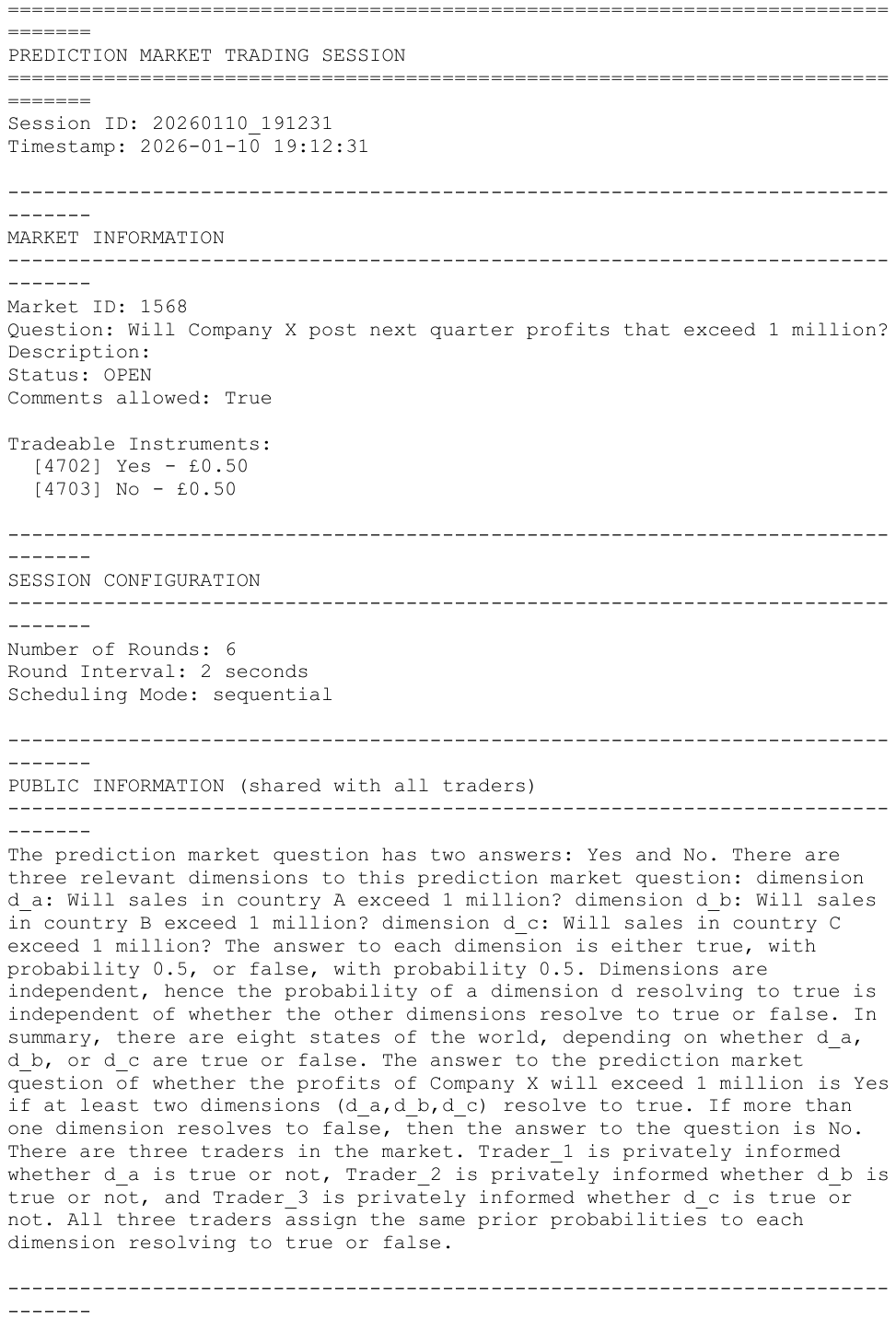}

\end{document}